\documentclass[twocolumn,reprint,floatfix,amsmath,amssymb,rmp]{revtex4}
\usepackage[T1]{fontenc}\usepackage{ae} \usepackage{aecompl}
\usepackage{graphicx,dcolumn,bm,color,comment,dsfont,amsmath,ulem,epsfig}

\renewcommand{\Re}{{\rm Re}}
\renewcommand{\Im}{{\rm Im}}
\newcommand{\ri}{{\rm i}}
\newcommand{\rd}{{\rm d}}

\newcommand{\kb}{k_{\rm B}}
\newcommand{\alphamat}{\underline{\underline{\alpha}}}

\newcommand{\chimat}{\underline{\underline{\chi}}}
\newcommand{\blockt}{\boldsymbol{T}^{-1}}
\newcommand{\I}{{\rm i}}

\def\bbm[#1]{\mbox{\boldmath $#1$}}
\newcommand{\ket}[1]{\displaystyle{|#1\rangle}}
\newcommand{\bra}[1]{\displaystyle{\langle #1|}}

\newcommand{\TT}{\mathbb{T}}

\newcommand{\GG}{\mathbb{G}}

\newcommand{\trace}[1]{{\operatorname{Tr}} \left[ #1 \right]}

\begin{document}

\preprint{APS/123-QED}

\title{Near-field Radiative Heat Transfer in Many-Body Systems}

\author{S.-A. Biehs}
 \affiliation{Institut f\"{u}r Physik, Carl von Ossietzky Universit\"{a}t,
D-26111 Oldenburg, Germany.}
\author{R. Messina}
\affiliation{Laboratoire Charles Fabry, UMR 8501, Institut d'Optique, CNRS, Universit\'{e} Paris-Saclay, 2 Avenue Augustin Fresnel, 91127 Palaiseau Cedex, France.}
\author{P. S. Venkataram}
\affiliation{Department of Electrical Engineering, Princeton University, Princeton, New Jersey 08544, USA.}
\author{A. W. Rodriguez}
\affiliation{Department of Electrical Engineering, Princeton University, Princeton, New Jersey 08544, USA.}
\author{J. C. Cuevas}
\affiliation{Departamento de F\'{\i}sica Te\'{o}rica de la Materia Condensada and Condensed Matter Physics Center (IFIMAC), Universidad Aut\'{o}noma de Madrid, E-28049 Madrid, Spain}
\author{P. Ben-Abdallah}
\affiliation{Laboratoire Charles Fabry, UMR 8501, Institut d'Optique, CNRS, Universit\'{e} Paris-Saclay, 2 Avenue Augustin Fresnel, 91127 Palaiseau Cedex, France.}
\affiliation{Universit\'{e} de Sherbrooke, Department of Mechanical Engineering, Sherbrooke, PQ J1K 2R1, Canada.}
\email{pba@institutoptique.fr}

\begin{abstract}
Many-body physics aims to understand emergent properties of systems made of many interacting objects. 
This article reviews recent progress on the topic of radiative heat transfer in many-body systems consisting of thermal
emitters interacting in the near-field regime. Near-field radiative heat transfer is a rapidly emerging 
field of research in which the cooperative behavior of emitters gives rise to peculiar effects which 
can be exploited to control heat flow at the nanoscale. Using an extension of the standard Polder and van 
Hove stochastic formalism to deal with thermally generated fields in $N$-body systems, along with their mutual
interactions through multiple scattering, a generalized Landauer-like theory is derived to describe 
heat exchange mediated by thermal photons in arbitrary reciprocal and non-reciprocal multi-terminal 
systems. In this review, we use this formalism to address both transport and dynamics in these systems 
from a unified perspective. Our discussion covers: (i) the description of non-additivity of heat 
flux and its related effects, including fundamental limits as well as the role of nanostructuring and 
material choice, (ii) the study of equilibrium states and multistable states, (iii) the relaxation 
dynamics (thermalization) toward local and global equilibria, (iv) the analysis of heat transport 
regimes in ordered and disordered systems comprised of a large number of objects, density and range 
of interactions, and (v) the description of thermomagnetic effects in magneto-optical systems and heat transport mechanisms in non-Hermitian many-body systems. We conclude this review by listing outstanding 
challenges and promising future research directions.
\end{abstract}

\maketitle

\tableofcontents

\section{Introduction}

Heat transfer in a given system is in its simplest sense (i.e. ignoring multiple irreversible transport 
processes~\cite{Onsager1931}) thermal energy in transit due to a spatial temperature difference \cite{Bergman2011}. There 
are three basic heat transfer modes: {\it conduction, convection}, and {\it radiation}. In the case of 
a stationary medium, which could be a solid or a fluid, {\it conduction} refers to heat transfer through 
local agitation of atoms or charges that occurs across the medium in response to a temperature gradient. 
Ultimately, the carriers responsible for heat conduction are phonons, molecular vibrations or electrons/ions 
in the case of electrical conductors. The second mode of transport is {\it convection}, and refers to 
heat transfer that occurs between a surface and a moving fluid when they are at different temperatures 
(or by advection inside the fluid itself). Finally, the third heat transfer mechanism is {\it thermal 
radiation}, which is the topic of this review. All bodies at a finite temperature emit energy in the form 
of electromagnetic waves (or photons). Hence, even in the absence of an intervening medium, there is always 
heat transfer via thermal radiation between bodies at different temperatures. This makes thermal radiation 
one of the most ubiquitous physical phenomena and its understanding of critical importance for many different 
areas of science and engineering \cite{Modest2013,Howell2016,Zhang2007}.

\begin{figure}
\includegraphics[width=0.6\columnwidth,clip]{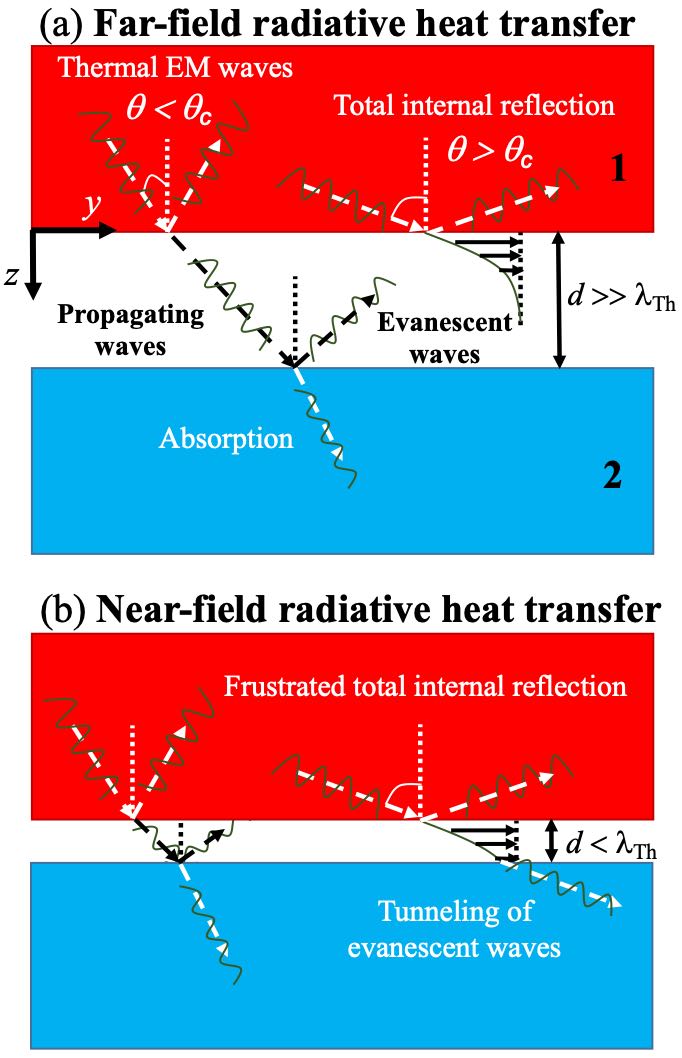}
\caption{\label{fig-NFRHT} (a) Far-field radiative heat transfer between two infinite parallel plates (media 1 and 2) separated by a vacuum gap. In this scenario, the gap size $d$ is much larger than the thermal wavelength, $\lambda_{\rm Th}$, and the two plates exchange heat only via propagating waves. The evanescent waves generated in the vacuum gap by total internal reflection are not able to reach the second plate and do not contribute to the heat transfer. (b) When $d < \lambda_{\rm Th}$ the tunneling of evanescent waves can give a significant contribution to the radiative heat transfer and in this way the Planckian (or black-body) limit can be greatly overcome in this near-field regime.}
\end{figure}

Traditionally, our understanding of thermal radiation is based on Planck's law \cite{Planck1914}, which 
establishes that a black body (an object that absorbs all the radiation that impinges on it) emits thermal 
radiation following a broadband distribution that only depends on the body's temperature. Planck's law 
provides a unified description of a variety of thermal radiation phenomena and, in particular, it sets an 
upper limit (Stefan-Boltzmann's law) for the radiative heat transfer (RHT) between bodies. However, Planck's 
law was derived using ray optics and hence, it is expected to fail when the spatial dimensions in a thermal 
problem are smaller than or comparable to the thermal wavelength $\lambda_\mathrm{Th}$ defined by Wien's 
displacement law ($\sim$10 $\mu$m at room temperature)~\cite{Planck1914}. In particular, Planck's law 
fails to describe RHT between objects separated by distances $\lesssim \lambda_{\rm Th}$ 
\cite{JP1999,Volokitin2007}. In this {\it near-field} regime, RHT can be dominated by evanescent waves 
(or photon tunneling), not taken into account in Planck's law, and the Planckian (or black-body) limit 
can be greatly overcome by bringing objects sufficiently close, see Fig.~\ref{fig-NFRHT}. This phenomenon 
was first predicted within the rigorous framework of fluctuational electrodynamics (FE) \cite{Rytov1989} 
by Polder and Van Hove in the early 1970s \cite{Polder1971}, see Sec.~\ref{sec-two-body}. This near-field 
radiative heat transfer (NFRHT) enhancement was first hinted in several experiments in the late 1960s
\cite{Hargreaves1969,Domoto1970}, but it was not firmly confirmed until the 2000s
\cite{Kittel2005,Hu2008,Narayanaswamy2008,Rousseau2009,Shen2009}. Since then, numerous experiments exploring 
different aspects of NFRHT have been reported and they have boosted the field of thermal radiation 
\cite{Ottens2011,Shen2012,Kralik2012,Zwol2012a,Zwol2012b,Guha2012,Worbes2013,Shi2013,SGetal2014,
Song2015b,Kim2015,Lim2015,KIetal2015,St-Gelais2016,Song2016,Bernardi2016,Langetal2017,KIetal2017,TKetal2017,
Ghashami2018,Fiorino2018a,AFetal2018,Fiorino2018c,DeSutter2019,VMetal2019}. These experiments have, in turn,
generated hope that NFRHT may have an impact on different technologies such as heat-assisted
magnetic recording, thermal lithography, scanning thermal microscopy, coherent thermal sources, 
near-field based thermal management, thermophotovoltaics, and other energy conversion devices,
see~\cite{Basu2009,Song2015a,Cuevas2018,SK2019} and references therein.

In parallel to these experimental advances, over the last two decades, there has been a huge amount 
of theoretical activity. Initially, attention was devoted to the importance of choice of materials and the 
elucidation of the different mechanisms of near-field thermal radiation. In that regard, polar 
dielectrics exhibiting polaritonic resonances that lead to surface modes have played a 
prominent role in this field \cite{Mulet2002}. Then, following nanophotonics concepts, 
a lot of work has been devoted to assess the possibility of further enhancing NFRHT and to 
tune its spectral properties by using nanostructures such as thin films and multilayer 
systems~\cite{Volokitin2007,Biehs2007a,Biehs2007b,Francoeur2008,SIMetal2013,PbaetalJAP2009}, 
photonic crystals and gratings~\cite{Ben-Abdallah2010,Biehs2011,Rodriguez2011,Guerout2012,
Messina2017}, and metasurfaces~\cite{Dai2016,Liu2015,Fernandez-Hurtado2017}. The 
investigation of the use of metamaterials for further enhancing NFRHT
\cite{Biehs2011,SABetal2012,Guo2012,JoulainetalPRB2010} or low-dimensional materials like graphene 
or phosphorene to tune NFRHT~\cite{Volokitin2011,VBSetAl2012,Ilic2012,PRetal2015,XLL2014,YZ2018,XLL2019} 
has also attracted significant attention. Another topic of great importance has been the study of the
active control of NFRHT by different means, including the use of phase-transition materials
\cite{Zwol2011a,Zwol2011b,FMetal2016}, the application of an external magnetic field
\cite{MoncadaEtAl2015}, or the regulation of chemical potentials for photons with an external
bias \cite{Chen2015}. There are also several theoretical proposals for functional
devices that make use of NFRHT for thermal management \cite{Otey2010,PBAandSAB2015},
thermophotovoltaics~\cite{AN2003,Laroche2006,Basu2007,Zhao2017}, and other energy applications 
\cite{Chen2015,Chen2016}. On a more fundamental level, quantum approaches based on the 
Huttner-Barnett model, quantum Langevin equations, non-equilibrium Green's function method, and 
the master-equation approach for open quantum systems have been 
proposed~\cite{MJetal2003,SAB2013,KSetal2014,GB2016,JSW2017,KS2018}.

From a broader perspective, a new general picture of RHT has emerged in recent years with
profound similarities to other heat and charge transport phenomena, including phonon
conduction in nanoscale systems and coherent electronic transport in mesoscopic 
devices \cite{Cuevas2017}. In particular, RHT is now routinely described in terms of the
Landauer formula, originally proposed in the context of electronic mesoscopic systems
\cite{Imry1999,Datta1997}, where the energy and charge transport are mainly determined by 
the transmission function describing the transfer probability of the carriers. Moreover, 
techniques employed to compute transmission functions (scattering approaches, 
Green's function techniques, etc.) are conceptually very similar in all those contexts. 
This connection between RHT and conduction allows us not only to profit from the experience 
in other fields, but can also serve as the starting point for a unified description of
different heat transfer modes in situations where different types of carriers may compete 
or even interfere. An example of this type of situation is realized in the context of the 
heat transfer in subnanometer gaps where recent experiments have reported conflicting
observations in an intermediate regime where the contribution of different carriers 
(photons, phonons, and electrons) may be comparable \cite{Cui2017a,Kloppstech2017}. 
While the situation seems to be clear in the limiting cases where either conduction
\cite{Cui2017b,Mosso2017,Cui2019} or NFRHT \cite{Kim2015} are clearly expected to 
dominate, the description of the crossover between them might require novel theories 
where different carriers are treated on an equal footing \cite{Chiloyan2015,VenkataramPRL2018}.

Conceptually speaking, a major advance in the field in the last decade has been 
the development of theoretical models of RHT in many-body systems, the central topic of 
this review. Such a theory deals with radiative heat exchange in systems composed of multiple
thermal emitters able to cooperatively interact. The collective behaviors in these systems 
give rise to singular phenomena that we discuss in the present manuscript. Until 2011, FE 
had been primarily used to describe RHT between two bodies, but the situation changed with
the report of the first version of a many-body theory of RHT describing a collection of small 
dipolar particles \cite{PBAetal2011}. Soon after, this many-body theory was generalized to deal 
with bodies of arbitrary size and shape \cite{RMandMA2011a,MKetal2012}, and new refinements 
of the theory are being constantly reported to deal with more complex optical
materials. Again, there is here a clear analogy with developments in mesoscopic physics, where B\"uttiker's
extension of the Landauer formalism to multi-terminal systems laid down the basis for the
understanding of numerous charge and energy transport phenomena in mesoscopic systems
\cite{Datta1997}. As we shall discuss in detail in this review, the many-body theory of 
NFRHT opened the door for predicting and analyzing a plethora of novel physical phenomena 
with no analogues in two-body systems. Thus, for instance, it became possible to explore
thermal analogues of intrinsic many-body phenomena like the Hall effect \cite{PBA2016}
or heat persistent current \cite{LZandSF2016}. It has also made it possible to propose a wide range
of thermal functional devices that are intrinsically many-body in nature, such as the thermal
transistor \cite{PBAandSAB2014}. This theory also allowed for the first time to understand 
the different heat propagation regimes in disordered systems involving large collection of
objects, and paved the way for hydrodynamic modelling of transport in these media. 
Although recent experimental works have explored the possibility to tune radiative heat
transfers in many-body systems \cite{Thompsonetal2020} by actively changing the relative
position of nearby objects, to our knowledge, many-body systems have yet to be
experimentally investigated in the purely near-field regime.

The field of NFRHT has been the subject of different reviews over the years. Thus, for instance, 
the reviews by \cite{Joulain2005} and \cite{Volokitin2007} covered the FE theory and basic concepts 
of NFRHT, but for obvious reasons do not include crucial theoretical and experimental 
advances in recent years. The reviews by \cite{Basu2007} and \cite{PBAZNA2019} focus on potential
applications of near-field thermal radiation in thermophotovoltaics. There are recent reviews like 
that of \cite{Song2015a} that already presents some of the most recent advances and, in particular, 
describes the main experimental techniques developed in recent years. The review by \cite{Cuevas2018}
provides an interesting and updated perspective of the field, but does not contain an in-depth 
description of theoretical developments. The present review article focuses on the theory of NFRHT in 
many-body systems, which has not been covered so far in a self-contained and unified framework. This topic 
is becoming a central focus of the field of thermal radiation, as it promises an 
entirely new generation of thermal radiation applications, and its understanding is likely to determine 
the future of RHT as a forefront research line.

The structure of the paper goes as follows. In
Sec.~\ref{sec-two-body}, we set the stage for this review by
discussing NFRHT in two-body systems. Here, we put the emphasis on the
modern view of NFRHT and review the most important theoretical
advances in this topic, as well as the experimental state of the
art. Specifically, we begin by briefly recalling the basics of the
theory of FE and then discuss its application to the important case of
two parallel plates (Sec.~\ref{sec-plate-plate}). This basic
configuration is used to illustrate the critical role of material
choice (Sec.~\ref{sec-metals}), including a preliminary discussion of
non-reciprocal materials in Sec.~\ref{sec-MO1}. Section~\ref{sec-Nano}
is devoted to the analysis of the role of nanostructuring in tailoring
and most importantly enhancing NFRHT, including recent works focused
on multilayer structures, photonic crystals, metamaterials, gratings,
metasurfaces, graphene sheets, and surface roughness. We then move
beyond planar structures in Sec.~\ref{sec-geometry} to discuss NFRHT
between objects of arbitrary size and shape. General-purpose numerical
methods developed so far for the description of NFRHT in arbitrary
geometries are then discussed in Sec.~\ref{sec-numerical}. We conclude this first part of the review in Sec.~\ref{sec-bounds} with an in-depth discussion of recently derived limits on the largest NFRHT rates that could ever be realized by an optimal choice of material and geometric configuration. Specifically, we highlight the prohibitive role that multiple scattering (a critical feature of many-body physics to be further discussed in subsequent sections) plays in limiting heat-transfer enhancements that may be achieved through nanostructuring, resulting in optimal flux rates not much larger than what is observed in planar polaritonic materials, at least in the context of two-body heat exchange.

Section III constitutes the bulk of this review and covers a great
variety of aspects of the theory of near-field thermal radiation in
many-body systems. We first discuss the problem of light absorption by
a set of non-emitting objects which collectively interact and show
that these systems can be treated as a whole with a dressed
susceptibility that takes into account both cooperative interactions
as well as the resonant response of individual objects. Next, a
generalized Landauer formula is derived to describe radiative heat
transfer in the general situation in which all objects are emitting,
using transmission coefficients describing the pairwise efficiency of
coupling between any two objects. Using this theoretical framework, we
highlight the singular aspects of heat transport in these systems
compared to those seen in two-body systems. We start to illustrate
these peculiarities in Secs.~\ref{Sec:NonAddDip} and
\ref{Sec:NonAddMac}, where we prove the non-additivity of heat flux, a
fundamental feature of these systems. We also show that $N$-body
interactions can amplify heat flux or lead to saturation mechanisms
close to the contact without the need to introduce non-locality in
material responsivity. In Sec.~\ref{steady sate}, we discuss
equilibrium conditions for any given system, and show that equilibrium
states are generally not unique and can be, along with their
stability, identified and characterized by standard perturbative
techniques. We also show that multistable systems can be exploited,
for instance, to make a boolean treatment of information with thermal
photons or build thermal self-oscillators.  In the subsequent section,
we address the problem of heat transport in various complex systems
using both a kinetic approach based on the approximate Boltzmann
transport equation for the resonant modes supported by the system, and
from a generalized Landauer theory that takes into account all modes
in the continuum. Several physical effects (radiative drag effect,
heat-flux focusing, heat pumping and long-range heat transport)
inherent to many-body systems are then introduced and discussed. In
Sec.~\ref{Sec:RelaxDyn}, we address the relaxation problem of
many-body systems and show that the temperature field can evolve at
different time scales, depending on the nature of
interactions. Furthermore, we discuss the current solutions proposed
to dynamically control the heat flux exchanged in these systems by
modulating either geometrical configuration, optical properties, or
via adiabatic control of their temperature. In Sec.~\ref{Sec:regimes},
we analyze various heat transport regimes in systems consisting of a
large number of objects, and show that RHT can be described as a
generalized random walk with a non-Gaussian probability distribution
function. Unlike what happens in solid-state physics for heat
conduction in bulk materials, we demonstrate the existence of
anomalous heat transport regimes and highlight that these regimes
closely depend on the system dimension, drastically changing from
dilute to dense systems. The next few sections are devoted to
non-reciprocal systems. Unlike reciprocal systems, in these
non-Hermitian systems the classical notion of Lorentz reciprocity is
violated, giving rise to specific heat-transfer mechanisms. After
extending in Secs.~\ref{non-reciprocal framework} and \ref{MONnano}
the theoretical framework to deal with heat exchange, we discuss in
Secs.~\ref{magnetoresistance} several thermomagnetic effects
(magnetoresistance, permanent currents, Hall effect) that take place
in magneto-optical systems and we underline in Sec.~\ref{supercurent
  and spin} the link between these effects and the topological
structure of the Poynting field. We also stress in
Sec.~\ref{rectification} the potential of these systems to efficiently
tune the direction of heat flow. Finally, we conclude this review by
listing outstanding challenges and a broader outlook of potential
future research directions.

\section{Two-body systems} \label{sec-two-body}

\begin{figure}
\includegraphics[width=\columnwidth,clip]{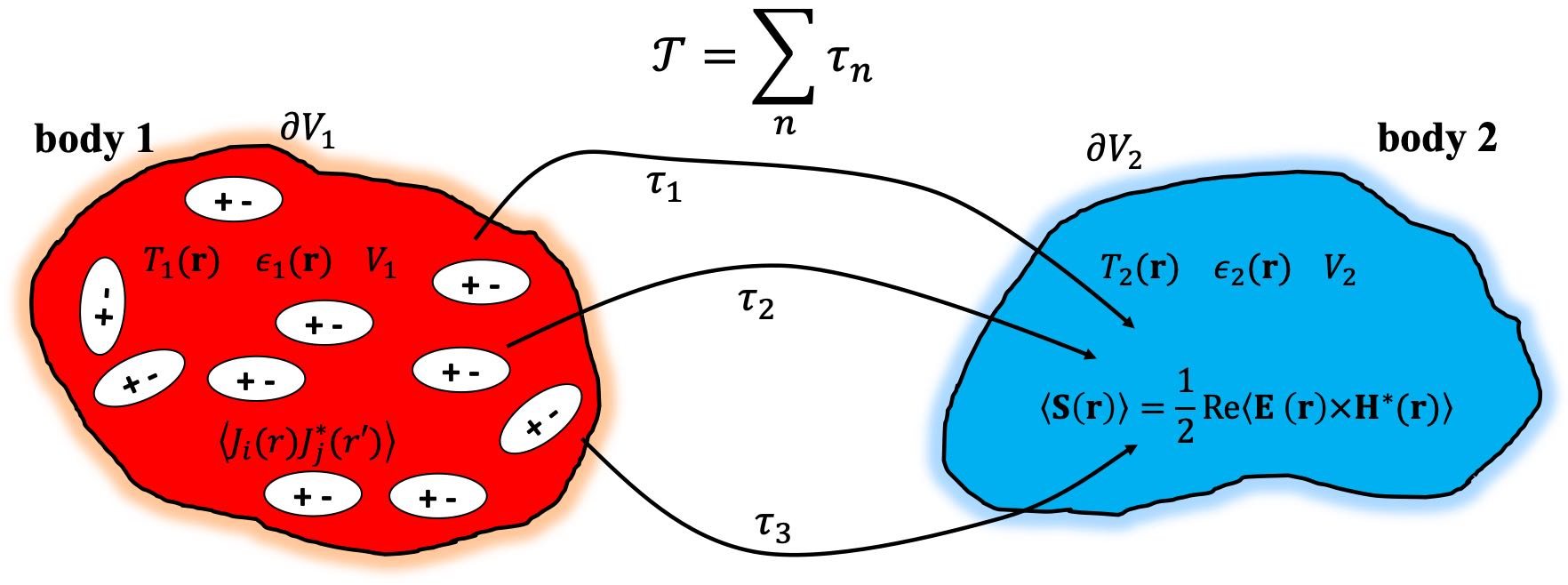}
\caption{\label{fig-FE} Fluctuational electrodynamics: Schematic of radiative heat transfer in a two-body system. The two bodies of volumes $V_1$ and $V_2$ have temperature profiles $T_1({\bf r})$ and $T_2({\bf r})$ and frequency-dependent dielectric functions $\epsilon_1({\bf r},\omega)$ and $\epsilon_2({\bf r},\omega)$. Electromagnetic fields ${\bf E}$ and ${\bf H}$ are generated by the random currents ${\bf J}$ in the bodies due to their non-vanishing correlations given by the fluctuation-dissipation theorem. The net power exchanged by the two bodies is determined by the total transmission ${\cal T}$ that can be expressed as a sum of individual transmission coefficients $\tau_n$.}
\end{figure}

Most theoretical work on the topic of NFRHT is primarily based on Rytov's FE theory. Developed 
in the 1950s \cite{Rytov1989}, FE is a semiclassical theory which assumes that thermal radiation 
is generated by random, thermally activated electric currents inside the bodies. Thus, the technical 
problem in the description of RHT between different objects boils down to the solution of the 
stochastic Maxwell's equations, with random electric currents as radiation sources. To illustrate 
the idea, let us consider two optically isotropic and non-magnetic bodies separated by a vacuum gap, 
see Fig.~\ref{fig-FE}. In the framework of FE, the RHT problem is completely specified 
by the temperature distributions $T_i({\bf r})$ ($i=1,2$) and the dielectric functions 
of the materials, $\epsilon_i({\bf r},\omega)$. The macroscopic Maxwell's equations 
to be solved adopt the following form in the frequency domain
\begin{eqnarray}
  \nabla \times {\bf E}({\bf r},\omega) & = & i \omega \mu_0 {\bf H}({\bf r},\omega), \\
  \nabla \times {\bf H}({\bf r},\omega) & = & -i \omega \epsilon_0 \epsilon({\bf r},\omega) 
  {\bf E}({\bf r},\omega) +  {\bf J}({\bf r},\omega) ,
\end{eqnarray}
where ${\bf E}$ and ${\bf H}$ are the electric and magnetic fields, 
${\bf r}$ is the position vector, and $\epsilon_0$ and $\mu_0$ are the vacuum 
permittivity and permeability, respectively. In the second equation, the fluctuating 
current density distributions ${\bf J}({\bf r},\omega)$ within the bodies are the sources 
of the thermal radiation. The statistical average of these currents vanishes, i.e., 
$\langle {\bf J} \rangle = 0$, but their correlations are given by the 
fluctuation-dissipation theorem~\cite{WE1984,Rytov1989,Joulain2005}
\begin{equation}
\begin{split}
\label{eq-FDT}
  \langle \mathbf{J}({\bf r},\omega) \otimes \mathbf{J}^{\ast}({\bf r}^{\prime},\omega) 
  \rangle & =
  \frac{4 \hbar \omega^2 \epsilon_0}{\pi} \mbox{Im} \{\epsilon({\bf r},\omega) \}  
    \\
  &\quad \times n(\omega,T({\bf r})) \delta({\bf r}-{\bf r}^{\prime}) ,
\end{split}
\end{equation}
where $\hbar$ is the Planck constant and $n(\omega,T) = 1/(\exp[\hbar 
\omega/k_{\rm B}T]-1)$ is the Bose function. In simple terms, the calculation of the
radiative power exchanged by bodies 1 and 2 is done by first solving the Maxwell equations
with the appropriate boundary conditions defined by geometries of the bodies and assuming 
that the random electric currents occupy the whole body 1. Then, with the solution for the
fields around body 2, the statistical average of the Poynting vector is computed: 
$ \langle {\bf S({\bf r},\omega)} \rangle = \mbox{Re} \langle {\bf E}({\bf r},\omega) 
\times {\bf H}({\bf r},\omega) \rangle/2$. Finally, the results are integrated over
frequency and over a closed surface enclosing body 2. Of course, to evaluate the net RHT, one 
needs to calculate in a similar way the heat transferred from body 2 to body 1.

This innocent-looking problem is, however, quite challenging in general, and analytical
solutions are only known in a handful of situations. One of the main goals of the rest of 
this section is to present the solution in cases of increasing complexity focusing on 
two-body systems. Let us say at this stage that, as mentioned in the introduction, the 
net power, $P_\mathrm{net}$, exchanged via thermal radiation between two objects of 
(homogeneous) temperatures $T_1$ and $T_2$ can always be expressed via means of the 
Landauer formula, as one can easily understand with the following heuristic argument. 
The net radiative power is the balance between the heat power transferred from one body 
to the other: $P_\mathrm{net} = P_{1 \to 2} - P_{2 \to 1}$, where the individual 
contributions are given by
\begin{equation}
  \mathcal{P}_{i \to j}  =  \int^{\infty}_0 \frac{d\omega}{2\pi} \hbar \omega n(\omega, T_i) {\cal T}_{ji}(\omega).
\end{equation}
Here, $\hbar \omega$ is the energy of an electromagnetic mode of frequency $\omega$ and the Bose function $n(\omega,T)$ is describing the thermal occupation of that mode, and 
${\cal T}_{ji}(\omega)$ is the total transmission coefficient
that correspond to the sum of the probabilities over all the modes of frequency $\omega$ 
that can be transferred from body $i$ to body $j$. In the case of a two-body system 
(with no environment), detailed balance imposes that ${\cal T}_{21}(\omega)$ = 
${\cal T}_{12}(\omega) = {\cal T}(\omega)$ and the expression of the net power reduces 
to the celebrated Landauer formula~\cite{Polder1971,PBAKJ2010,SABJJ2010}
\begin{equation}
  \label{eq-Landauer}
  \mathcal{P}_\mathrm{net} = \int^{\infty}_0 \frac{d\omega}{2\pi} \hbar \omega 
  \left[ n(\omega, T_1) - n(\omega, T_2) \right] {\cal T}(\omega) .
\end{equation}
Following the spirit of the Landauer approach in mesoscopic physics, the total transmission 
can be analyzed in terms of radiation channels and it can be expressed as 
\begin{equation}
  \label{eq-channels}
  {\cal T}(\omega) = \sum_n \tau_n(\omega) ,
\end{equation} 
where the $\tau$'s are the individual transmission probabilities of the different open channels 
(bounded between 0 and 1). This point is particularly useful to establish simple upper bounds for 
RHT, as we shall discuss later in this review.

\subsection{Parallel plates} \label{sec-plate-plate}

As mentioned in the introduction, the importance of the contribution of evanescent waves 
in the RHT between two objects and the possibility to overcome the Planckian limit in the
near-field regime was first put forward by Polder and van Hove \cite{Polder1971}. These
authors calculated the NFRHT rate between two infinite parallel plates, a geometry that has 
become the workhorse of NFRHT and that is schematically represented in Fig.~\ref{fig-NFRHT}.
We shall refer to the upper plate as medium 1 and the lower plate as 2, and assume that 
they are at constant temperatures $T_1$ and $T_2$, respectively. In the case of optically
isotropic and nonmagnetic materials, Polder an Van Hove showed that the radiative power
per unit area, i.e.\ the heat flux $\Phi$, between the parallel plates is given by 
Eq.~(\ref{eq-Landauer}) with the following replacement of the transmission coefficient 
by a transmission coefficient per unit area:
\begin{equation}
  \label{eq-pp}
  {\cal T}(\omega) \longrightarrow \int^{\infty}_0 \frac{d \kappa}{2\pi} \; \kappa \, 
  \tau(\omega,\kappa,d) .
\end{equation}
Here, $\kappa = \sqrt{k^2_x+k^2_y}$ is the magnitude of the wave vector parallel to the
plates, see coordinate system in Fig.~\ref{fig-NFRHT}(a), $d$ is the gap size, and 
$\tau(\omega,\kappa,d)$ is the total (sum over polarizations) transmission probability of 
an electromagnetic mode of frequency $\omega$ and parallel wave vector $\kappa$. In the 
case of isotropic materials, this total transmission is equal $\tau(\omega,\kappa,d) = 
\tau_s(\omega,\kappa,d) + \tau_p(\omega,\kappa,d)$, where the contributions of $s$- and 
$p$-polarized waves (or alternatively TE- and TM-waves) are given by ($\alpha=s,p$)
\begin{equation}
\label{eq-trans-pp}
  \tau_{\alpha}(\omega, \kappa,d) = \left\{ \begin{array}{ll}
    \frac{(1 - |r^{\alpha}_{1}|^2) ( 1 - |r^{\alpha}_{2}|^2)}{|D^{\alpha}|^2},
     & \kappa < k_0 \\
    \frac{4 \mbox{Im}(r^{\alpha}_{1}) \mbox{Im}( r^{\alpha}_{2}) 
    e^{-2|q_\mathrm{v}|d} }{|D^{\alpha}|^2} , & \kappa > k_0 \, , 
    \end{array} \right.
\end{equation}
where $k_0 = \omega/c$ is the wavenumber in vacuum and $D^{\alpha} = 1 - r^{\alpha}_{1} 
r^{\alpha}_{2} e^{2 i q_\mathrm{v}d}$, $c$ is the speed of light, $q_\mathrm{v} = 
\sqrt{\omega^2/c^2 - \kappa^2}$ is the perpendicular component of the wave vector in the
vacuum gap, and $r^{\alpha}_{i}$ are Fresnel (or amplitude reflection) coefficients given by 
\begin{equation}
  r^{s}_{i} = \frac{q_\mathrm{v} - q_i}{q_\mathrm{v} + q_i}, \;\;\; 
  r^{p}_{i} = \frac{\epsilon_i q_\mathrm{v} - q_i}{\epsilon_i q_\mathrm{v} + q_i} .
\label{eq-Fresnel}
\end{equation}
Here, $\epsilon_i(\omega)$ is the dielectric function of medium $i=1,2$, assumed to only
depend on frequency (local media), and $q_i = \sqrt{\epsilon_i k_0^2 - \kappa^2}$.

The key point in this result is that the integral in Eq.~(\ref{eq-pp}) is carried out 
over all possible values of $\kappa$ and therefore, it includes the contribution of both
propagating waves ($\kappa < k_0$) and evanescent waves ($\kappa > k_0$). These latter 
ones are not taken into account in Stefan-Boltzmann's law. The contribution of the
evanescent waves decays exponentially with the gap size, see Eq.~(\ref{eq-trans-pp}), 
and it becomes negligible in the far-field regime ($d \gg \lambda_{\rm Th}$). However, 
in the near-field regime ($d <\lambda_{\rm Th}$) the contribution of evanescent waves, 
often referred to as photon tunneling, can become very significant and for sufficiently 
small gaps, it may completely dominate the heat transfer. The black-body result is obtained
from Eq.~(\ref{eq-pp}) by ignoring the evanescent waves and assuming perfect transmission 
for the propagating waves for all frequencies and wave vectors. In that case, the radiative
power per unit area is given by Stefan-Boltzmann's law: $\Phi_{\rm BB} = \sigma(T^4_1 - 
T^4_2)$, where $\sigma = 5.67 \times 10^{-8}$ W/(m$^2$K$^4$).

\subsection{Metals vs.\ dielectrics} \label{sec-metals}

\begin{figure}
\includegraphics[width=\columnwidth,clip]{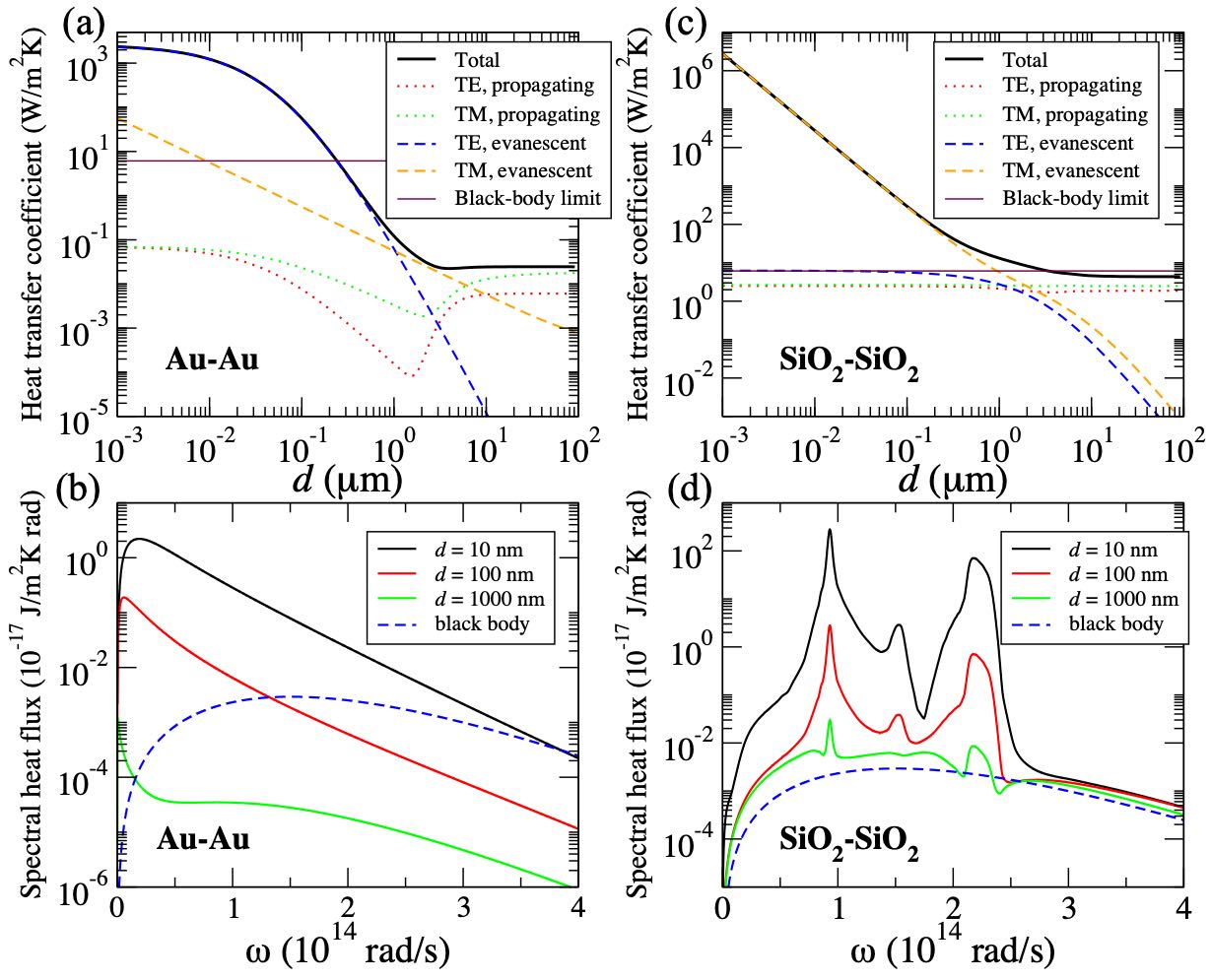}
\caption{\label{fig-plate-plate} (a) Heat transfer coefficient at room temperature (300 K) as a function of the gap size for two infinite parallel plates made of Au. The different lines correspond to the total contribution (black solid line) and to the contributions of propagating and evanescent waves for TE and TM polarizations. The horizontal line shows the result for two black bodies: 6.124 W/(m$^2$ K). (b) The spectral heat flux (or conductance per unit area and frequency) as a function of the radiation frequency corresponding to the case of panel (a). The solid lines correspond to three different values of the gap size in the near-field regime, while the blue dashed line is the result for two black bodies. (c,d) The same as in panels (a,b) for SiO$_2$.}
\end{figure}

The parallel-plate configuration allows us to illustrate not only the impact of evanescent waves in the near-field regime, but also the importance of the choice of materials.
There are two main classes of materials when it comes to NFRHT, namely metals (or related
materials with free carriers like doped semiconductors) and dielectrics (especially polar
dielectrics that exhibit polaritonic resonances like SiO$_2$, SiN, SiC, etc.). As an example
of the results for these two types of materials, we show in Fig.~\ref{fig-plate-plate}(a,c)
the gap dependence of the room-temperature heat-transfer coefficient, i.e.\ the radiative heat conductance per unit area, for two parallel plates made of Au and SiO$_2$. 
In those panels we also show the individual contributions of propagating and evanescent 
waves for TE and TM polarizations. Notice that in both cases the Planckian limit (indicated
with an horizontal line) is greatly overcome for sufficiently small gaps. This is
particularly remarkable in the silica case, where for $d=1$ nm the heat flux is almost 5
orders of magnitude larger than the black-body limit. Notice also that there are clear
differences between Au and SiO$_2$. For Au, the NFRHT rate is dominated by TE evanescent waves,
which originate from eddy currents inside the Au plates \cite{Chapuis2008}. This typically leads to a saturation of the heat transfer coefficient for small gaps. On the contrary, 
in the silica case, NFRHT is dominated by TM evanescent waves that can be shown to stem from 
surface phonon polaritons (SPhPs): quasiparticle excitations that arise from the strong coupling of electromagnetic fields with the 
optical phonon modes of polar dielectrics \cite{Mulet2002}. These surface electromagnetic 
waves are hybrid or cavity modes that reside in both plates and have a penetration depth that 
is on the order of the gap size~\cite{SBZM2009}, which implies that they are more and more 
confined to the surfaces as the gap is reduced \cite{Song2015b}. The increase of the density 
of the states of theses modes~\cite{SABJJ2010,PBAKJ2010} upon reducing the gap size is reflected 
in a characteristic $1/d^2$ dependence of the heat transfer coefficient for polar dielectrics.

Apart from enhancing NFRHT, evanescent waves are also responsible for a 
drastic modification of the spectral heat flux (or heat conductance per unit frequency), see Fig.~\ref{fig-plate-plate}(b,d). Thus, for instance, in the SiO$_2$ case 
the spectral heat flux is dominated by two peaks that appear at the frequencies of the 
optical modes of this polar dielectric. This is dramatically different as compared to 
the broadband Planck's distribution and it is also due to the fact that NFRHT in 
this case is dominated by SPhPs. 

\begin{figure}
\includegraphics[width=0.7\columnwidth,clip]{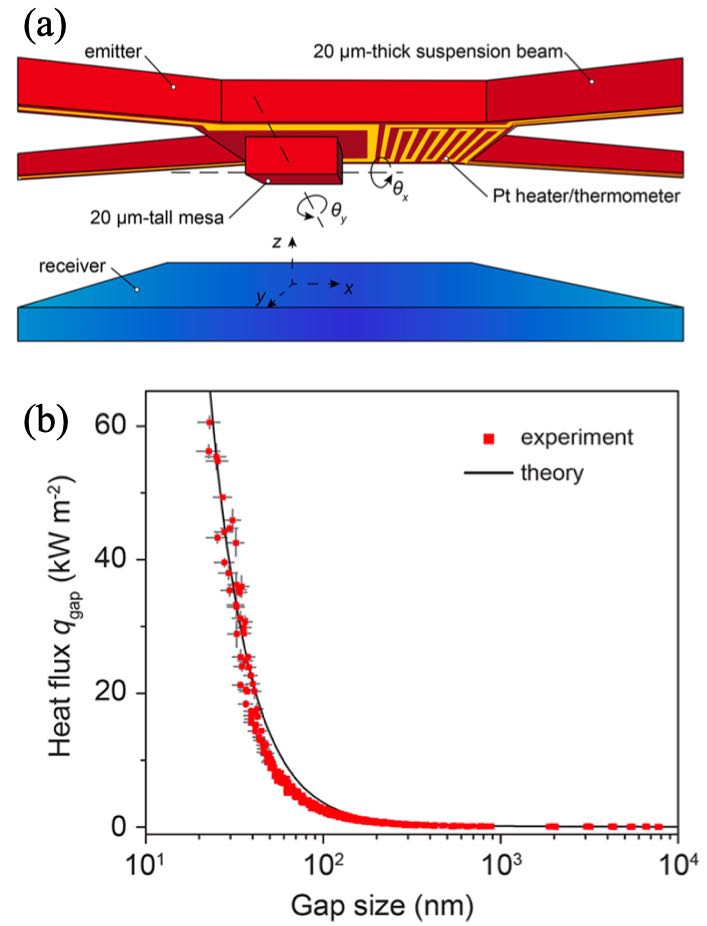}
\caption{(a) Schematic illustration of NFRHT measurement configuration used by \cite{Fiorino2018a}. The emitter microdevice is comprised of a square mesa and Pt heater/thermometer suspended on a thermally isolated island. The receiver is a macroscopically large (1 cm $\times$ 1 cm) plate. (b) The corresponding heat flux versus gap size in the case of an emitter and an receiver made of SiO$_2$. Measured data (red squares) is compared to the theoretical result (solid black line) obtained within FE. Reprinted with permission from \cite{Fiorino2018a}. Copyright 2018 ACS.}
\label{fig-NFRHT-exp}
\end{figure}

In principle, the plate-plate configuration discussed above is ideally suited 
to experimentally investigate NFRHT because some of the largest enhancements in this 
regime are expected to occur in this setting. However, this configuration is very
difficult to realize in practice because it is very complicated to achieve and maintain 
good parallelism between macroscopic plates at nanometer separations. In recent years
several groups have overcome this hurdle and have developed novel techniques to explore 
the plate-plate configuration in the near-field regime and they have been able to 
confirm the results of the FE theory. Some of those experiments have made use of 
macroscopic ($\sim$cm $\times$ cm) planar surfaces \cite{Hu2008,Ottens2011,Bernardi2016,
Ghashami2018,DeSutter2019}, while others are based on microscopic plates (50 $\mu$m $\times$
50 $\mu$m) \cite{SGetal2014,Song2016,St-Gelais2016,Fiorino2018a}. The use of macroscopic
planar surfaces is conceptually simple, but in practice it is more difficult to ensure the 
parallelism and to have clean and smooth surfaces over such large areas. For this 
reason, the smallest gaps achieved with this strategy are still above a hundred nanometers \cite{DeSutter2019}. On the other hand, the use of microdevices
facilitates the parallelization of the systems and the characterization of the surfaces. 
With this approach, it has become possible to explore gaps as small as 30 nm 
\cite{Fiorino2018a}, as we illustrate in Fig.~\ref{fig-NFRHT-exp}. In this 
example, a microdevice comprising a Pt resistor, which heats up the emitter and measures
its temperature, was used to measure the NFRHT rate between two SiO$_2$ surfaces down to
gaps of about 30 nm. For these tiny gaps, it was found that the heat conductance was 
about 1200 times larger than in the far-field regime and about 700 times larger than the
black-body limit, in excellent agreement with the theory results based on FE.

\subsection{Non-reciprocal materials} \label{sec-MO1}

A special class of materials that has attracted a lot of attention in the context of 
thermal radiation is that of non-reciprocal materials. These materials do not satisfy 
Lorentz reciprocity~\cite{CCetal2018} and, in practice, are optically anisotropic
materials with dielectric tensors which are non-symmetric. A paradigmatic example is that 
of magneto-optical (MO) materials where the non-reciprocity is induced either by an internal
magnetization like in ferromagnets or by an external magnetic field like in doped
semiconductors. Part of the attention is due to the suggestion that these materials might
violate Kirchhoff's law \cite{LZandSF2014}, which establishes the equality of thermal
emissivity and absorptivity. Although it has been shown that this is not case in a 
two-body situation (one body could be an environment) \cite{RMAEetal2017}, this class of
materials does give rise to countless novel thermal-radiation phenomena in the context of
many-body systems, as it will be amply discussed later in this review.

In the context of NFRHT in two-body non-reciprocal systems, most of the work so far has
focused on the analysis of MO materials and, in particular, on the study of the use of an
external magnetic field as a way to actively control thermal radiation. Special attention 
has been devoted to doped semiconductors, which in the presence of an external magnetic 
field exhibit a very strong MO activity in the infrared. The first theoretical study of 
this kind was reported by \cite{MoncadaEtAl2015} who analyzed the magnetic-field dependence 
of the heat-transfer coefficient of two parallel plates made of doped semiconductors 
(InSb or Si). These materials become optically anisotropic and non-reciprocal in the 
presence of an external magnetic field. Thus, the problem is to compute the RHT between
between two anisotropic parallel plates. This generic problem was addressed
by~\cite{GB2009,Biehs2011} and, similarly to the isotropic case discussed in 
Sec.~\ref{sec-plate-plate}, the net power per unit area or heat flux $\Phi$ is given 
by the Landauer formula of Eq.~(\ref{eq-Landauer}) with the substitution
\begin{equation}
\label{eq-plate-plate-MO}
   {\cal T}(\omega) \longrightarrow \int\!\! \frac{d \boldsymbol{\kappa}}{(2\pi)^2} 
   \tau(\omega,\boldsymbol{\kappa},d) .
\end{equation}
Here, $\boldsymbol{\kappa} = (k_x,k_y)^t$ is the wave vector parallel to the surface 
planes, and $\tau(\omega,\boldsymbol{\kappa},d)$ is the transmission probability of the
individual electromagnetic waves. Notice that the integral in Eq.~(\ref{eq-plate-plate-MO}) 
is now carried out over all possible directions of $\boldsymbol{\kappa}$ (the RHT is no 
longer isotropic) and, as usual, it includes the contribution of both propagating and
evanescent waves. The transmission coefficient $\tau(\omega,\boldsymbol{\kappa},d)$ can 
be expressed as
\begin{equation}
\begin{split}
  &\tau(\omega,\boldsymbol{\kappa},d) =  \\ & \begin{cases}
    \mbox{Tr} \left\{ [\mathds{1} - \mathds{R}_{1} \mathds{R}^{\dagger}_{1}] 
    \mathds{D}^{\dagger} [\mathds{1} - \mathds{R}^{\dagger}_{2} \mathds{R}_{2} ] 
    \mathds{D} \right\}, & \kappa < k_0 \\
    \mbox{Tr} \left\{ [\mathds{R}_{1} - \mathds{R}^{\dagger}_{1} ] \mathds{D}^{\dagger} 
    [\mathds{R}^{\dagger}_{2} - \mathds{R}_{2} ] \mathds{D} \right\} e^{-2|q_\mathrm{v}| d}, 
    & \kappa > k_0
   \end{cases} 
\end{split}
\label{eq-trans-MO}
\end{equation}
where the $2 \times 2$ matrices $\mathds{R}_{i}$ (with $i=1,2$) are the reflection matrices
characterizing the two interfaces. These matrices have the following generic structure
\begin{equation}
  \label{eq-refl-mat}
  \mathds{R}_{i} =  \begin{pmatrix} r^{ss}_{i} & r^{sp}_{i} \\ 
  r^{ps}_{i} & r^{pp}_{i} \end{pmatrix} ,
\end{equation}
where $r^{\alpha\beta}_{i}$ with $\alpha,\beta =s,p$ is the reflection amplitude for the
scattering of an incoming $\alpha$-polarized plane wave into an outgoing $\beta$-polarized
wave. In particular, the off-diagonal elements describe the polarization conversation, 
which does not occur for isotropic materials. Finally, the $2\times 2$ matrix $\mathds{D}$ 
in Eq.~(\ref{eq-trans-MO}) is defined as
\begin{equation}
  \mathds{D} = [ \mathds{1} - \mathds{R}_{1} \mathds{R}_{2} e^{2iq_\mathrm{v} d} ]^{-1}.
\end{equation}
The different reflection matrices appearing in Eq.~(\ref{eq-refl-mat}) can be computed 
within standard approaches for anisotropic multilayer systems.

\begin{figure}
  \includegraphics[width=0.8\columnwidth,clip]{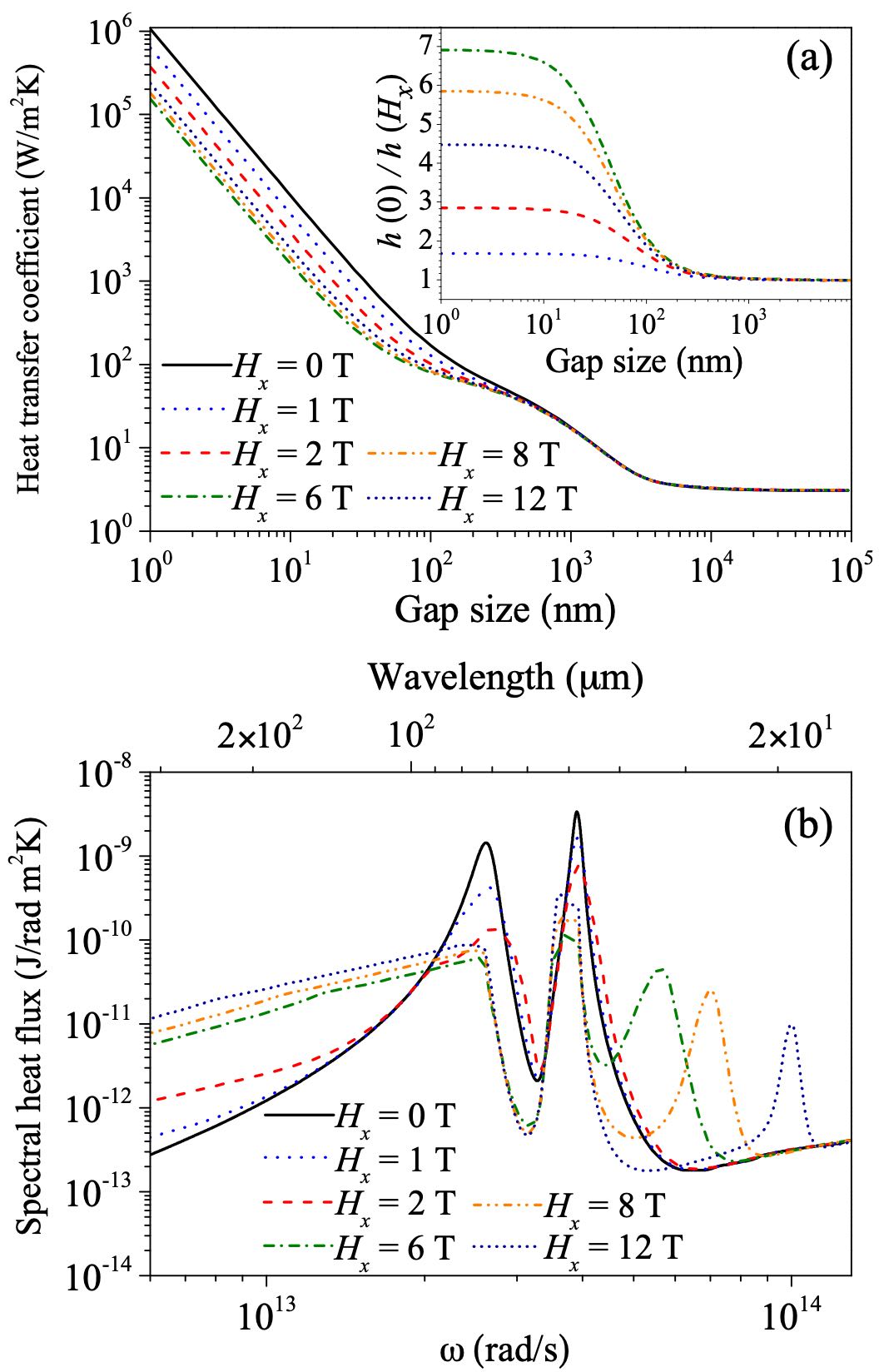}
  \caption{(a) Heat-transfer coefficient for two parallel plates made of $n$-doped InSb at 
  room temperature (300 K) as a function of the gap size for different values of a 
  magnetic field applied parallel to the surfaces of the plates ($x$-direction). The 
  inset shows the ratio between the zero-field coefficient and the coefficient for 
  different values of the field in the near-field region. (b) The corresponding spectral 
  heat flux as a function of the frequency (and wavelength) for a gap of $d=10$ nm and
  different values of the parallel field. Reprinted with permission from
  \cite{MoncadaEtAl2015}. Copyright 2015 American Physical Society.}
\label{fig-Hx}
\end{figure}

This formalism was used in \cite{MoncadaEtAl2015} to show that the NFRHT rate between two 
parallel plates made of doped InSb and Si can be strongly affected by the application 
of a static magnetic field, and relative changes of up to 700\% were predicted for fields 
of a few Teslas. These results are illustrated in Fig.~\ref{fig-Hx} for the case of a 
magnetic field oriented parallel to the plates. More recently, the same authors have 
also shown that NFRHT between two parallel plates made of MO materials can also be
modulated by simply changing the orientation of the external magnetic field 
\cite{Moncada-Villa2020}, which is the thermal analogue of well-known phenomenon of
anisotropic thermal magnetoresistance in the field of spintronics. This and other
thermomagnetic phenomena in the context small MO particles will be discussed in more 
detail in Sec.~\ref{Sec:NonReciprocal}.

\subsection{Nanostructuring and Roughness} \label{sec-Nano}

Following ideas and concepts of nanophotonics, many groups have explored nanostructuring 
as a strategy to further enhance NFRHT and to tune its spectral properties. In this 
subsection, we shall briefly review some of the ideas put forward in recent years in the 
context of NFRHT in nanostructured planar systems and also discuss the impact of
deviations from planarity.

\subsubsection{Multilayer structures and photonic crystals}

\begin{figure}
\includegraphics[width=0.9\columnwidth,clip]{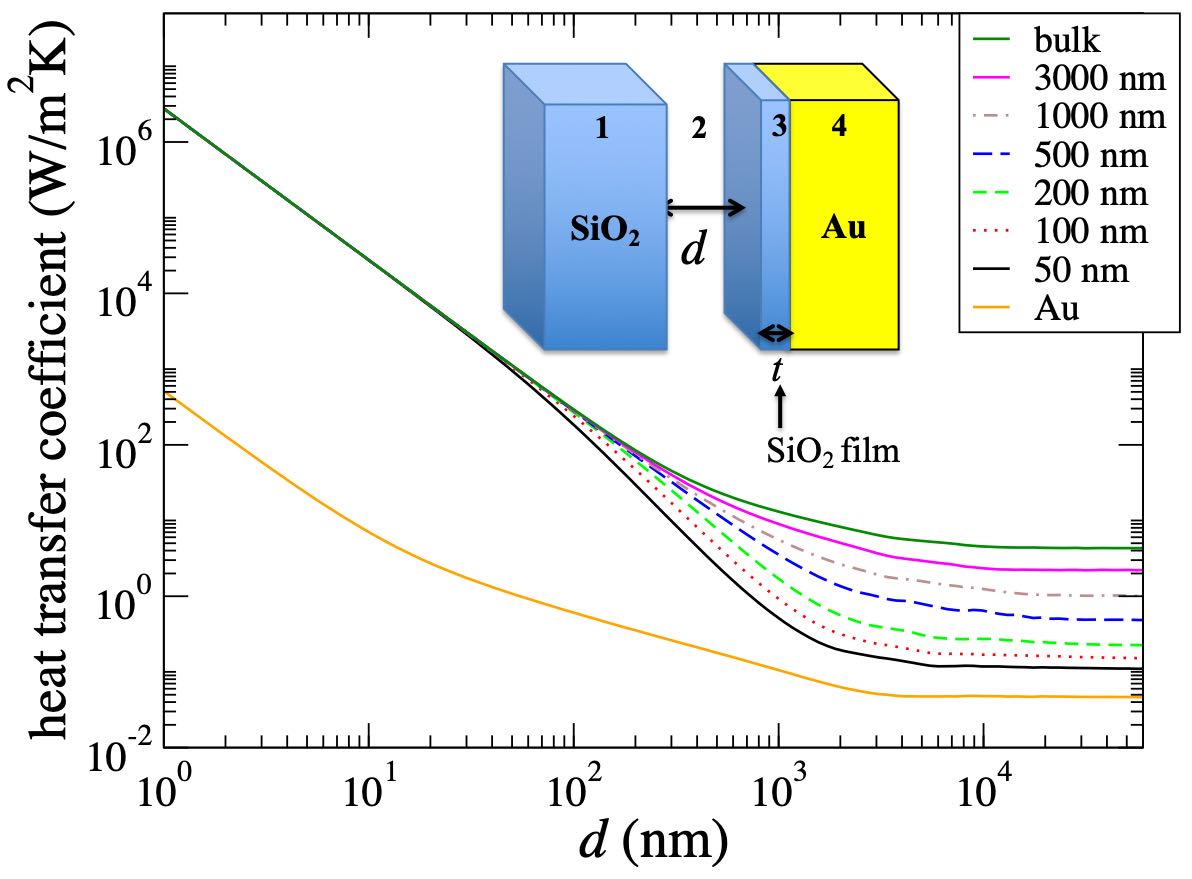}
\caption{Computed heat transfer coefficient as a function of gap size for the multilayer
  system shown in the inset at room temperature (300 K). This structure comprises a thick, 
  semi-infinite silica surface separated by a vacuum gap of size $d$ from a silica thin film 
  coating on a semi-infinite Au surface. Different curves correspond to different thicknesses 
  of the silica coating. Adapted from \cite{Song2015b}.}
\label{fig-thin-film}
\end{figure}

A natural extension of the plate-plate configuration discussed above is to replace 
the plates by planar multilayer structures or 1D photonic crystals 
\cite{Biehs2007a,Biehs2007b,Francoeur2008,PbaetalJAP2009,Pbaetal_apl2009,Ben-Abdallah2010,Francoeur2010,Francoeur2011,Basu2011,SIMetal2013,Miller2014,Jin2017,Iizuka2018}. A central idea in this
case is to incorporate thin films in layered systems to make better use of surface
electromagnetic modes. In practice, the RHT rate between two planar multilayer bodies comprised of an arbitrary number of layers can be formally described with the same formulas as in the 
plate-plate case, see Eqs.~(\ref{eq-pp}) and (\ref{eq-trans-pp}), but in this case $r^{\alpha}_1$ 
and $r^{\alpha}_2$ have to be interpreted as the reflection coefficients of the two subsystems 
(including their complete layered structures), see \cite{GB2009,Ben-Abdallah2010}. To give a 
concrete example, let us follow \cite{Song2015b} and consider the multilayer structure shown 
in the inset of Fig.~\ref{fig-thin-film} where the first body is an infinite SiO$_2$ plate
(medium 1) and the second body features a SiO$_2$ film of thickness $t$ (medium 3) deposited on 
a semi-infinite layer of Au (medium 4), while the medium 2 is the vacuum gap of size $d$. In this 
case, $r^{\alpha}_2$ in Eq.~(\ref{eq-trans-pp}) has to be replaced by~\cite{Biehs2007b}
\begin{equation}
R^{\alpha} = \frac{r^{\alpha}_{23} + r^{\alpha}_{34} e^{2iq_3t}}
{1-r^{\alpha}_{34} r^{\alpha}_{32} e^{2iq_3t}} ,
\end{equation}
which is the reflection coefficient of the subsystem formed by media 3 and 4. Here, as
usual, the $r^{\alpha}_{ij}$ are the Fresnel coefficients of the different interfaces: 
\begin{equation}
 r^{s}_{ij} = \frac{q_i-q_j}{q_i+q_j} \;\; \mbox{and} \;\;
 r^{p}_{ij} = \frac{\epsilon_j q_i - \epsilon_i q_j}
 {\epsilon_j q_i + \epsilon_i q_j} ,
 \end{equation}
 where $q_i = \sqrt{\epsilon_i k_0^2 - \kappa^2}$. Finally, the Fabry-P\'erot
 denominator in Eq.~(\ref{eq-trans-pp}) adopts now the form $D^{\alpha} = 1 - 
 r^{\alpha}_{21} R^{\alpha} e^{2iq_2d}$. 

In Fig.~\ref{fig-thin-film} we show representative results of the gap dependence of the-heat transfer coefficient of this multilayer structure for different values of the thickness of the silica film, ranging from 50 nm to bulk. We also show the result with no SiO$_2$ film for comparison. Notice that for small gaps ($d<100$ nm), the results are independent of the silica film thickness, which shows that the extraordinary NFRHT enhancements that occur in the bulk systems made of polar dielectrics are also possible in thin-film structures as long as the gap size is smaller than the film thickness~\cite{Biehs2007a,Biehs2007b}. As explained above,
the physical origin of these results can be traced back to the fact that NFRHT is 
dominated by electromagnetic cavity modes arising from SPhPs whose penetration depth scales with the gap size. Thus, when the gap is sufficiently small, all the heat transfer comes from a shallow region on the surface of the two bodies and NFRHT becomes independent of the film thickness. These qualitative predictions were subsequently experimentally confirmed by
\cite{Song2015b} using a 53-$\mu$m-diameter silica sphere as an emitter, instead of the silica plate used in the calculations of Fig.~\ref{fig-thin-film}. The finite curvature of the sphere results in smaller NFRHT enhancements, as compared to the planar structure, as it is easy to understand with the standard proximity approximation, see \cite{Song2015b}
for details.

To increase NFRHT beyond bulk systems, different groups have proposed to combine several
thin films to make use of the hybridization of the surface modes in different interfaces 
\cite{Biehs2007b,Francoeur2008,PbaetalJAP2009,Francoeur2011,Jin2017,Iizuka2018}. Another proposed
strategy to outperform bulk systems relies on the use of 1D photonic crystals 
\cite{Ben-Abdallah2010,MTetal2012}. In this case the heat transfer mechanism involves the surface 
Bloch states coupling supported by these media.

\subsubsection{Metamaterials}

Another topic that has been extensively studied in the context of NFRHT between nanostructured 
systems is the use of metamaterials, i.e., artificial structures with subwavelength features 
designed to exhibit complex optical properties that are difficult to find in naturally 
occurring (bulk) materials. In particular, special attention has been devoted to hyperbolic
metamaterials, which are a special class of highly anisotropic media whose electromagnetic
modes have an hyperbolic dispersion relation. To be precise, they are uniaxial materials 
for which one of the principal components of either the permittivity or the permeability
tensor is opposite in sign to the other two principal components. These systems have been
primarily fabricated based on designs involving hybrid metal-dielectric superlattices and metallic 
nanowires embedded in dielectric hosts \cite{Poddubny2013}. The interest in these metamaterials 
in the context of NFRHT lies in the fact that they have been predicted to behave as 
broadband super-Planckian thermal emitters \cite{Nefedov2011,SABetal2012,Guo2012}. This
behavior originates from the fact that these metamaterials can support electromagnetic 
modes that are evanescent in a vacuum gap, but which are propagating inside the material. 
This leads to broadband enhancement of the transmission efficiency of the evanescent modes
\cite{SABetal2012}. From the computational point of view, the heat transfer between 
hyperbolic metamaterials can be described using either the scattering approach for multilayer media described in the previous subsection, or the more general method discussed in the following subsection and applicable to laterally periodic patterned structures. In this latter case, and for appropiate (subwavelength) periodicities, it is typical to expoit an effective medium theory in order to reduce the problem to one involving planar but optically anisotropic materials, allowing application of the approach described in Sec.~\ref{sec-MO1}.

\begin{figure}
\includegraphics[width=0.8\columnwidth,clip]{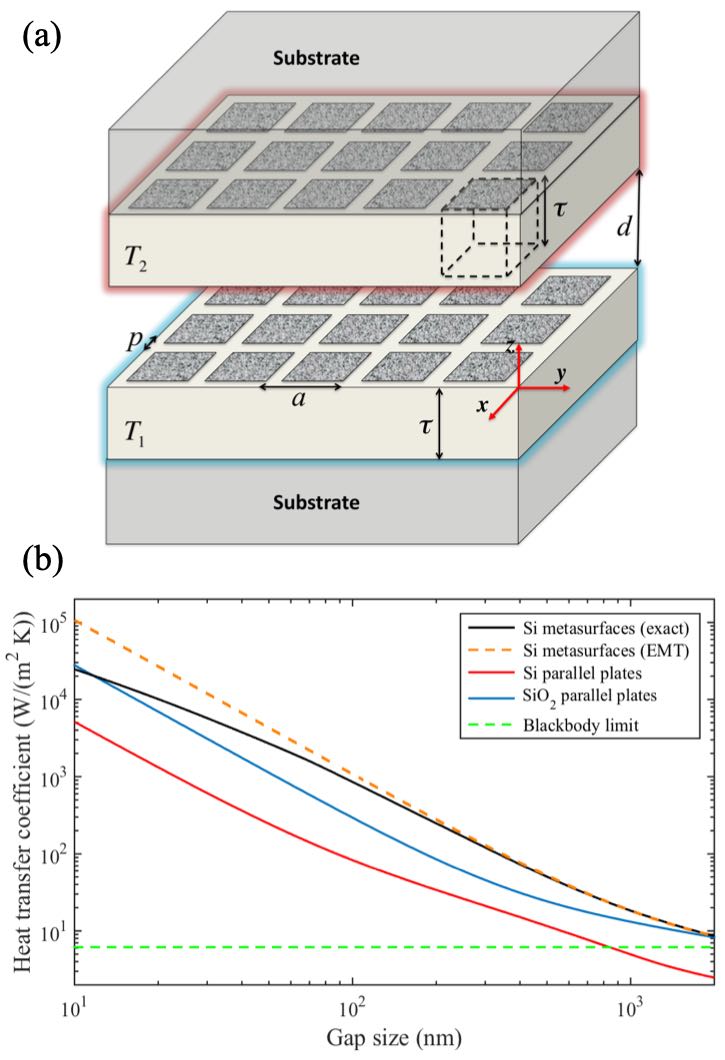}
\caption{(a) Schematic of two doped-Si metasurfaces made of 2D periodic arrays of square 
  holes placed on semi-infinite planar substrates and held at temperatures $T_1$ and $T_2$. 
  (b) Room-temperature heat transfer coefficient as a function of the gap size for the 
  doped-Si metasurfaces of panel (a) with $a= 50$ nm and a filling factor of 0.9 (black line).
  For comparison, the plot also includes the results for the Si metasurfaces computed with an
  effective medium theory (orange dashed line), SiO$_2$ parallel plates (blue line), and 
  doped-Si parallel plates (red line). The horizontal dashed line shows the black-body limit.
  Reprinted with permission from ref~\cite{Fernandez-Hurtado2017}. Copyright 2017 by the APS.}
\label{fig-metasurface}
\end{figure}

The special properties of hyperbolic metamaterials have spurred many theoretical 
investigations of their use in the context of NFRHT \cite{Biehs2013,Tschikin2013,Guo2013,Simovski2013,Liu2013,
Liu2014,Guo2014,SLetal2014,Miller2014,MTEtAl2015}. These works have in turn demonstrated that 
metamaterials do not outperform thin-film-based structures exhibiting SPhPs, as their increased 
density of states is compensated for by a decrease in the strength of the evanescent fields 
\cite{Miller2014}. Nevertheless, metamaterials exhibit other interesting properties; for 
instance, the long penetration depth of the hyperbolic modes can be advantageous for applications 
in near-field thermophotovoltaics.

\subsubsection{Gratings and Metasurfaces}

Also inspired by nanophotonic concepts, NFRHT between periodically patterned systems
has been intensively investigated from a theoretical point of view, both in 1D (gratings)
and in 2D (photonic crystals and periodic metasurfaces). Again, the goal of such nanostructuring is to tune the spectral heat transfer and enhance net NFRHT. Technically speaking,
the Landauer formula of the previous subsections can be straightforwardly generalized to 
deal with periodic systems by making use of Bloch's theorem. This was first done
by Bimonte and we refer to \cite{GB2009} for technical details. Using that generalized formula 
in combination with different techniques for the computation of reflection coefficients in 
periodic systems, typically via the rigorous coupled wave analysis (RCWA) method,
several groups have reported calculations of NFRHT between periodic metallic nanostructures 
in both 1D \cite{Guerout2012,Dai2015,Dai2016,Messina2017} and 2D \cite{Dai2016b,Jin2019}. 
The key idea in this case is to use nanostructuring to create new surface modes,
referred to as spoof plasmons \cite{Pendry2004}, whose frequencies can be adjusted by tuning
the length scales of these periodic systems so that their surface modes can be thermally 
populated at the desired working temperature. The reported results have clearly demonstrated 
the possibility of enhancing NFRHT over the corresponding planar (bulk) materials. However, 
NFRHT in these periodically patterned metallic structures continues to be smaller than that observed in simple (unstructured) planar polar dielectrics, with few exceptions \cite{Jin2019}.

There has also been significant theoretical work on the topic of NFRHT between dielectric 
photonic crystals and metasurfaces \cite{Rodriguez2011,Liu2015,Liu2015b}. Again,
these structured systems exhibit enhanced NFRHT with respect to their bulk counterparts,
but the resulting NFRHT rates are again much smaller than those of planar polar dielectrics.
In this regard, it is worth mentioning that it has been predicted that metasurfaces can 
indeed provide a way to enhance NFRHT between extended structures 
\cite{Fernandez-Hurtado2017}. To be precise, it has been shown that Si-based metasurfaces
featuring two-dimensional periodic arrays of holes, see Fig.~\ref{fig-metasurface}, can
exhibit a room-temperature near-field radiative heat conductance larger than any 
unstructured material to date. This enhancement relies on the possibility of largely 
tuning the spectral properties of the surface plasmon polaritons that dominate NFRHT in these
structures. In particular, nanostructuring enables the appearance of broadband and lower-frequency 
surface modes, increasing their contribution and occupation at room temperature, which constitutes 
one of the main strategies being pursued to enhance NFRHT. We conclude this subsection by noting that, to our knowledge, no experiment has thus far probed NFRHT between patterned structures.

\subsubsection{Graphene}

Two-dimensional materials are revolutionizing material science and they also hold
promise in the field of NFRHT. In particular, graphene has attracted
much attention as it can support delocalized surface plasmon polaritons (SPPs) that can contribute to NFRHT in spite of graphene's ultrasmall (one-atom) thickness \cite{Volokitin2011,Ilic2012}. What makes these surface modes so attractive, as compared to SPhPs in polar dielectrics,
is the possibility of modulating them electronically \cite{MessinaetalPhysRevB.87.085421}, which can be achieved by controlling graphene's chemical potential by means of a nearby gate electrode. Such a mechanism provides an ideal strategy to actively control NFRHT in graphene-based structures \cite{Papadakis2019}. On the other hand, several theoretical studies have shown that coating structures with graphene sheets may lead to a substantial increase in NFRHT \cite{VBSetAl2012,Lim2013,Messina2017b}. In this case, the idea is that appropriate engineering of the coupling of graphene's SPPs with other surface modes, like SPPs in doped Si or 
SPhPs in polar dielectrics, may increase the efficiency of heat exchange in the 
near-field regime. Another topic of great interest that has been theoretically 
investigated is the use of graphene-based structures in thermophotovoltaics
\cite{Ilic2012b,Messina2013,Svetovoy2014}. Furthermore, the role of graphene in NFRHT has
been theoretically studied in a wide variety of hybrid structures
\cite{Liu2014,XLL2015,Shi2017,Shi2018,Shi2019,Zhao2017b}.

\begin{figure}
\includegraphics[width=\columnwidth,clip]{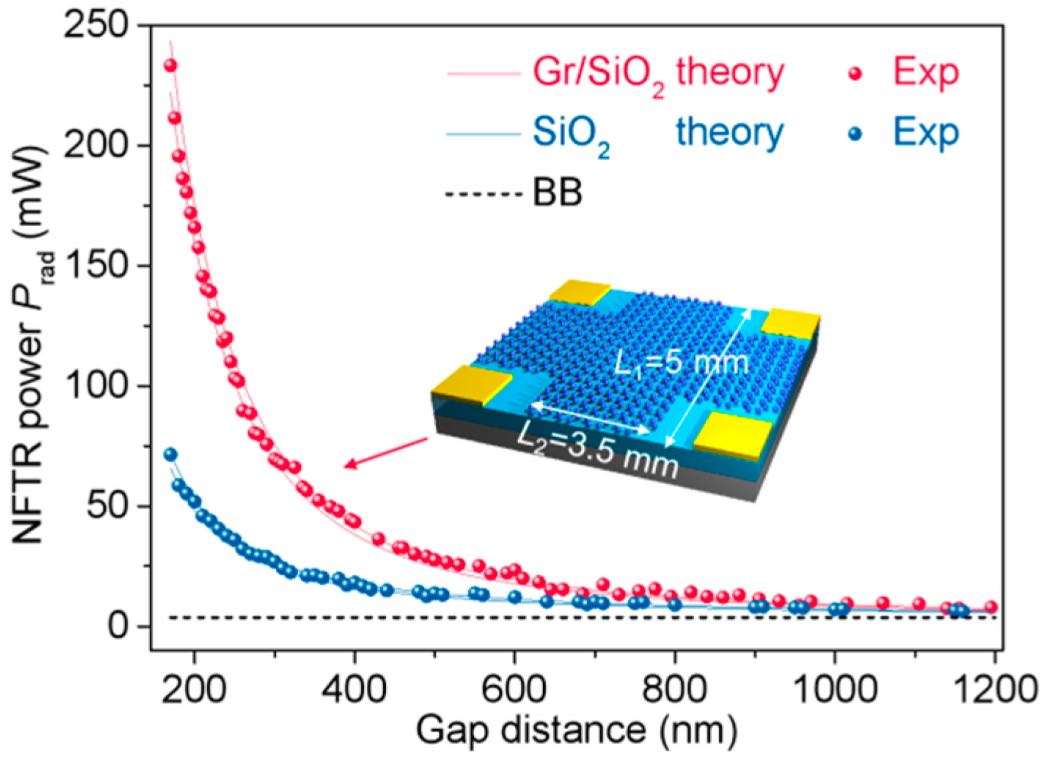}
\caption{Comparison of the NFRHT rate between 
  graphene(Gr)/SiO$_2$ pair (red-solid line) and SiO$_2$ pair (blue-solid line) with
  various gap sizes. The temperatures of the emitter and
  the receiver are 323.2 K and 301.5, respectively. Lines show the calculated values
  and spheres are the average values of four repeated measurements at each point. The inset shows a schematic illustration of the Gr/SiO$_2$ heterostructure. The black-body limit 
  has been plotted for comparison (black-dashed line). Reprinted with permission 
  from~\cite{Shi2019b}. Copyright 2019 ACS.}
\label{fig-graphene}
\end{figure}

From an experimental perspective, recent works have confirmed that graphene enables enhanced 
NFRHT between polar dielectrics \cite{Zwol2012b,Shi2019b} and between Si substrates (both 
insulating and conductive) \cite{Yang2018}. In particular, Shi \emph{et al.}\ \cite{Shi2019b} 
measured the NFRHT flux between two identical graphene-coated SiO$_2$ heterostructures with 
millimeter-scale surface area and reported a 64-fold enhancement compared to the corresponding black-body limit 
for a gap size of 170 nm, see Fig.~\ref{fig-graphene}. Moreover, these authors showed 
theoretically that the physical mechanism behind this large NFRHT enhancement is indeed the 
coupling between graphene's SPPs and silica's SPhPs. It is also worth mentioning that the first 
experimental demonstration of NFRHT modulation by electronic gating of a graphene field-effect 
heterostructure was just recently reported \cite{NHTetal2019}.

\subsubsection{Surface roughness}

Most calculations of NFRHT in planar structures assume that the
corresponding surfaces are perfectly flat. Such an idealization, for
instance, ignores practical considerations such as surface
roughness. The impact of surface roughness on NFRHT was addressed
theoretically by Biehs and Greffet \cite{Biehs2010}, in a plate-plate
configuration. Using a form of perturbation theory, they showed that
assuming reasonable values for the height of the roughness profile
($\sim 5$ nm), corrections to the heat transfer coefficient due to
roughness can lead to roughly order of magnitude differences compared
to perfectly flat surfaces when the gap size is on the order of a few
tens of nm, both for metals and polar dielectrics.  Moreover, they
showed that proximity approximations previously used for describing
rough surfaces are highly innacurate when gap sizes become much
smaller than the correlation length of the surface roughness, even
when the heat transfer is dominated by the coupling of surface
modes. We also note that the influence of surface roughness has also
been studied by way of the finite-difference time-domain method in
combination with the Wiener chaos expansion approach~\cite{ChenY2015},
along with its interplay with surface curvature~\cite{MKetal2013}.

\subsection{Impact of geometry}\label{sec-geometry}

Thus far, we have mainly discussed NFRHT in planar geometries in which the translational symmetry greatly simplifies the resolution of Maxwell's equations. In what follows, we turn to the analysis of the impact of geometry (heat exchange between structured bodies) and briefly discuss how the aforementioned RHT formulas can be generalized to handle objects of arbitrary size and shape.


The Polder-van Hove formula expressing $\mathcal{T}(\omega)$ in terms of Fresnel reflection coefficients or generalized reflection matrices is well-suited for calculations of heat transfer in systems with translational symmetry, including the aforementioned uniform planar slabs, thin films, gratings, photonic crystals, and periodic metamaterials. However, this leaves out a large class of systems of experimental and theoretical interest that do not exhibit such translational symmetries, particularly compact bodies like spheres or structured nanoparticles whose finite dimensions are relevant to the analysis of radiative heat transfer. Typically, in such cases, it is incumbent to exploit general-purpose techniques to compute field response quantities entering $\mathcal{T}(\omega)$, for the geometry in question, in terms of the system's Green's function. One such powerful general scattering formalism was developed by Kr\"{u}ger and coworkers~\cite{MKetal2012,BimonteARCMP2017}, arriving at the general formula (for reciprocal media),
\begin{equation} \label{eq:Krugertransmission}
  \mathcal{T}(\omega) = 4~\trace{\mathbb{R}^{*}_{2} \mathbb{W}_{1,2}
    \mathbb{R}_{1} \mathbb{W}_{2,1}^{*}}
\end{equation}
in terms of the radiation operator $\mathbb{R}_{p} = \GG_{0} (\Im(\TT_{p}) - \TT_{p} \Im(\GG_{0}) \TT_{p}^{*})\GG_{0}^{*}$ and scattering operator $\mathbb{W}_{pq} = \GG_{0}^{-1} (\mathds{1} - \GG_{0} \TT_{p} \GG_{0} \TT_{q})^{-1}$ for bodies $p, q \in \{1, 2\}$ defined in terms of the scattering T-operators $\TT_{p}$, which depend on the material properties and shape of the bodies and the Green's function operator $\GG_{0}$ in vacuum. The strength of this formulation lies in its broad applicability, as it generalizes beyond systems with discrete or continuous translational symmetry: it can in principle be used for arbitrary geometries, including compact bodies whose finite sizes in each dimension are relevant, with faster numerical convergence for appropriate choices of basis functions. Additionally, while this T-operator formalism casts thermal radiation in terms of volumetric scattering quantities, related contemporaneous surface integral equation formulations~\cite{RodriguezPRB2013} can similarly recover known
semi-analytical results for uniform planar media and be computationally amenable to general compact or extended geometries by casting thermal radiation purely in terms of surface unknowns, vastly reducing the computational complexity of calculations.

Furthermore, beyond simply aiding in generalizations of computations beyond extended media with translational symmetry, the T-operator formalism can shed further light on the number of contributing transmission channels to $\mathcal{T}(\omega)$. In the operators of Eq.~(\ref{eq:Krugertransmission}), an operator of particular interest~\cite{MillerAO2000,MillerJOSAB2007,MillerPRL2015,MoleskyPRB2020,
VenkataramPRL2020} is the off-diagonal block $\GG_{0(2, 1)}$ of the Green's function connecting points $\mathbf{r}'$ restricted to the volume of body 1 and $\mathbf{r}$ restricted to the volume of body 2. At first glance, the ability of electromagnetic fields to propagate through vacuum, or equivalently the coupling of all pairs of volumetric degrees of freedom in each of the different bodies, suggests 
that the number of channels will scale like the volume of each body. However, the electromagnetic surface equivalence theorem~\cite{HarringtonJEWA1989,RengarajanIEEE2000,ReidPRA2013,RodriguezPRB2013,
OteyJQSRT2014,SCUFF1} shows that the electromagnetic fields radiated by any volumetric polarization distribution to the exterior of some fictitious bounding surface can be exactly reproduced in that exterior region by an equivalent surface current distribution, therefore suggesting that the rank of $\GG_{0(2, 1)}$ actually scales with the {\it surface area} of each body; as shown by Polimeridis 
et al.~\cite{PolimeridisPRB2015}, it is indeed the effective rank of this off-diagonal scattering operator that determines the number of contributing transmission channels $\tau_{n}$. 
\begin{figure}
  \includegraphics[width=\columnwidth,clip]{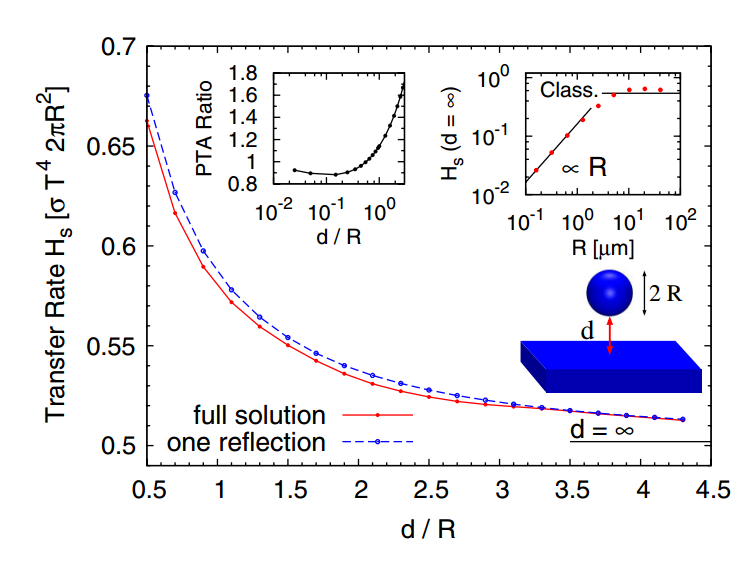}
  \caption{~\cite{MKetal2011} Transferred power $H_s$ by NFRHT between a SiO$_2$ sphere with radius $R = 5\mu{\rm m}$ at $300\,{\rm K}$ and a SiO$_2$ plane at $0\,{\rm K}$ as a function of distance $d$. The transferred power is normalized to the power emitted by a black body with a surface area given by the cross section of the sphere. From~\cite{MKetal2011}.}
  \label{fig-sphereplane}
\end{figure}

Based on the scattering approach and standard Green's function formalism, there have been many studies of the heat flux between a sphere and a plane, as shown in Fig.~\ref{fig-sphereplane}, and between two spheres~\cite{Narayanaswamy2008,MKetal2011,OteyPRB2011,KS2011}. Reviews highlighting other studies of NFRHT in non-planar geometries can be found in ~\cite{OteyJQSRT2014,BimonteARCMP2017}. Early studies of heat transfer between compact bodies typically focused on high-symmetry objects with simple shape. However, there have been far fewer studies of NFRHT in nanostructured compact bodies compared to the preponderance of examples for extended media (including the previously-discussed gratings, photonic crystals, and metasurfaces) because the former, unlike the latter, does not easily succumb to semianalytical expressions for arbitrary geometries in the absence of symmetries like continuous or discrete translational invariance. With this in mind, the next section discusses the development of various numerical methods to compute radiative heat transfer in a broad array of systems.

\subsection{Numerical methods}\label{sec-numerical}

\begin{figure*}
  \includegraphics[width=0.8\textwidth]{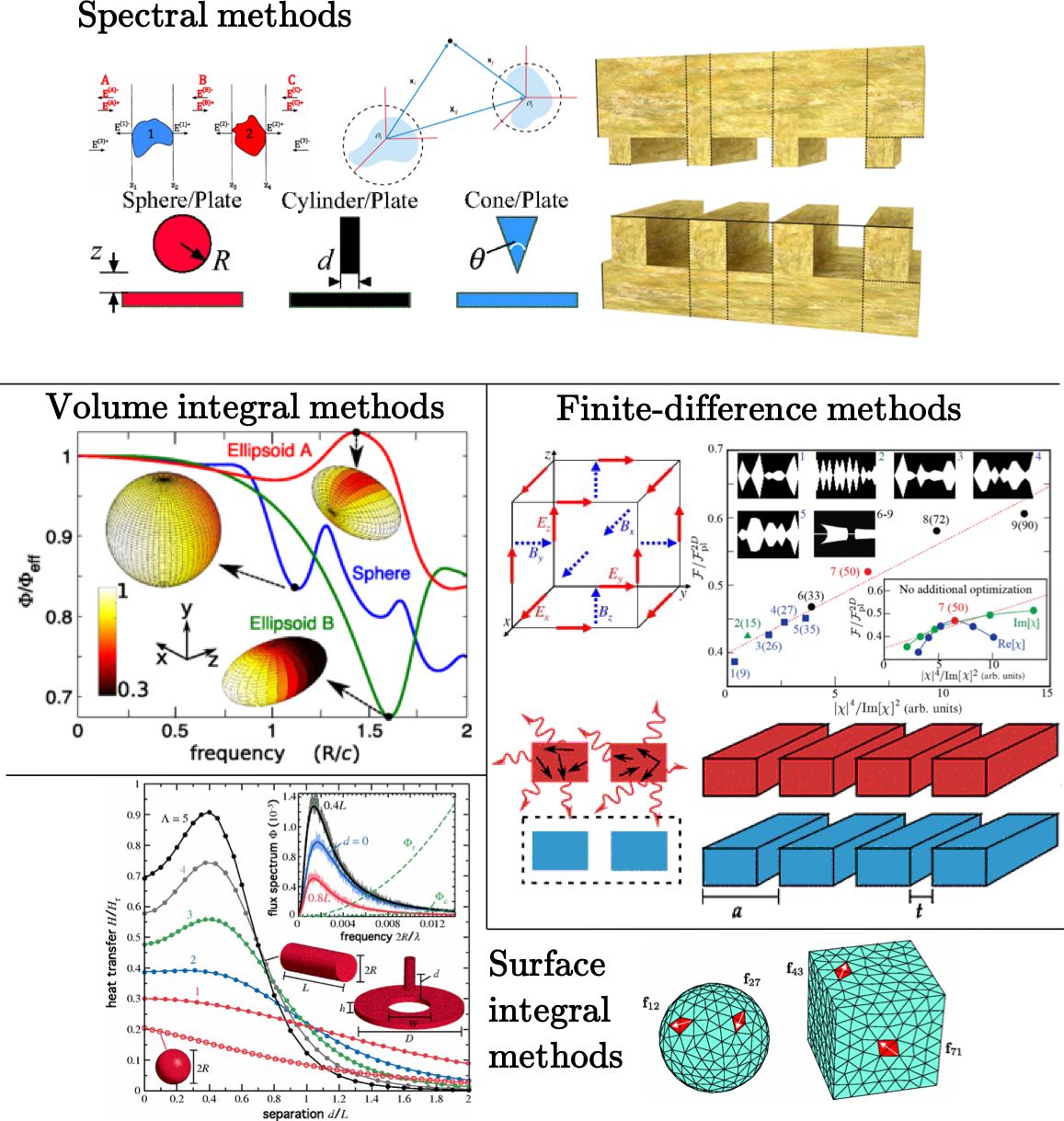}
  \caption{Collage of selected computational methods. Schematics of basis functions, along 
   with selected results, for spectral~\cite{BimonteARCMP2017,McCauleyPRB2012,Messina2017,
   RMandMA2011b,MKetal2012}, finite-difference~\cite{Jin2019,Rodriguez2011,WernerJCP2013}, 
   volume integral~\cite{PolimeridisPRB2015}, and surface
   integral~\cite{RodriguezPRL2013,RodriguezPRB2013,ReidPRA2013} methods.}
  \label{fig-computationalEM}
\end{figure*}

Advances in computational hardware and numerical algorithms have led to an explosion of computational methods to study radiative heat transfer. Notably, the facts that the Landauer form of the radiative heat transfer power depends only on the Bose function $n(\omega, T)$ and the Landauer energy
transmission spectrum $\mathcal{T}(\omega)$, and that the latter in Eq.~(\ref{eq:Krugertransmission}) only depends on classical electromagnetic scattering quantities, means that standard computational 
techniques may be readily applied to studying radiative heat transfer. These methods, illustrated schematically with examples in Fig.~\ref{fig-computationalEM}, essentially fall into one of two categories, depending on the choice of either a spectral or localized basis expansion~\cite{ReidPROCIEEE2013,OteyJQSRT2014,Song2015a,BimonteARCMP2017,Cuevas2018}, each of which brings a set of benefits and drawbacks.

\subsubsection{Spectral methods}\label{spec_methods}

Techniques based on spectral expansions~\cite{MKetal2012,BimonteARCMP2017} express the T-operators of each individual body in terms of delocalized spectral functions (e.g.\ the spherical vector waves discussed above). These basis functions 
include but are not limited to plane waves (Fourier basis)~\cite{GB2009,RMandMA2011b,Messina2013,MessinaPRB2016, 
Jin2017}, Bloch waves~\cite{NarayanaswamyJQSRT2005,MFetal2009,Ben-Abdallah2010,MTetal2012,Messina2017}, and spherical or cylindrical harmonics~\cite{Narayanaswamy2008,OteyPRB2011,MKetal2011,McCauleyPRB2012}.
The use of these basis functions is most convenient when the geometries involved exhibit discrete or continuous symmetries, like translation or rotation, as that can make the resulting matrix expressions for the relevant operators nearly diagonal, making computations far more efficient. However,
in the absence of such symmetries, or when different bodies have shapes of different symmetries, not only are the resulting matrices
dense, but the convergence with respect to increasing numbers of basis functions slows dramatically. Furthermore, we note that with few exceptions, such as work on graphene sheets~\cite{CastroNetoRMP2009,SerneliusPRB2012,Ilic2012,WunschNJP2006}, most applications of these spectral techniques have in practice focused on simple local isotropic homogeneous susceptibilities $\chi(\omega)$.
\subsubsection{Decomposition methods}\label{decomp_methods}
By contrast, techniques based on localized expansions~\cite{OteyJQSRT2014,Cuevas2018,Song2015a} 
express either T-operators or Maxwell Green's functions in terms of localized basis functions. 
One such technique is the finite-difference frequency domain method~\cite{WenJHT2010,
Jin2019}, in which Maxwell's equations in the frequency domain are discretized on a lattice of grid points. In the context of RHT, fields in response to individual dipolar sources embedded in the radiating objects can be computed independently and then summed according to weights determined by the fluctuation--dissipation theorem; alternatively, the uncorrelated nature of dipolar sources at different spatial positions means that all such fluctuating sources can be simultaneously introduced and modelled as stochastic, random sources with correlation functions given by the fluctuation--dissipation theorem (requiring ensemble averages over many source realizations to reduce noise, as in Monte-Carlo integration). The latter interpretation lends itself to a direct Langevin or stochastic time-domain 
simulation of Maxwell's equations~\cite{Rodriguez2011}. This last class of time-domain method has 
the added benefit that discretized spatial differential operators are represented as sparse matrices, 
and allows representations of broad classes of nonlocal (spatially dispersive) susceptibility models 
in terms of spatial differential operators, such as the hydrodynamic model~\cite{KlimchitskayaPRB2015,XiaoFOP2016}, 
all the while being applicable to arbitrary body shapes. On the other hand, multiscale or large 
problems become particularly challenging to simulate as the propagation of electromagnetic fields 
through vacuum means that the entire space between bodies must also be discretized, even if the 
separation is much larger than relevant body feature sizes, so the resulting convergence with respect 
to resolution can be prohibitively slow.

A related class of technique is the so-called volume-integral formulation of Maxwell's 
equations~\cite{PolimeridisPRB2015,JinPRB2016,JinPRB2017}, of which the discrete dipole approximation
(DDA)~\cite{EdalatpourJQSRT2016,SEandMF2016,SEandMF2014,RMAEetal2017} may be thought of 
as a special case. In general, volume integral formulations use various classes of localized basis
functions as basis expansions for T-operators and $\GG_{0}$. Unlike finite-difference methods, 
these techniques have the advantage of only requiring basis functions within the volumes of 
material bodies, with the full scattering problem represented by expressing the full Green's function 
in terms of the individual materials' scattering matrices and the analytically known free-space 
Green's function of the corresponding intervening medium. As expected, however, different choices 
of basis functions offer challenges and tradeoffs with respect to numerical convergence. As further 
elucidated below, DDA is effectively a volume integral formulation in which each body is discretized 
into point dipolar particles with equivalent Clausius--Mossotti polarizabilities: this approximation typically 
yields accurate results for dielectric media, but suffers from poor convergence when simulating metals 
with highly delocalized plasmons. In contrast, volume-integral formulations guaranteed to converge 
require a so-called Galerkin discretization of the problem based on use of either voxel~\cite{PolimeridisIEEE2015} 
or Schaubert-Wilton-Glisson (tetrahedral)~\cite{ReidARXIV2017} basis functions. In either case, the basis functions may be identical 
and displaced on a regular grid/lattice covering each body, in which case the matrix representation of 
$\GG_{0}$ may be sparse (and therefore computationally easier to handle) due to the translational 
symmetries inherent in $\GG_{0}$, though this often comes at the costs of computing matrix elements 
of $\GG_{0}$ for regions where no materials are present, or of losing flexibility over discretizing
certain regions more finely than others~\cite{PolimeridisPRB2015}. Exactly the opposite tradeoff 
occurs if the volumes are discretized in an irregular manner, with different weights given to different
basis functions~\cite{ReidARXIV2017}: it then becomes possible to discretize certain regions more 
finely than others, which is of particular relevance to near-field radiative heat transfer between 
large bodies where only a few fine features are very close to one another, but at the cost of the matrix
representation of $\GG_{0}$ becoming dense due to the loss of obvious symmetries in the representation.
Furthermore, in all cases, volume integral formulations can model inhomogeneous and anisotropic 
susceptibilities and even temperature gradients~\cite{PolimeridisPRB2015,JinPRB2016}, but modeling nonlocal
susceptibilities has proved to be more of a challenge.

A class of techniques related to the volume integral formulation are those based on the surface integral
formulation~\cite{RodriguezPRL2013,RodriguezPRB2013} of Maxwell's equations. These techniques compute 
the Landauer energy transmission spectrum $\mathcal{T}$ according to a formula that looks superficially similar 
to Eq.~(\ref{eq:Krugertransmission}) but whose derivation and implementation requires a
different set of techniques. In particular, surface-integral formulations make consistent use of the surface 
equivalence theorem~\cite{HarringtonJEWA1989,RengarajanIEEE2000,ReidPRA2013,RodriguezPRB2013,OteyJQSRT2014,SCUFF1}
to recast all free polarization sources and total electromagnetic fields in terms of equivalent surface currents,
with the relevant operators being the Green's functions of the homogeneous susceptibilities comprising
each body, as well as the surface integral operator relating incident fields to induced equivalent
surface currents. In principle, the operators relevant to the surface integral formulation can be
expanded in a spectral basis~\cite{RodriguezPRB2013}, but as in the T-operator formulation, convergence
suffers for bodies that do not exhibit requisite symmetries. Instead, it is more common to expand the
relevant operators in a localized basis like the Rao-Wilton-Glisson basis~\cite{RodriguezPRL2013,RodriguezPRB2013} of tetrahedral functions.

Finally, we point out that any of these frequency domain methods could have instead been cast in the
time domain. In the context of computational electromagnetism, this is most commonly achieved by using 
the finite-difference time domain method~\cite{Rodriguez2011,LuoPRL2004}. This has many of the same benefits 
and detriments of the aforementioned finite-difference frequency domain method. Techniques based on molecular dynamics have also been 
used to compute radiative heat transfer in systems comprising nanoparticles~\cite{DominguesPRL2005},
though the scaling of the volume with the cube of the number of atoms makes computations unwieldy in 
practice for large nanoparticles. For both of these time domain techniques, the main advantages are their 
generality with respect to materials, the simple computational implementation (as the temporal evolution 
operators are represented as sparse matrices), the ability to extract dynamical information, and their 
ability in principle to incorporate {\it nonlinear} material response. In the case of molecular dynamics,
susceptibilities can be simulated fairly generally as the method is based on simulating classical Newtonian 
particle dynamics, though interactions other than harmonic or Coulomb couplings are typically based on 
empirical rather than ab-initio models. The main disadvantages for both sets of techniques are losses 
in computational efficiency from needing to explicitly simulate fluctuating polarization sources obeying fluctuation--dissipation statistics, which requires that averages be taken 
over a large ensemble of calculations.

\subsection{Upper bounds on near-field heat transfer}\label{sec-bounds}

As noted above, the Stefan--Boltzmann formula or blackbody limit was
derived over a century ago under the assumptions of ray optics, and
consequently fails to provide an upper bound of the maximum heat flux
that can be extracted from an object in the near-field regime. While
it is known that, as in far-field emission, appropiate choice of
object geometry (nanostructuring) and materials can enhance NFRHT, the
lack of such a limit applicable in the near field begs the question:
how much more room for improvement can be expected from either of
these design criteria? Over the past few decades, there have been
several succesful attempts at addresing this fundamental question,
starting with analyses of maximum NFRHT achievable in planar
geometries (where the main design criterion is the choice of
material)~\cite{AIV2004,PBAKJ2010,SABetal2012} and followed more
recently by limits applicable to arbitrary nanostructures and
materials~\cite{MillerPRL2015,VenkataramPRL2020}. Technically
speaking, it is clear that upper limits to the heat flux are
determined by bounds on the transmission coefficient
$\mathcal{T}(\omega)$ per unit area in Eq.~(\ref{eq-Landauer}), which
is itself determined by the per-channel transmission factors
$\tau_n(\omega)$ entering Eq.~(\ref{eq-channels}). The aim of arriving
at a bound on RHT is therefore to discern the maximum number and
contribution of tranmission channels that may be excited by a yet
unknown optimal choice of material and geometry.

In the case of two planar bodies, the maximum heat flux is determined
by the bounds on the transmission coefficient $\mathcal{T}(\omega)$
per unit area in Eq.~(\ref{eq-pp}), which is determined by the
transmission factor $\tau_\alpha (\omega,\kappa) \in [0,1]$
corresponding to transversal waves of frequency $\omega$, lateral
wavevector $\kappa$, and polarization $\alpha = s,p$. It is then clear
that $\mathcal{T}(\omega)$ can be maximized if the transmission factor
$\tau_\alpha (\omega,\kappa)$ is maximal over a broad frequency and
lateral wavevector range. For example, when assuming that at a given
frequency, all transversal waves contribute a maximal transmission
factor of unity up to some threshold value $\kappa_{\rm max}$, the
upper bound for the transmission coefficient per unit area between two
planar bodies can be written as
\begin{equation}
  \mathcal{T}_{\mathrm{pl}}(\omega) \leq  2 \int_{0}^{\kappa_{\max}} \frac{\mathrm{d}\kappa}{2\pi} \kappa = N(\omega), 
\end{equation}
where $N(\omega)$ may be interpreted as the number of contributing
transmission modes or channels per unit
area~\cite{PBAKJ2010,SABJJ2010}. By definition, the contribution of
propagating waves is restricted to $\kappa < k_0$. Hence, setting
$\kappa_{\rm max} = k_0$, one obtains the maximum value of
$\mathcal{T}(\omega) = k_0^2/2\pi$ for propagating waves. Inserting
this maximum value in Eq.~(\ref{eq-pp}), one finds that the largest
heat flux $\Phi_{\rm pr}^{\max}$ that can ever be carried by
propagating waves is precisely the black-body value $\Phi_{\rm BB}$
given by Stefan-Boltzmann's law~\cite{Bergman2011,Planck1914}. Thus,
it is the additional contribution coming from evanescent waves with
$\kappa \geq k_0$ and not accounted for in Stefan-Boltzmann's law that
allows NFRHT to surpass the blackbody limit.

At first glance, it may appear that there is no upper bound to
$\kappa_{\rm max}$ in the evanescent sector, at least within the scope
of local continuum electromagnetism, suggesting that $\mathcal{T}_{\rm
  pl}(\omega)$ is unbounded. However, even simple considerations imply
otherwise. For instance, inside a dielectric, the largest possible
lateral wavevector allowed is given by the edge of the Brillouin zone
$\pi/a$, where $a$ is the lattice constant of the medium. Hence, only
waves up to wavevectors $\kappa_{\rm max} \approx \pi/a$ contribute
heat flux. Ignoring possible band degeneracies and physical
constraints imposed by material and geometric considerations, this
gives the following idealized upper bound on the maximum possible heat
flux between two dielectrics~\cite{AIV2004}:
\begin{equation}
    \Phi^{\max}_{\rm pl, ideal} \approx \frac{\kb^2 \pi^2}{24 \hbar a^2}(T^2_1 - T_2^2).
\end{equation}
Assuming a wavevector cutoff set by a lattice constant on the order of
the atomic scale ($a \approx 10^{-10}\,{\rm m}$), and room-temperature
operation ($T_1 = 300\,{\rm K}$ and $T_2 = 0\,{\rm K}$), yields a heat
flux of $10^{13}\,{\rm W}{\rm m}^{-2}$ that is unrealistically large
compared to the black-body value of about $460\,{\rm W}{\rm
  m}^{-2}$. Taking into account the nature of evanescent waves within
the vacuum gap between the two planar materials, one may derive a more
sensible upper bound. For instance, the field amplitude of evanescent
waves of a given $\kappa$ in the quasi-static regime drops
exponentially as $\exp(- \kappa z)$ with respect to the distance $z$
from the interface. As a consequence, one can expect that only
evanescent waves having $1/\kappa \approx z > d$ or $\kappa < 1/d$ can
meaningfully contribute to the heat flux between two planar interfaces
a distance $d$ apart, suggesting that $\kappa_{\max} \approx 1/
d$. In~\cite{PBAKJ2010}, it is argued that only evanescent modes with
$1/\kappa \approx z > d/2$ overlap significantly and contribute, so a
distance-dependent cutoff $\kappa_{\max} \approx 2/ d$ is used to
provide an estimate of the upper limit for $\mathcal{T} \leq 2/ \pi
d^2$, leading to the following gap-dependent upper bound on the net
heat flux~\cite{PBAKJ2010}:
 \begin{equation}
    \Phi^{\max}_{\rm pl, gap} = \frac{\kb^2}{6 \hbar d^2}(T^2_1 - T_2^2).
\end{equation}
The choice of $\kappa_{\max} = 1/d$ would decrease this estimate by a
factor of $1/4$. Note that this cutoff is consistent with the fact
that $\mathcal{T}$ scales as $\exp(-2\kappa d)$ with the separation
distance $d$. A similar simple and general, albeit material
independent expression for the upper limit of the heat flux
contribution has also be found for the case of two hyperbolic
metamaterials~\cite{SABetal2012}.

Material considerations further constrain the allowed heat flux
between planar media. In particular, Biehs and
Greffet~\cite{SABJJ2010} derived a more realistic frequency-dependent
cutoff $\kappa_{\max} = \ln[2/\Im(\chi)]/d$ that accounts for the
impact of material absorption through the material-specific loss rate
$\Im[\chi(\omega)]$, where $\chi$ is the medium's susceptibility. In
particular, knowledge of the analytical form of the reflection
coefficients at an interface can be used to show that the maximum flux
occurs for materials satisfying the surface-mode resonance condition,
$\Re(1/\chi) = -1/2$. The fact that in the quasi-static regime the
heat flux scales like $1/d^2$ can be understood from the fact that the
number of contributing evanescent modes per unit area scales like
$1/d^2$~\cite{PBAKJ2010,SABJJ2010,SABetal2012}. Generalizations of
related analysis to bound the performance of planar metasurfaces
(nanstructured materials with subwavelength systems) have recently
been made~\cite{SABetal2012, Miller2014}, showing for instance that
metasurfaces cannot significantly enhance NFRHT beyond planar thin
films.


Efforts aimed at identifying the number and relative contribution of
transmission channels that may arise in non-planar media require a
different framework. Recently, Miller et al~\cite{MillerPRL2015}
recast radiative heat transfer between two bodies as a series of
independent absorption and emission problems (ignoring additional
constraints posed by the presence of multiple scattering among the two
objects) to obtain bounds that only depend on the bodies' material
susceptibilities and separation. In particular, recent work showed
that given an incident field on an object of susceptibility
$\chi(\omega)$, the maximum polarization field that can arise at any
point inside the object at a frequncy $\omega$ depends on the
``material response factor''~\cite{MillerOE2016},
\begin{equation}
  \zeta(\omega) = \frac{|\chi(\omega)|^{2}}{\Im[\chi(\omega)]}
\end{equation}
Such a figure of merit yields a measure of the resistivity or
dissipation of the medium and thereby captures the impact of losses on
the resonant optical response of a body. The material response factor
arises from the optimal magnitude of the T-operator for maximal
absorption in isolation~\cite{MillerOE2016}, and encodes
electromagnetic many-body and multiple scattering effects within the
body in isolation; this optimal magnitude is achievable at a
polaritonic resonance, determined by the value of $\Re(1/\chi)$, which
in turn can be tailored through nanostructuring. Exploiting the
maximum polarization responsivity of a medium in combination with
electromagnetic reciprocity, Miller et al found an upper bound on the
net transmission $\mathcal{T} \leq 4\zeta_{1} \zeta_{2} \int_{V_{1}}
\mathrm{d}\mathbf{r}' \int_{V_{2}} \mathrm{d}\mathbf{r} \sum_{i,j}
|G_{0}(\omega, \mathbf{r}, \mathbf{r}')|^{2}$ that depends
quadratically on the effective loss rate of the system $\zeta =
\sqrt{\zeta_{1} \zeta_{2}}$, with $\zeta_{1}$ and $\zeta_{2}$ denoting
the material factors of the bodies, and on the integral of the vacuum
Green's function over volumes $V_{1}$ and $V_{2}$ representing {\it
  any convenient domain} that may contain bodies 1 and 2,
respectively. Such a double integral may be cast as a Frobenius norm
of the off-diagonal matrix $\GG_{0(2, 1)}$, which was previously
identified in related works by D.~A.  Miller et al~\cite{MillerAO2000,
  MillerJOSAB2007} on optical communication limits. However, such an
analysis depends crucially on the assumption that each body is capable
of simultaneously and optimally emitting electromagnetic fields in the
absence of the other, and of optimally absorbing electromagnetic
fields in the presence of the other, which effectively neglects
additional physical constraints arising from the unavoidable impact of
multiple scattering between the two bodies. As a result, the limits
have been shown to be tight in situations where multiple scattering
can be neglected, namely quasistatic media subject to relatively large
material losses~\cite{Jin2019}. This problem becomes particularly
acute in the context of bounds on extended structures, where the
inability to account for tighter bounds on the transmission
eigenvalues causes the quadratic dependence on $\zeta$ to far outstrip
the observed logarithmic dependence on $\zeta$ seen in polaritonic
planar media near the resonance condition $\Re(1/\chi) = -1/2$ (and
predicted by the above planar bounds), suggesting more room for
enhancements in NFRHT through nanostructuring than has been observed
in practice.

%
%

In recent work, Venkataram et
al~\cite{MoleskyPRB2020,VenkataramPRL2020} developed a set of
algebraic techniques to derive tighter bounds on NFRHT that
incorporate not only constraints on material response but also
multiple scattering. Specifically, the transmission coefficient for
two arbitrarily shaped bodies at any given frequency $\omega$ was
found to be bounded above by,
\begin{align}
  \mathcal{T}_{\rm arb}(\omega) &= \sum_n \tau_n(\omega) \nonumber \\
  &\leq \sum_{n} 
  \begin{cases}
    1, & \zeta_{1} \zeta_{2} g_{n}^{2} \geq 1 \\
    \frac{4\zeta_{1} \zeta_{2} g_{n}^{2}}{(1 + \zeta_{1}\zeta_{2} g_{n}^{2})^{2}}, & \zeta_{1} \zeta_{2} g_{n}^{2} < 1
  \end{cases}
\end{align}
where the dependence on $\omega$ inside the various factors has been
deprecated. These bounds depend not only on the resistivity
$\zeta_i(\omega)$ of each body $i=\{1,2\}$ at the given frequency, but
also on a set of ``radiative efficacy" coefficients $g_{n}(\omega)$
denoting the singular values of the {\it vacuum} off-diagonal Maxwell
Green's function $\GG_{0(2, 1)}$ connecting dipoles in one object to
the resulting fields on the other. Moreover, the bounds move beyond
simply identifying the set of channels able to contribute to heat
transfer, previously estimated on the basis of the effective rank of
$\GG_{0(2, 1)}$, and instead exploit the specific singular values of
$\GG_{0(2,1)}$ in combination with the loss rate of the medium to
quantitatively determine the maximum possible transmission for each
channel. Once the set of channels that could possibly contribute
(having nonzero radiative coupling $g_{n}$) is identified, the ability
of each transmission channel to saturate the Landauer upper bound of
unity ($\tau_n \leq 1$) is determined by the degree to which the
radiative efficacies are able to overcome material losses, captured by
the condition $\zeta_{1} \zeta_{2} g_{n}^{2} \geq 1$; the per-channel
bound is less than unity for those channels unable to meet such a
condition. In addition to correctly reproducing the transition and
eventual saturation in the growth of NFRHT between dipolar
nanoparticles, from material-loss-dominated growth in the polarization
response to the Landauer tranmission bounds of unity, these limits
reveal that extended nanostructured bodies cannot significantly
outperform resonant planar polaritonic slabs even in
principle. Specifically, evaluation of the radiative efficacies for
any set of nanostructures contained within semi-infinite half-space
domains yields a limit on the net transmission of
\begin{multline}
  \mathcal{T}_{\rm arb}(\omega) \times d^{2} / A \leq \\ \frac{1}{2\pi}
  \begin{cases}
    \ln\left(1 + \frac{\zeta_{1} \zeta_{2}}{4}\right), & \zeta_{1} \zeta_{2} < 4 \\
    \frac{1}{2} \ln(\zeta_{1} \zeta_{2}) + \frac{1}{8} 
    \left[\ln\left(\frac{\zeta_{1} \zeta_{2}}{4}\right)\right]^{2}, & \zeta_{1} \zeta_{2} \geq 4
\end{cases}
\end{multline}
which exhibits a weak squared-logarithmic dependence on $\zeta$, in
line with the observed logarithmic peak value of $\mathcal{T}$ for
planar slabs at a polaritonic resonance~\cite{SABJJ2010,MillerPRL2015}.

Based on this recent analysis, it is evident that the observed
inability of nanostructuring to significantly enhance the amplitude of
$\mathcal{T}$ at any given frequency beyond what is achievable with
resonant planar materials is a ``feature" of the underlying physics of
NFRHT, and not a ``bug" in sampling a limited design space: the
maximum channel able to saturate the Landauer transmission limit of
unity for {\it any} nanostructure scales logarithmically as
$\frac{1}{2d} \ln\left(\frac{\zeta_{1} \zeta_{2}}{4}\right)$ provided
the system is in the underdamped (resonant) regime $\zeta_{1}
\zeta_{2} \geq 4$. Intuitively, this result may be seen as dissonant
with the established utility of nanostructuring for enhancing
far-field electromagnetic absorption and scattering, and the
significantly stronger enhancements of local densities of states that
can arise in the vicinity of structured materials. However, the
channels of radiative heat transfer between two separable bodies in
proximity have little to do with the channels that carry energy away
from a body (or an aggregate two-body system), so there is no reason
to believe that enhancement of the latter transmission channel
contributions would necessarily increase the former. 

The transition from a quadratic~\cite{MillerPRL2015} to a much weaker
logarithmic~\cite{VenkataramPRL2020} dependence of the bounds on
material conductivity once multiple-scattering constraints are
introduced illustrates the restricted and prohibitive nature of
nanostructuring in tailoring mutual scattering across a wide range of
resonant channels. Such a tradeoff precisely explains why the success
of nanostructuring in enhancing local fields does not readily
translate into equivalent enhancements in NFRHT.  As reviewed in
Sec.~\ref{sec-metals} and Sec.~\ref{sec-Nano}, metallic nanostructures
can indeed greatly enhance heat exchange compared to their planar
counterparts, but as these limits suggest, not much more than what may
be achieved with planar polar dielectrics. Finally, while multiple scattering ultimately hampers the maximum heat exchange that any two bodies can experience, as we shall see in the next section, it underlies several important transport effects in many-body systems.



\section{Many-body systems}

Until this last decade, theoretical and experimental work in the topic
near-field radiative heat transport was primarily relegated to the
study of heat exchange between two objects, while transport in systems
composed of objects in mutual interactions remained largely unexplored
and out of the reach of classical FE. In 2011, Ben Abdallah et
al.~\cite{PBAetal2011} laid out the theoretical foundations for
studying NFRHT in simple many-body systems made of small interacting
objects in the dilute regime, paving the way for a new research
direction on the topic of nanoscale heat transfer. Since then,
numerous works have revealed new many-body effects, including the
emergence of new physical and transport behaviors, and unraveling a
large number of potential applications in domains such as nanoscale
thermal management, energy-conversion technology, and information
processing. In the following sections, we describe these
peculiarities.

\subsection{Heat flux in dipolar many-body systems}

Understanding the mechanisms that drive light matter interactions is one of the main goal in optics. In the following, we address the problem of light absorption and thermal emission by a set of small objects in which cooperative interactions as well as heat exchange take place in these systems.

\subsubsection{Light absorption in dipolar systems}\label{Sec:Absorption}

To start let us consider the case of non-emitting objects which are only able to scatter and absorb light from an external source, i.e.\ we are neglecting thermal radiation at this stage. In the simplest case of a small isolated particle located at position $\mathbf{r}'$ in vacuum, the optical response of this particle can be described by the response to a simple permanent dipolar electric moment $\mathbf p(\mathbf{r}')$.

The electric field produced at point $\mathbf{r}$ around this dipole takes the following form
\begin {equation}
   \mathbf E_{\rm p}(\mathbf{r})=\omega^2 \mu_0\mathds{G}_0(\mathbf{r},\mathbf{r}')\mathbf p(\mathbf{r}').\label{Eq:dipole_field}
\end{equation}
Here~\cite{Novotny}
\begin{equation}
\begin{split}
     \mathds{G}_{\mathrm{0}}(\mathbf{r},\mathbf{r}') &=\frac{\exp({\rm i}k_0\rho)}{4\pi \rho}
     \biggl[\left(1+\frac{{\rm i}k_0\rho-1}{k^{2}_0\rho^{2}}\right)\mathds{1} \\
     &\qquad +\frac{3-3ik_0\rho-k^{2}_0\rho^{2}}{k^{2}_0\rho^{2}}\widehat{\boldsymbol{\rho}}\otimes\widehat{\boldsymbol{\rho}}\biggr]
\end{split}
\label{Eq:GreenVac}
\end{equation}
is the free space Green tensor defined with the unit vector $\widehat{\boldsymbol{\rho}}\equiv\boldsymbol{\rho}/\rho$,
$\boldsymbol{\rho}=\mathbf{r}-\mathbf{r}'$, $k_0 = \omega/c$ is the wave vector while $\mathds{1}$ denotes the unit dyadic tensor and $\mu_0$ denotes the vacuum permeability. When this particle is illuminated by an incident field $\mathbf E_{\rm inc}$, the local electric field $\mathbf{E}_{\rm loc}$ measured at any point $\mathbf{r}$ is the superposition of the incident field and the field generated (scattered) by the dipole. Therefore, according to expression (\ref{Eq:dipole_field}), this field decomposes into
\begin {equation}
    \mathbf E_{\rm loc}(\mathbf{r})=\mathbf E_{\rm inc}(\mathbf{r})+\omega^2 \mu_0\mathds{G}_0(\mathbf{r},\mathbf{r}')\mathbf p(\mathbf{r}').
\label{Eq:local_field_dipole}
\end{equation}
The electromagnetic power $\mathcal{P}$ dissipated in the particle can be calculated from the rate of work
\begin {equation}
   \mathcal{P}_{\rm abs}=\frac{1}{2}\intop_{V}\!\!\rd V\, \Re\bigl(\mathbf{j^* \cdot E_{\rm loc}}\bigr) 
\label{Eq:rate_work}
\end {equation}
done by the electromagnetic field in a volume $V$ including the particle. Here $\mathbf{j}$ denotes the local electric current density in the volume $V$. In the dipolar approximation $\mathbf{j}(\mathbf{r}')=- {\rm i}\omega\mathbf p \delta(\mathbf{r}-\mathbf{r}')$ so that
\begin {equation}
    \mathcal{P}_{\rm abs}= \frac{1}{2} \Re \bigl( {\rm i}\omega\mathbf{p}^* \cdot \mathbf{E}_{\rm loc}\bigr) = -\frac{\omega}{2}\Im\bigl(\mathbf{p}^*\cdot\mathbf{E}_{\rm loc}\bigr).
\label{Eq:rate_work2}
\end {equation} 
Using the following relation 
\begin {equation}
   \mathbf{p}(\mathbf{r}')=\epsilon_0 \alpha \mathbf{E}_{\rm inc}(\mathbf{r}')\label{Eq:moment_polarizability}
\end {equation} 
between the incident field and the dipolar moment, where $\alpha$ is the electric polarizability, the power dissipated in the particle reads~\cite{ST2014}
\begin {equation}
   \mathcal{P}_{\rm abs}=\frac{\omega|\mathbf{E}_{\rm inc}|^2\epsilon_0}{2}\biggl(\Im[\alpha] - \frac{k_0}{6\pi}
   |\alpha|^2\biggr).
\label{Eq:power_single}
\end {equation}
It is common to quantify light absorption using the absorption cross-section defined as the ratio 
\begin {equation}
   \sigma_{\rm abs}=\frac{\mathcal{P}_{\rm abs}}{\mathcal{F}_{\rm inc}}\label{Eq:cross_section_def}
\end {equation}
of this dissipated power by the incident flux
\begin {equation}
  \mathcal{F}_{\rm inc}=\frac{c\epsilon_0}{2}\mid\mathbf{E}_{\rm inc}\mid^2.
\label{Eq:incident_flux}
\end {equation}

\begin{figure}
\includegraphics[width=\columnwidth]{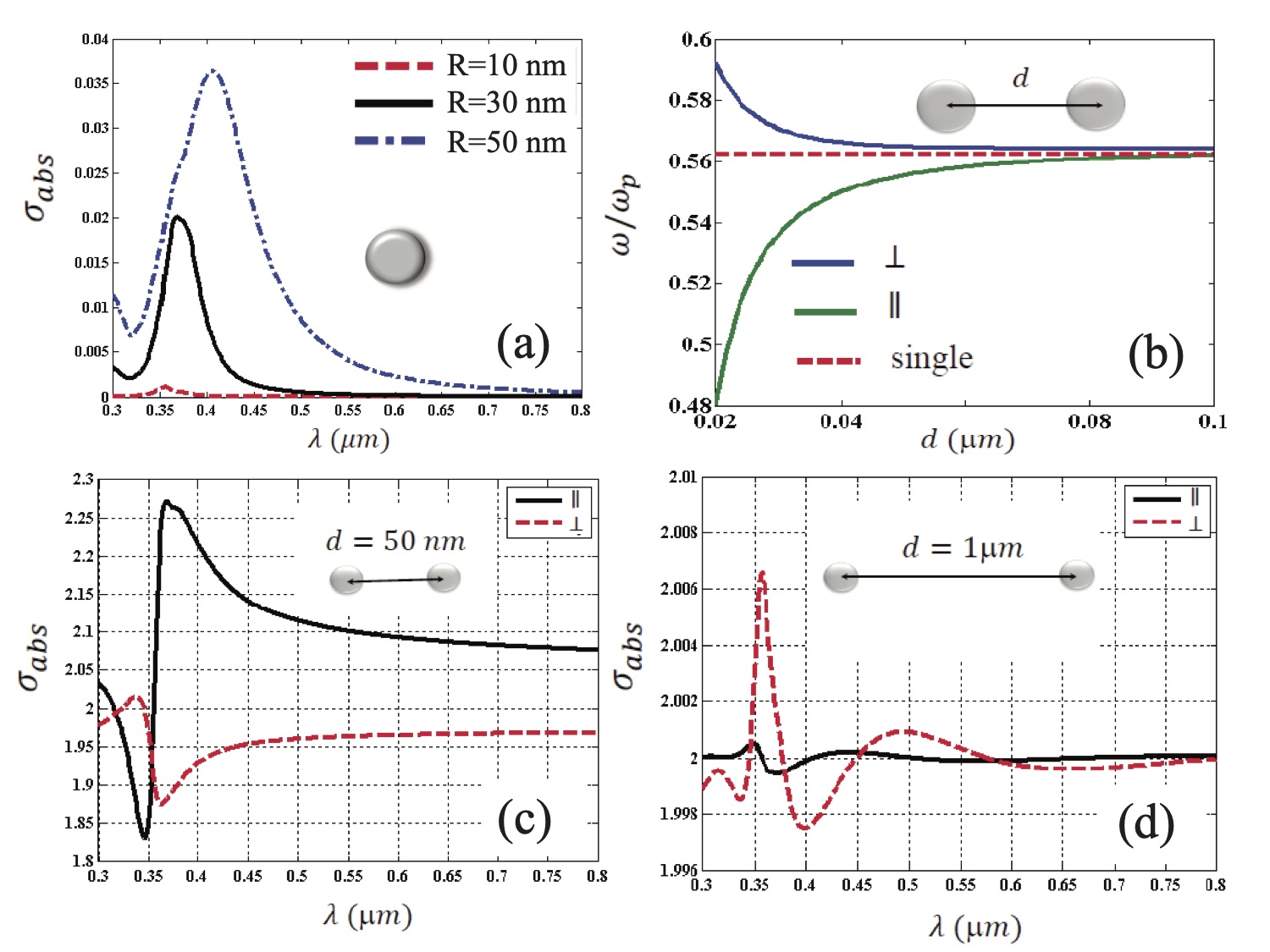}
   \caption{(a) Absorption cross-section of spherical silver nanoparticles with respect to the wavelength. (b) Orthogonal and parallel configurational resonance frequencies for a dimer of silver nanoparticles ($R = 10\:{\rm nm}$) in vacuum with respect to their separation distance $d$. The red horizontal line represents the plasmon resonance of an isolated particle.(c)-(d) Absorption cross-sections for a dimer of silver nanoparticles ($R = 10\,$nm) and normalized by the absorption of a single particle. From~\cite{Raj}}.
\label{dressing}
\end{figure}

For a collection of dipoles located at the position $\mathbf{r}_i$ ($i = 1, \ldots, N$) the multi-scattering process between the particles must be taken into account \cite{Langlais2014}. Under an external illumination by an incident field $\mathbf E_{\rm inc}$, the local electric field $\mathbf{E}_{\rm loc}$ measured at any point results from the superposition of the incident and all scattered fields as
\begin {equation}
  \mathbf{E}_{\rm loc}(\mathbf{r}) = \mathbf{E}_{\rm inc}(\mathbf{r}) + \omega^2 \mu_0\sum_{j = 1}^N \mathds{G}_0(\mathbf{r},\mathbf{r}_j)\mathbf{p}_{j}.
\label{Eq:external_field1}
\end{equation}
By introducing the notation $ \mathbf{p}_i = \mathbf{p}(\mathbf{r}_i)$, $ \mathbf{E}_{{\rm loc},i}= \mathbf{E}_{\rm loc}(\mathbf{r}_i)$, $\mathbf{E}_{{\rm inc},i} = \mathbf{E}_{\rm inc}(\mathbf{r}_i)$ the total power absorbed by this set of dipoles takes the general form \cite{Hugoninetal2015} 
\begin{equation}
\begin{split}
   \mathcal{P}_{\rm abs} &=  \frac{\omega}{2}\biggl( \sum_{i = 1}^N \Im\bigl(\mathbf{p}_i\cdot\mathbf{E_{\rm inc,i}}^*\bigr) \\
   &\qquad - \sum_{i,j = 1}^N  \Im\bigl(\mathbf{p^*}_i \boldsymbol{D}_{ij} \mathbf{p}_j \bigr)\biggr)
\end{split}
\label{Eq:power2}
\end{equation}
where we have introduced the $N\times N$ block matrix
\begin{equation}
   \boldsymbol{D}_{ij} = \mu_0 \omega^2 \mathds{G}_{0}(\mathbf{r}_i,\mathbf{r}_j).
\label{Eq:Dblock}
\end{equation}
This relation generalizes expression (\ref{Eq:power_single}) to arbitrary systems of coupled dipoles. For isotropic and homogeneous particles the generalized vector field of dipolar moments reads
\begin {equation}
  \begin{pmatrix} \mathbf{p}_{1} \\ \vdots \\ \mathbf{p}_{N} \end{pmatrix} = \boldsymbol{A} \begin{pmatrix} \mathbf{E}_{{\rm loc}, 1} \\ \vdots \\ \mathbf{E}_{{\rm loc},N} \end{pmatrix}
\label{Eq:dipole_moment}
\end{equation}
introducing the block matrix
\begin{equation}
   \boldsymbol{A}_{ij} = \epsilon_0\delta_{ij}\alphamat_i
\label{Eq:Ablock}
\end{equation}
where $\alphamat_i$ is the electric polarizability tensor associated to the $i^{th}$ particle. Using Eq.~(\ref{Eq:external_field1}), this expression can be reformulated with respect to vectorial incident field as
\begin {equation}
  \begin{pmatrix} \mathbf{p}_{1} \\ \vdots \\ \mathbf{p}_{N} \end{pmatrix} = \boldsymbol{A} \tilde{\boldsymbol{T}}^{-1}\begin{pmatrix} \mathbf{E}_{{\rm inc}, 1} \\ \vdots \\ \mathbf{E}_{{\rm inc},N} \end{pmatrix}
\label{Eq:dipole_moment2}
\end{equation}
with
\begin{equation} 
  \tilde{\boldsymbol{T}}_{ij} = \delta_{ij}\mathds{1}-(1-\delta_{ij})k_0^2 \mathds{G}_{0}(\mathbf{r}_i,\mathbf{r}_j) \alphamat_j.
\label{Eq:Tblocktilde}
\end{equation}
This block matrix $\tilde{\boldsymbol{T}}^{-1}$ defines the interplay between all dipoles and the block matrix
\begin{equation}
  \boldsymbol{\alpha}_{\rm dr} = \frac{1}{\epsilon_0} \boldsymbol{A} \tilde{\boldsymbol{T}}^{-1}  
  \label{Eq:dressed_polarizability}
\end{equation}
also called dressed polarizability~\cite{ECetal2012} results from the multi-scattering process in the set of dipoles. Using the slightly different block matrix
\begin{equation} 
  \boldsymbol{T}_{ij} = \delta_{ij}\mathds{1}-(1-\delta_{ij})k_0^2\alphamat_i \mathds{G}_{0}(\mathbf{r}_i,\mathbf{r}_j).
\label{Eq:Tblock}
\end{equation}
it can also be expressed as
\begin{equation}
  \boldsymbol{\alpha}_{\rm dr} = \frac{1}{\epsilon_0} \boldsymbol{T}^{-1} \boldsymbol{A},
  \label{Eq:dressed_polarizability2}
\end{equation}
because $\boldsymbol{T A} = \boldsymbol{A} \tilde{\boldsymbol{T}}$ and $\boldsymbol{T}^{-1} \boldsymbol{A} = \boldsymbol{A} \tilde{\boldsymbol{T}}^{-1}$. This dressed polarizability shows that two types of resonances play a role in the interaction of light with the set of coupled dipoles. The first ones are the resonances of the isolated particles themselves (i.e.\ the poles of $\alphamat_i$) while the second (i.e.\ the poles of the determinant of $\boldsymbol{\alpha}_{\rm dr}$ or $\boldsymbol{T}^{-1}$) are configurational resonances (see Fig.~\ref{dressing}) and they depend on the spatial distribution of dipoles. So that the $3N$ dipolar resonance which are degenerate for spherical nanoparticles, for instance, couple and form a band of $3 N$ resonances in general. Depending on the symmetry in the configuration some of the resonances remain degenerate despite the coupling. A simple example is a chain of nanoparticles. There one finds $N$ two-fold degenerate vertical and $N$ longitudinal resonances~\cite{FandW2004} forming bands of coupled modes. A general consequence of this dressing due to the coupling is a broadening of the absorption spectrum in a coupled $N$-dipole system.
%

\subsubsection{Exchanged Power and Poynting vector}
\label{Sec:NDipol}

Now we consider the most general situation where the particles are also emitting heat radiation. The fundamental relations to describe heat exchange in a system of $N$ dipoles having temperatures $T_1, \ldots, T_N$ within the framework of the FE have first been derived in \cite{PBAetal2011}. In Ref.~\cite{RMetal2013} the relations for the heat exchange were generalized to treat also the interaction of the $N$ dipolar objects with an environment or background in thermal equilibrium at some temperature $T_b$, but only for isotropic dipolar object. Subsequently, these expressions have been extended to anisotropic and non-reciprocal systems taking also the radiation correction into account~\cite{MN2014,RMAEetal2017} and the expression for the mean Poynting vector of such an $N$-dipole system have been determined to quantify its far-field thermal emission~\cite{RMAEetal2017}. Finally, in Refs.~\cite{AOetal2019b,Ott2020} the method from Ref.~\cite{RMetal2013} was used to determine the general expressions for the mean Poynting vector and the exchanged heat in a system of $N$ dipoles immersed in an environment at temperature $T_b$ which can also be non-reciprocal. A further generalization which takes the possibility of magnetic polarizabilities into account can be found in Ref.~\cite{AM2012,JDetal2017}. Here we review mainly the derivation of the heat exchange and the mean Poynting vector for $N$ dipolar objects described by an electric polarizability tensor $\alphamat$ within the framework of~\cite{RMetal2013}. This approach is valid for nanoparticles with a size much smaller than the thermal wavelength and for inter-particle distances and distance between the particles and interfaces of the environment larger than twice the diameter~\cite{ANGC2008,DBCN2019,COSF2011}.

To derive the exchanged power and the mean Poynting vector in an N-dipole system we consider the total electric and magnetic fields
\begin{align}
	\mathbf{E}(\mathbf{r},\omega) &= \omega^2\mu_0 \sum_{i = 0}^N \mathds{G}^{\rm EE}(\mathbf{r},\mathbf{r}_i) \mathbf{p}_i + \mathbf{E}^b(\mathbf{r},\omega), \label{Eq:Efielddipol} \\
	\mathbf{H}(\mathbf{r},\omega) &= \omega^2\mu_0 \sum_{i = 0}^N \mathds{G}^{\rm HE}(\mathbf{r},\mathbf{r}_i) \mathbf{p}_i + \mathbf{H}^b(\mathbf{r},\omega) \label{Eq:Hfielddipol}
\end{align}
which are generated by the fluctuational background fields $\mathbf{E}^b(\mathbf{r})$ and $\mathbf{H}^b(\mathbf{r})$ and the induced and fluctuational dipoles of all particles ($i = 1, \ldots, N$)
\begin{equation}
  \mathbf{p}_i = \mathbf{p}_i^{\rm ind} + \mathbf{p}_i^{\rm fl}
\end{equation}
where the induced dipole moments 
\begin{equation}
  \mathbf{p}_i^{\rm (ind)} = \epsilon_0\underline{\underline{\alpha}}_i \mathbf{E}(\mathbf{r}_i)
\label{Eq:induceddipolemoment}
\end{equation}
can be expressed in terms of the polarizability tensor $\underline{\underline{\alpha}}_i$ of the i$^{\rm th}$ dipole.
Here we have introduced the electric and magnetic Green functions $\mathds{G}^{\rm EE}$ and $\mathds{G}^{\rm HE}$ generated by electric dipole moments as defined in Ref.~\cite{WE1984} which are now not necessarily the vacuum Green functions, but the general Green functions taking the geometry and material properties of the backround into account. As a consequence the total electric field $\mathbf{E}_i = \mathbf{E}(\mathbf{r}_i)$ at the position of the $i$-th dipole is given by the field contributions due to the fluctuating dipole moments $\mathbf{p}_j^{\,\,\rm fl}$ of all other dipoles $j \neq i$ and the background field $\mathbf{E}_i^{b} = \mathbf{E}^b(\mathbf{r}_i)$ including direct thermal emission and multiple scattering. It can be written as~\cite{RMetal2013}
\begin{equation}
    \begin{pmatrix} \mathbf{E}_{1} \\ \vdots \\ \mathbf{E}_{N} \end{pmatrix} =\boldsymbol{DT}^{-1}\begin{pmatrix} \mathbf{p}_1^{\rm fl} \\ \vdots \\ \mathbf{p}_N^{\rm fl}\end{pmatrix}+ (\boldsymbol{1}+\boldsymbol{DT}^{-1}\boldsymbol{A})\begin{pmatrix} \mathbf{E}_1^{\rm b} \\ \vdots \\ \mathbf{E}_N^{\rm b}\end{pmatrix}.
	    \label{Eq:electricfieldBlock}
\end{equation}
Similarly the induced dipole moments $\mathbf{p}_i$ for each particle $i$ can be expressed in terms of the fluctuating dipole moments of all other particles and the background field~\cite{RMetal2013}
\begin{eqnarray}
  \begin{pmatrix} \mathbf{p}_{1} \\ \vdots \\ \mathbf{p}_{N} \end{pmatrix} =\boldsymbol{T}^{-1}\begin{pmatrix} \mathbf{p}_1^{\rm fl} \\ \vdots \\ \mathbf{p}_N^{\rm fl}\end{pmatrix}+ (\boldsymbol{T}^{-1}\boldsymbol{A})\begin{pmatrix} \mathbf{E}_1^{\rm b} \\ \vdots \\ \mathbf{E}_N^{\rm b}\end{pmatrix}.
\label{Eq:dipolBlock}
\end{eqnarray}
The auxilliary $3N\times3N$-block matrices $\boldsymbol{D}$, $\boldsymbol{A}$, and $\boldsymbol{T}$ are defined as in Eqs.~(\ref{Eq:Dblock}), (\ref{Eq:Ablock}), (\ref{Eq:Tblock}) but with the vacuum Green function $\mathds{G}_0(\mathbf{r}_i,\mathbf{r}_j)$ replaced by $\mathds{G}_{ij}^{\rm EE} = \mathds{G}^{\rm EE}(\mathbf{r}_i, \mathbf{r}_j)$ and by $\boldsymbol{1}_{ij} = \delta_{ij}\mathds{1}$.

Equipped with this set of expressions it is now possible to derive the dissipated heat in a given dipole $i$ and the mean Poyting vector in a general $N$ dipole system. Analogous to (\ref{Eq:rate_work}) the mean power received by the i$^{\rm th}$ dipole is defined as the power dissipated in dipole $i$
\begin{equation}
\begin{split}
	\mathcal{P}_{i} &= \biggl\langle\frac{d\mathbf{p}_i(t)}{dt} \cdot \mathbf{E}_{i}(t)\biggr\rangle \\
		      &= 2{\rm Im}\int_{0}^{\infty}\!\!\frac{{\rm d}\omega}{2\pi}\, \omega \langle\mathbf{p}_i(\omega) \cdot \mathbf{E}_{i}^*(\omega)\rangle.  
\end{split}
\label{Eq:powerNdipoldef}
\end{equation}
Hence by definition the dissipated power inside dipole $i$, i.e.\ the heat flowing into that dipole, is positive. The mean Poynting vector due to the dipoles and the background fields is given by
\begin{equation}
\begin{split}
	\langle \mathbf{S}(\mathbf{r}) \rangle &= \langle \mathbf{E}(t) \times \mathbf{H}(t) \rangle \\
    &= 2{\rm Re}\int_{0}^{\infty}\!\!\frac{{\rm d}\omega}{2\pi}  \, \langle\mathbf{E}(\mathbf{r},\omega) \times \mathbf{H}^*(\mathbf{r},\omega)\rangle . 
\end{split}
  \label{Eq:PVNdipoldef}
\end{equation}
These expressions already include the fact that the fluctuational fields and dipole moments are stationary so that the mean power and mean Poynting vector do not depend on time. They can be evaluated by assuming that the fluctuational dipole moments and the background fields are in local thermal equilibrium at temperatures $T_i$ ($i = 1,\ldots,N$) and $T_b$. Then the mean values for the power and Poynting vector which are obviously given by the correlation functions of the fields and the dipole moments can be evaluated by employing the fluctuation-dissipation theorem~\cite{RK1966} and assuming that the background fields and the dipole moments are statistically independent, i.e.\ correlation functions between the background field and the fluctuating dipoles $\langle \mathbf{E}^{\rm b}\otimes \mathbf{p}_i \rangle$ vanish. For the fields the fluctuation-dissipation theorems are~\cite{GSA1975}
\begin{align}
	\langle\mathbf{E}^{\rm b}_{i} \otimes \mathbf{E}^{\rm b^*}_{j}\rangle &= 2\omega^2\mu_0\hbar\Big(n_b+\frac{1}{2}\Big)\frac{\mathds{G}_{ij}^{\rm EE}-{\mathds{G}_{ji}^{\rm EE}}^\dagger}{2\I},
	\label{Eq:FDTfielda} \\
	\langle\mathbf{E}^{\rm b}_i \otimes \mathbf{H}^{\rm b}_j \rangle &= 2\omega^2\mu_0\hbar\Big(n_b+\frac{1}{2}\Big)\frac{\mathds{G}^{\rm EH}_{ij}-\mathds{G}^{\rm HE^\dagger}_{ji}}{2\ri}
  \label{Eq:FDTfieldb}
\end{align}
using the notation $\mathds{G}_{ij}^{\rm EH} = \mathds{G}^{\rm EH}(\mathbf{r}_i, \mathbf{r}_j)$ and $\mathds{G}_{ij}^{\rm HE} = \mathds{G}^{\rm HE}(\mathbf{r}_i, \mathbf{r}_j)$. Analoguously, for the dipole moments the fluctuation-dissipation theorem is determined by~\cite{RMetal2013}
\begin{equation}
  \langle\mathbf{p}_i^{\rm fl}\otimes {\mathbf{p}_j^{\rm fl}}^* \rangle = 2\epsilon_0\hbar \delta_{ij} \Big(n_i +\frac{1}{2}\Big)\chimat_i .
  \label{Eq:FDTdipol}
\end{equation}
The generalized susceptibility of the $i^{\rm th}$ particle is given by~\cite{RMetal2013,RMAEetal2017,FHandSAB2019}
\begin{equation}
  \chimat_{i} = \frac{\alphamat_i-\alphamat^\dagger_i}{2\I}-k_0^2\alphamat_i\frac{\mathds{G}_{ii}^{\rm EE}-{\mathds{G}_{ii}^{\rm EE}}^\dagger}{2\I}\alphamat^\dagger_i.
	\label{Eq:generalizedSusceptibility}
\end{equation}
The first term of the generalized susceptibility describes simply the intrinsic absorptivity of the dipole, whereas the second term is a radiation correction taking into account that the dipole is coupled to the environment which modifies its absorptivity. In free space this second term simply reads $- k_0^3/(6 \pi) \alphamat\,\alphamat^\dagger$~\cite{RMAEetal2017}. Hence, by comparing with Eq.~(\ref{Eq:power_single}) we see that with $\chimat_i$ we retrieve the absorptivity of a dipole $i$ placed in vacuum for the isotropic case $\alphamat_i = \alpha_i \mathds{1}$.

Inserting the expressions for the fields and dipole moments into the definitions (\ref{Eq:powerNdipoldef}) one obtains for the mean power received by particle $i$~\cite{Ott2020}
\begin{equation}
	\mathcal{P}_{i} = 3 \int_{0}^{\infty}\frac{{\rm d}\omega}{2\pi}\hbar\omega \sum_{j = 1}^{N}(n_j-n_b) \mathcal{T}_{ij} 
\label{Eq:PowerDipolGeneral}
\end{equation}
where the transmission coefficients are defined as
\begin{equation}
	{\mathcal{T}}_{ij} = \frac{4}{3} \epsilon_0 \Im {\rm Tr}\Big[\blockt_{ij}\chimat_j(\boldsymbol{DT}^{-1})_{ij}^\dagger\Big].
\label{Eq:Transmissionij}
\end{equation}
Equation~(\ref{Eq:PowerDipolGeneral}) is the general expression for the dissipated power or heat flowing into a dipole at temperature $T_i$ surrounded by $N-1$ dipoles at temperatures $T_j$ ($j \neq i$) described by an anisotropic or even non-reciprocal polarizability immersed in a general environment or background at temperature $T_b$ which can itself be anisotropic or non-reciprocal, properties which are taken into account via the polarizability and the Green function. 
In general, if either the dipole or the background or both are non-reciprocal one has $\mathcal{T}_{ij} \neq \mathcal{T}_{ji}$~\cite{LZetal2018,FHandSAB2019}. It should be noted that in the literature a variety of different equivalent expressions for the transmission coefficients $\mathcal{T}_{ij}$ can be found as for instance in \cite{PBAetal2011,RMetal2013,MN2014,RMAEetal2017,AOetal2019b,Ott2020}. Finally, when replacing $n_j - n_b$ by $n_j - n_i + n_i - n_b$ Eq.~(\ref{Eq:PowerDipolGeneral}) can be recast into the more intuitive form~\cite{RMetal2013}
\begin{equation}
	\mathcal{P}_{i} = 3 \int_{0}^{\infty}\frac{{\rm d}\omega}{2\pi}\hbar\omega \biggl( \sum_{j\neq i} (n_j-n_i) {\mathcal{T}}_{ij} + (n_i - n_b) \mathcal{T}_{ib} \biggr)
\label{Eq:PowerDipolGeneralb}
\end{equation}
with $\mathcal{T}_{ib} =\sum_{j}{\mathcal{T}}_{ij}$. This formula has the advantage that it clearly expresses the power dissipated into dipole $i$ by the power exchanged between dipole $i$ and all the other dipoles and the power of dipole $i$ exchanged with the environment. 

Similarly, by starting with the definition of the mean Poynting vector in (\ref{Eq:PVNdipoldef}) one obtains for the spectral heat flux for the $N$ fluctuating dipoles immersed in a background~\cite{Ott2020}
\begin{equation}
\begin{split}
	\langle S_{\omega,\alpha}\rangle &=4\hbar\omega^2\mu_0k_0^2 \sum_{\beta,\gamma = x,y,z} \epsilon_{\alpha\beta\gamma} {\rm Re}\Bigg[\\
	&\quad \sum_{j = 1}^N(n_j-n_b) \sum_{i = 1}^N \bigl(\mathds{G}^{\rm EE}_{0i}\blockt_{ij} \bigr) \chimat_{j} \sum_{k = 1}^N \bigl(\mathds{G}^{\rm HE}_{0k}\blockt_{kj}\bigr)^\dagger \\
	&\quad +\frac{n_b}{2 \ri} \sum_{i,j=1}^{N} \biggl(\mathds{G}^{\rm EE}_{0i}\blockt_{ij}\alphamat_j\mathds{G}^{\rm EH}_{j0} 
	-\big(\mathds{G}^{\rm HE}_{0i}\blockt_{ij}\alphamat_j\mathds{G}^{\rm EE}_{j0}\big)^\dagger \biggr) \\
&\quad +\frac{n_b}{k_0^2}\Big(\frac{\mathds{G}^{\rm EH}_{00}-\mathds{G}^{\rm HE^\dagger}_{00}}{2\ri}\Big) \Bigg]_{\beta\gamma}
\end{split}
\label{Eq:PVDipolgeneral}
\end{equation}
where $\epsilon_{\alpha\beta\gamma}$ is the Levi-Civita tensor and $\mathds{G}_{0i}^{\rm EE} = \mathds{G}^{\rm EE}(\mathbf{r},\mathbf{r}_i), \mathds{G}_{00}^{\rm EE} = \mathds{G}^{\rm EE}(\mathbf{r},\mathbf{r})$, etc. The first term describes the heat flux emitted by the particles into the background, the last term describes the heat flux of the background fields without the dipoles, and the second term describes the interference of the background fields due to the presence of the dipoles. In the case that the background geometry fulfills Lorentz reciprocity~\cite{CCetal2018} the last terms vanishes since then ${\mathds{G}^{\rm EH}_{ij}}^\dagger = - {\mathds{G}^{\rm HE}_{ji}}^*$. This simply means that if we have no dipoles the mean heat flux in the background $\langle \mathbf{S}^b \rangle = \langle \mathbf{E}^b(t) \times \mathbf{H}^b(t) \rangle$ which is at local thermal equilibrium vanishes. On the other hand, as shown by Silvereinha~\cite{MGS2017} for a non-reciprocal background there can be a non-vanishing mean Poynting vector even in thermal equilibrium.

In certain cases the heat flux between the dipolar objects is dominant so that the emission into the background is negligibly small. If for example the dipoles are placed into a vacuum at temperature $T_b$ then the power exchanged between the dipoles is for distances much smaller than the thermal wavelength, i.e.\ in the near-field regime, much larger then the power exchange with the background~\cite{RMetal2013}. When placing the dipolar objects, for instance, close to a substrate then the inter-dipole heat exchange is still dominating if the distance between the dipoles is much smaller than the distance to the substrate~\cite{Ott2020}. In such situations, the $N$-dipole system can also be treated as a closed system. This can be done by neclegting in the above expressions the heat exchange between the dipoles and the background and the heat flux due to the background fields so that
\begin{equation}
	\mathcal{P}_{i} = 3 \int_{0}^{\infty}\frac{{\rm d}\omega}{2\pi}\hbar\omega \sum_{j\neq i} (n_j-n_i) \mathcal{T}_{ij}
\label{Eq:PowerDipolGeneralNoBG}
\end{equation}
and
\begin{equation}
\begin{split}
	\langle S_{\omega,\alpha}\rangle &=4\hbar\omega^2\mu_0k_0^2 \!\! \sum_{\beta,\gamma = x,y,z}\!\! \epsilon_{\alpha\beta\gamma}\sum_{j = 1}^{N} n_j \\
	&\quad \times{\rm Re}\Bigg[ \sum_{i = 1}^N \bigl( \mathds{G}^{\rm EE}_{0i}\blockt_{ij} \bigr) \chimat_{j} \sum_{k = 1}^N \bigl(\mathds{G}^{\rm HE}_{0k}\blockt_{kj}\bigr)^\dagger \Bigg]_{\beta\gamma}.
\end{split}
\label{Eq:PVDipolgeneralNoBG}
\end{equation}
Note that, even though $\mathcal{P}_i$ contains only the power dissipated in dipole $i$ due to the heat exchange with all other dipoles, the mean Poynting vector includes also the thermal radiation of all dipoles into their background which is assumed to have zero temperature. To be fully consistent with the assumption that the background is simply removed from the description the second term in the generalized susceptibility $\chimat_j$ in Eq.~(\ref{Eq:generalizedSusceptibility}) might be neglected. For systems where the dipole approximation is valid this term is typically very small and can therefore often be neglected anyway. 

The same equations can be obtained by neglecting in the derivation right from the start any contribution from the background fields. In this case $\mathcal{P}_{i}$ can also be obtained by considering the power exchanged between all pairs of dipoles, only, as originally done in many works as for instance in Ref.~\cite{PBAetal2011}. To this end, the heat dissipated in dipole $i$ due to a fluctuational field $\mathbf{E} _{ij} = \bigl(\boldsymbol{DT}^{-1}\bigr)_{ij} \mathbf{p}^{\rm fl}_j$ generated by a fluctuational dipole $\mathbf{p}^{\rm fl}_j$ is considered as the power flow from dipole $j$ to $i$ yielding
\begin{equation}
\begin{split}
    \mathcal{P}_{j \rightarrow i} &= \biggl\langle\frac{d\mathbf{p}_i(t)}{dt}\cdot\mathbf{E}_{ij}(t)\biggr\rangle \\
              &= 3\int_{0}^{\infty}\frac{{\rm d}\omega}{2\pi} \hbar \omega n_j \mathcal{T}_{ij}(\omega).
\end{split}
\label{Eq:waermestromNij}
\end{equation}
Then the power dissipated by the $i^{\rm th}$ dipole is just the sum of the power flowing between dipole $i$ and the other objects 
\begin{equation}
\begin{split}
	\mathcal{P}_{i} &= \sum_{j \neq i} \bigl( \mathcal{P}_{j \rightarrow i} - \mathcal{P}_{i \rightarrow j}\bigr) \\
	                &= \sum_{j \neq i} 3\int_{0}^{\infty}\frac{{\rm d}\omega}{2\pi} \hbar \omega \Big( n_j \mathcal{T}_{ij}(\omega)- n_i\mathcal{T}_{ji}(\omega)\Big).
\end{split}
\label{Eq:waermenetto}
\end{equation}
Since in thermal equilibrium $\mathcal{P}_i = 0$ we can derive the condition~\cite{ILandPBA2017,AOetal2019b}
\begin{equation}
  \sum_{j \neq i} \mathcal{T}_{ij}(\omega) =  \sum_{j \neq i} \mathcal{T}_{ji}(\omega).
\label{eq:TijTjiEq}
\end{equation}
This condition simply expresses the fact that even though $T_{ij} \neq T_{ji}$ in general, the heat flux from $i$ to all other dipoles [rhs of (\ref{eq:TijTjiEq})] must be the same as the heat flow from all other dipoles to $i$ [lhs of (\ref{eq:TijTjiEq})] in equilibrium. By inserting this equilibrium condition into the second term of (\ref{Eq:waermenetto}) we retrieve (\ref{Eq:PowerDipolGeneralNoBG}).

\subsubsection{Non-additivity in many-dipole systems}
\label{Sec:NonAddDip}

Before we discuss the non-additivity of the power exchange in a $N$-dipole system based on Eq.~(\ref{Eq:waermestromNij}), let us focus on the power exchange between two dipoles ($N = 2$). The first derivation of the heat exchange between two dipolar objects within the framework of FE was given in \cite{AIV2001} and extended to take magnetic dipole moments into account~\cite{POCEtAl2008,AM2012} as well as multipolar contributions \cite{APMEtAl2008,DBCN2019}. A quantum dynamical description can be found in~\cite{SAB2013,GB2016} and a discussion of different prefactors found in the literature in \cite{Dedkov2011,KS2019}. Using our expression in Eq.~(\ref{Eq:waermestromNij}) for $N=2$ and temperatures $T_1 \neq 0\,{\rm K}$ and $T_2 = 0\,{\rm K}$ we obtain
for the power received by dipole $2$ 
\begin{equation}
   \mathcal{P}_{1 \rightarrow 2} = 3 \int_{0}^{\infty}\frac{{\rm d}\omega}{2\pi}\hbar\omega n_1 \mathcal{T}_{21}.
	\label{Eq:P12dipole}
\end{equation}
The transmission coefficient $\mathcal{T}_{12}$ can be expressed as
\begin{equation}
	{\mathcal{T}}_{21} = \frac{4}{3} k_0^4 \Im {\rm Tr}\Big[ \mathds{D}^{-1} \mathds{G}_{21} \chimat_1 \bigl(\mathds{D}^{-1} \mathds{G}_{21}\bigr)^\dagger \tilde{\chimat}_2 \Big].
\label{Eq:Transmission12b}
\end{equation}
with $\mathds{D} = (\mathds{1} + k_0^4 \mathds{G}_{21} \alphamat_1 \mathds{G}_{12} \alphamat_2)$
introducing the generalized susceptibility
\begin{equation}
   \tilde{\chimat}_{2} = \frac{\alphamat_2-\alphamat^\dagger_2}{2\I}-k_0^2\alphamat_2^\dagger \frac{\mathds{G}_{22}-\mathds{G}_{22}^\dagger}{2\I}\alphamat_2.
\label{Eq:generalizedSusceptibilityTilde}
\end{equation}
Note that this general susceptibility only differs slightly from the definition (\ref{Eq:generalizedSusceptibility}), whereas for isotropic dipoles both definitions coincide. This is the most general expression of the transmission coefficient for two dipolar objects in a given environment of any shape. The appearance of the terms $\mathds{D}^{-1}$ in the transmission coefficient are due to multiple interactions between the dipoles. Therefore the hybridization of any localized dipole resonance due to the strong coupling for small distances is accounted for in this expression. Note that Eq.~(\ref{Eq:Transmission12b}) resembles Eq.~(36) of~\cite{RMAEetal2017} but with the slight difference that in that work $\chimat_2$ is used instead of $\tilde{\chimat}_2$. On the other hand, the form of the transmission coefficient (\ref{Eq:Transmission12b}) has also been found in \cite{MKetal2012,FHandSAB2019} within the scattering approach of \cite{MKetal2012}. However, within the range of validity of the dipole approximation the second term in $\chimat$ or $\tilde{\chimat}$ typically can be neglected and many works
simply use
\begin{equation}
   \chimat_i \approx \tilde{\chimat}_i = \frac{\alphamat_i-\alphamat^\dagger_i}{2\I}.
\end{equation}

\begin{figure}
	\begin{center}\includegraphics[width=0.3\textwidth]{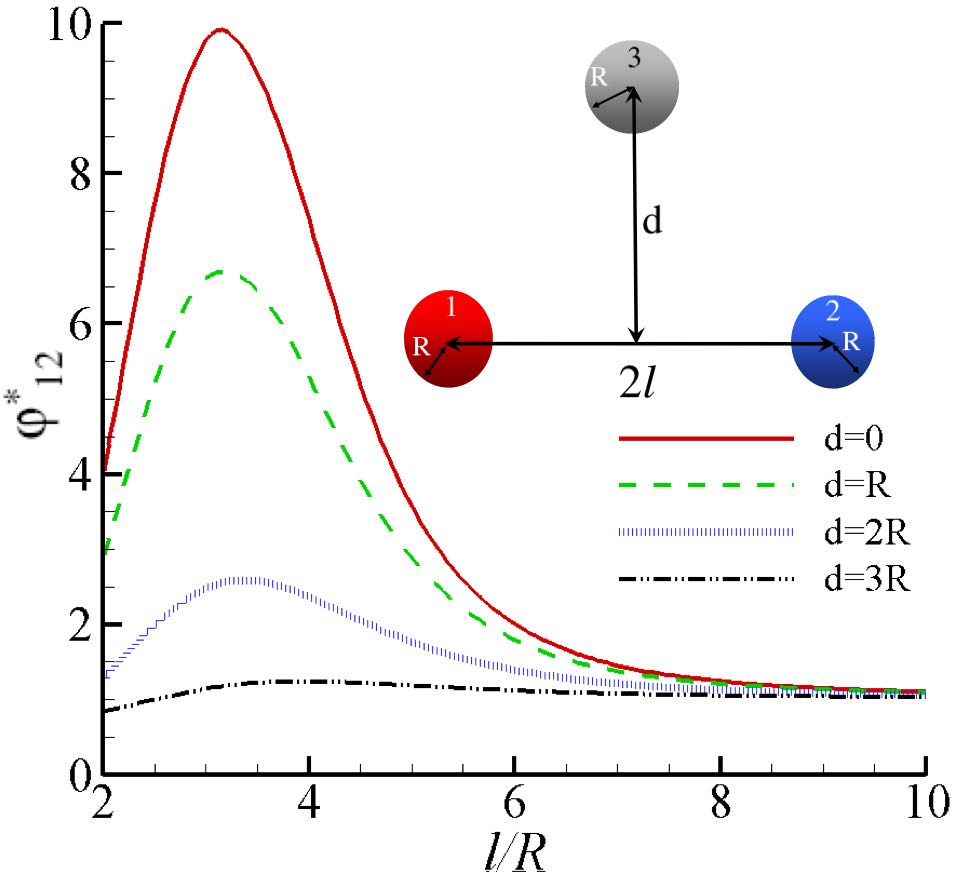}
	\end{center}
	\caption{Power flow exchanged between two SiC nanoparticles at $T_1 = 300\,{\rm K}$ (red) and at $T_2 = 0\,{\rm K}$ (blue) in presence of a third SiC nanoparticle at temperature $T_3=0\,{\rm K}$ (grey) and normalized by the power exchanged between two isolated particles, i.e.\ $\varphi_{12}^{*} = \mathcal{P}_{1 \rightarrow 2} (T_1,T_2,T_3)/\mathcal{P}_{1 \rightarrow 2} (T_1,T_2)$. From~\cite{PBAetal2011}. }
	\label{fig:HFthreeparticles}
\end{figure}

\begin{figure}
	\begin{center}\includegraphics[angle=-90,width=0.45\textwidth]{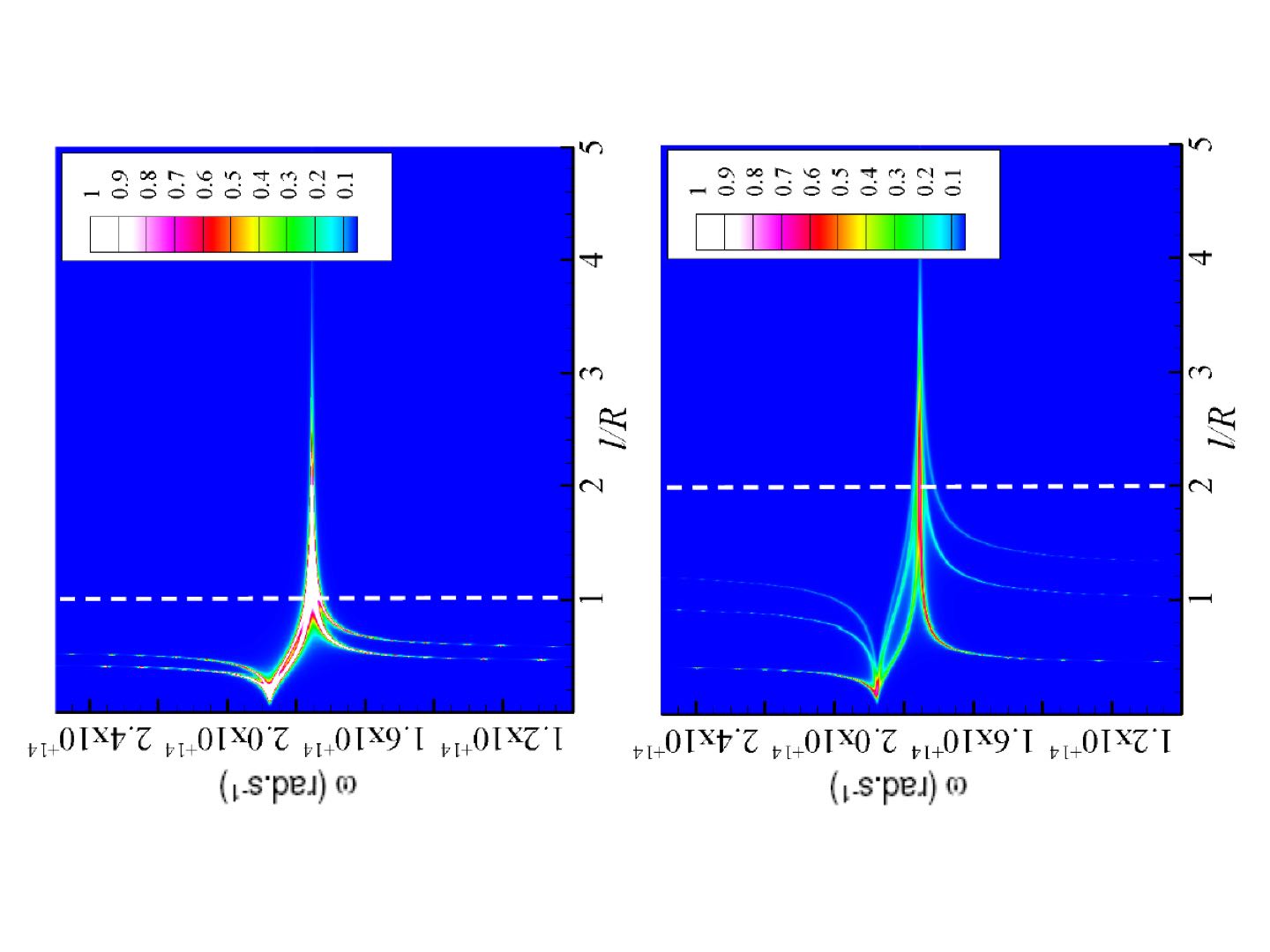}
	\end{center}
	\caption{Transmission coefficient $\mathcal{T}_{21}$ between (a) two SiC nanoparticles and (b) between two SiC nanoparticles in presence of a third SiC nanoparticles as in Fig.~\ref{fig:HFthreeparticles} for $d = 0$. The dashed line marks the region where the particles would touch. The unphysical region beyond this line is shown to illustrate the hybridization mechanism of dipolar resonances, which can be nicely seen in that region. From~\cite{PBAetal2011}. }
	\label{fig:TCthreeparticles}
\end{figure}

Now, when adding a third dipole at $T_3 = 0\,{\rm K}$ then we still can use (\ref{Eq:P12dipole}) and (\ref{Eq:Transmission12b}) to quantify the power exchanged between dipole $1$ and $2$. The main difference is that $\boldsymbol{T}^{-1}_{12}$ now also contains the coupling with the third dipole. Hence the sheer presence of the third particle changes the transmission coefficients due to the fact that it changes the mode structure which is for dipoles with a localized resonance again due to the hybridization for three dipoles this time (see Fig.~\ref{fig:TCthreeparticles}) responsible for the broadening of the absorption spectrum as discussed in Sec.~\ref{Sec:Absorption}. As a consequence, the presence of a third dipole changes the power exchange $\mathcal{P}_{1 \rightarrow 2}$ proving that the heat exchange in an $N$-dipole system is non-additive. This formalism is only valid for interparticles distances larger than $4R$, $R$ being the radius of the particles. It can be extended to smaller distances by including multipolar contributions~\cite{BCandAN2019}.

This many-body effect can be exploited to enhance for example the exchanged power between two dipolar objects $1$ and $2$ by bridging the distance via a third dipole which is placed between $1$ and $2$ as shown in~\cite{PBAetal2011} (see Fig.~\ref{fig:HFthreeparticles}). However, it should be kept in mind that the heat flux between two dipoles in a $N$-dipole system cannot be arbitrarily enhanced. As discussed in~\cite{PBAetal2011} it can be easily shown that each of the conductance between two dipoles can be at most 3 times the quantum of thermal conductance. Nonetheless, this upper limit is difficult to achieve leaving much space for optimmizations. Several works have shown that it is possible to tailor the inter-dipole heat flux via a third dipole or third object. For example, \cite{RMetal2013} has studied the relaxation dynamics for the three-body configuration and \cite{JDetal2017} have also included the possibility to have a magnetic polarizability as needed to describe metallic nanoparticles in the infrared. Furthermore, using prolate~\cite{MN2014,RIetal2014,MN2015} or oblate~\cite{ORCandMN2016} spheroidal nano-particles it has been demonstrated that by changing the relative orientation of the nano-particles and in particular an intermediate nanoparticle the heat flux can be switched and enhanced efficiently (see also Fig.~\ref{Fig:Switch}. Furthermore, the coupling of two nanoparticles via the surface modes of an interface or intermediate medium has been studied as discussed in detail in Sec.~\ref{Sec:Guiding}. Finally, the non-additivity of the heat exchange has consequences for the transport properties in nano-particle chains and complex nanoparticle networks as discussed in detail in Sec.~\ref{Sec:ComplexNetworks}.

\subsubsection{T-DDA (as example of application)}

The expressions for the heat exchange in systems with $N$ dipolar objects in (\ref{Eq:PowerDipolGeneralNoBG}) without the contribution of the background as derived by~\cite{PBAetal2011} have been employed first by~\cite{SEandMF2014} to determine the heat exchange between macroscopic objects with isotropic and later by~\cite{RMAEetal2017} for macroscopic objects with anisotropic and magneto-optical material properties. The idea is to replace the macroscopic objects by a great number $N$ of small cubes of volume $V_i$ ($i = 1,\ldots,N$) which can be approximated as dipoles with the corresponding polarizabilities. In Ref.~\cite{RMAEetal2017} the polarizability including the radiative corrections, as rederived by~\cite{SAetal2010} and originally also used by~\cite{BTD1988}, writes
\begin{equation}
	\alphamat_i = \biggl( \mathds{1} - i\frac{k^3_0}{6\pi}
	\alphamat_{0i}\biggr)^{-1} \alphamat_{0i}
\end{equation}
 in terms of the quasistatic polarizability
\begin{equation}
  \alphamat_{0i} = 3 V_i (\boldsymbol{\varepsilon} - \mathds{1})(\boldsymbol{\varepsilon} + 2 \mathds{1})^{-1}.
  \label{Eq:quasistaticPolarizability}
\end{equation}
Note that in Ref.~\cite{SEandMF2014} another expression for the dressed polarizability has been used known as the strong form of the coupled dipole method. A detailed discussion on the different expressions of the dressed polarizabilities in the context of classical coupled dipole method~\cite{PP1973} has been given by~\cite{AL1992}. 

\begin{figure}
	\begin{center}
	  \includegraphics[width=0.45\textwidth]{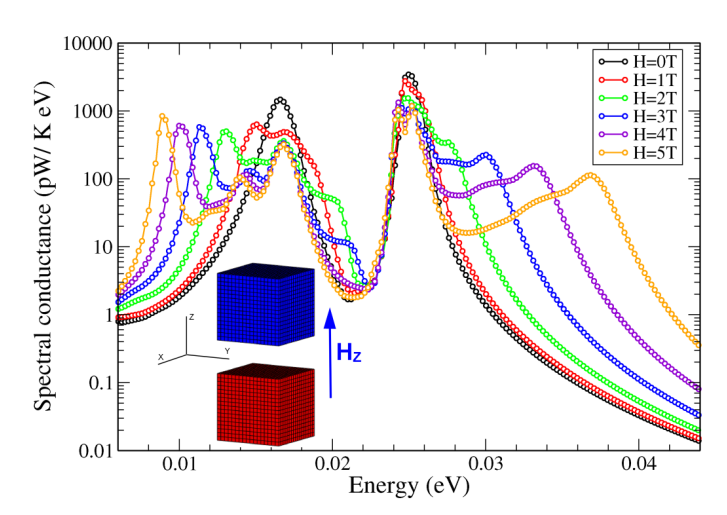}
	  \includegraphics[width=0.45\textwidth]{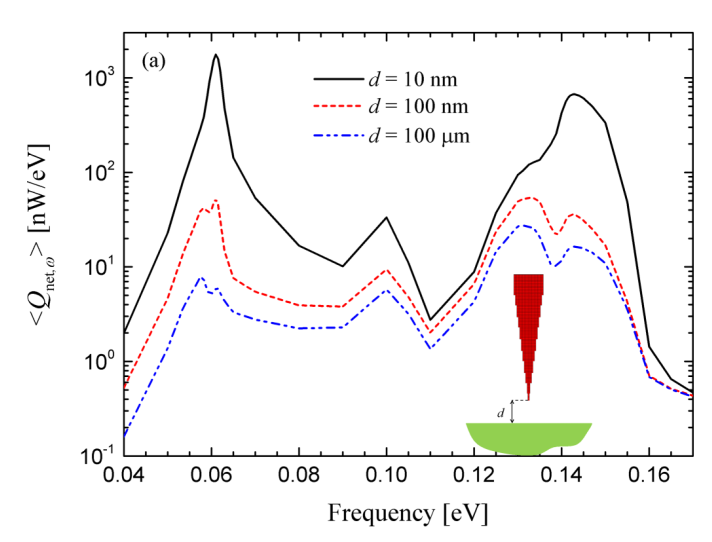}
	\end{center}
	\caption{(a) Spectral conductance as a function of the energy for InSb
             cubes with a cube side of 1 micron separated by a 500 nm gap, at
             T = 300 K, and for various values of the magnetic field H applied
             along the z direction. The inset shows the discretization geometry: the number of dipoles per cube is 4913 (each one has an edge
             length of 59 nm). From~\cite{RMAEetal2017}.
            (b) Spectral heat flux between a silica probe and silica surface. From~\cite{SEandMF2016}.}
	\label{fig:DDAcubes}
\end{figure}

This method known as discrete dipole approximation (DDA) for describing thermal radiation phenomena between macroscopic objects has been coined~\cite{SEandMF2014} thermal discrete dipole approximation (T-DDA). It has been succesfully employed to determine the heat flux between macroscopic reciprocal and non-reciprocal cubes and spheres~\cite{SEandMF2014,SEetal2015,RMAEetal2017,RMAEetal2018}, and also for the heat flux between a sharp conical tip and a planar substrate~\cite{SEandMF2016}. In principle this method can also be used to determine the heat flux between two macroscopic objects in arbitrary many-body systems. As discussed in \cite{SEetal2015} in detail, the large number of dipolar subvolumes needed to describe macroscopic objects or have a convergent numerical result sets a certain limit to this numerical method. See also the discussion in Sec.~\ref{sec-numerical}.

Finally, the TDDA method also allows for determining the thermal emission of macroscopic objects by calculation of the mean Poynting vector from Eq.~(\ref{Eq:PVDipolgeneralNoBG}) in the far-field regime~\cite{RMAEetal2017}. This can also be done with a standard DDA by determing the absorptivity as discussed in Sec.~\ref{Sec:Absorption} of the macroscopic object modelled by an assembly of dipoles and then using the Kirchhoff law to determine the emissivity. Now, the main advantage of the TDDA is that it allows to attribute to each volume element a given temperature. Hence, TDDA opens up the possibility to calculate thermal emission of macroscopic objects with a given temperature distribution, whereas the standard DDA can only handle emission of isothermal objects or dipolar assemblies. Note, that the assumption of local thermal equilibrium sets strict bounds to the spatial variation of temperature distributions~\cite{WE1984}.

\subsection{Heat flux in macroscopic many-body systems}\label{Sec:HFmany}

In the last section we have described a formalism allowing to account for the heat exchange in an arbitrary set of dipolar particles. As clarified above, although formally and computationally simpler, this framework is limited in terms of distance between the particles. For this reason, in the last decade several theoretical schemes have been developed to account for the heat transfer in configurations of two or more macroscopic bodies. The purpose of these techniques is to address bodies with in principle arbitrary geometry and optical properties. As we have seen in Secs.~\ref{sec-geometry} and \ref{sec-numerical}, several techniques have been introduced to successfully treat this problem. We are going to focus here on scattering-matrix techniques, where each macroscopic body is described in terms of its scattering operators, accounting for its response to an incoming electromagnetic field.

\subsubsection{Scattering-matrix formalism}

\begin{figure}
	\begin{center}\includegraphics[width=0.4\textwidth]{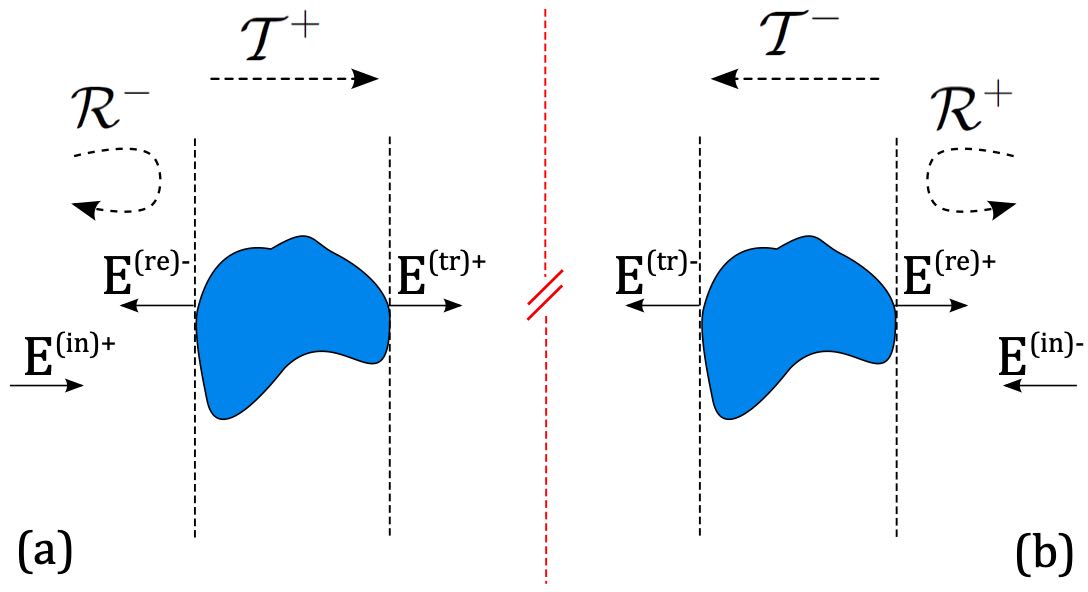}
	\end{center}
	\caption{Definition of reflection and transmission operators associated with an individual body. From \cite{RMandMA2011b}.}
	\label{fig:DefRT}
\end{figure}

Two closely-related formalisms based on this approach have been introduced between 2009 and 2011 by Bimonte~\cite{GB2009}, Kr\"{u}ger and collaborators~\cite{MKetal2011,MKetal2012} and Messina and Antezza~\cite{RMandMA2011a,RMandMA2011b}. The main difference between the these works is that Kr\"{u}ger et al. derive expressions which are suitable to any choice of basis for the electromagnetic field, while Messina and Antezza explicitly use a plane-wave basis, thus providing more explicit (albeit less general) expressions in terms of the individual scattering operators. In order to define these operators, the electric field in any region of the system is decomposed in plane waves as
\begin{equation}
\begin{split}
   \mathbf{E}^{\phi}(\mathbf{r},t)&=2\text{Re}\Biggl[\sum_p\int_0^{+\infty} \!\! \frac{\rd\omega}{2\pi} \int\frac{\rd^2\boldsymbol{\kappa}}{(2\pi)^2}\,\exp[\ri\mathbf{k}^\phi\cdot\mathbf{r}]\\
   &\qquad\,\times\exp[-\ri \omega t]\hat{\bbm[\epsilon]}_p^\phi(\boldsymbol{\kappa},\omega)E_p^{\phi}(\boldsymbol{\kappa},\omega)\Biggr],
\end{split}
\end{equation}
where $\omega$ is the frequency, $\boldsymbol{\kappa}=(k_x,k_y)$ the projection of the wavevector on the $x$-$y$ plane, $p$ the polarization index, taking values 1 (transverse electric) and 2 (transverse magnetic), 
$\phi$ the propagation direction along the $z$ axis. Moreover, $\mathbf{k}^\phi=(\boldsymbol{\kappa},\phi k_z)$ is the full wavevector, while the unit polarization vectors are defined as follows:
\begin{equation}
\begin{split}
  \hat{\bbm[\epsilon]}_\text{TE}^\phi(\boldsymbol{\kappa},\omega)&=\hat{\mathbf{z}}\times\hat{\boldsymbol{\kappa}}=\frac{1}{\kappa}(-k_y\hat{\mathbf{x}}+k_x\hat{\mathbf{y}})\\
  \hat{\bbm[\epsilon]}_\text{TM}^\phi(\boldsymbol{\kappa},\omega)&=\frac{c}{\omega}(-\kappa\hat{\mathbf{z}}+\phi
k_z\hat{\boldsymbol{\kappa}}).
\end{split}
\end{equation}
Each body is described in terms of four scattering operators $\mathcal{R}^\phi (\omega)$ and $\mathcal{T}^\phi (\omega)$ ($\phi=+,-$), connecting the amplitudes $E_p^{\phi}(\boldsymbol{\kappa},\omega)$ of the incoming and scattered fields, as (suppressing the frequency arguments)
\begin{equation}
  \begin{split}
   E_p^{\text{(re)}\phi}(\boldsymbol{\kappa})&=\sum_{p'}\int \!\!\frac{\rd^2\boldsymbol{\kappa}'}{(2\pi)^2} \, \bra{p,\boldsymbol{\kappa}}\mathcal{R}^\phi\ket{p',\boldsymbol{\kappa}'}E_{p'}^{\text{(in)}-\phi}(\boldsymbol{\kappa}'),\\
   E_p^{\text{(tr)}\phi}(\boldsymbol{\kappa})&=\sum_{p'}\int \!\! \frac{d^2\boldsymbol{\kappa}'}{(2\pi)^2}\,\bra{p,\boldsymbol{\kappa}}\mathcal{T}^\phi\ket{p',\boldsymbol{\kappa}'}E_{p'}^{\text{(in)}\phi}(\boldsymbol{\kappa}'),
   \end{split}
\end{equation}
where each mode $(\omega,\boldsymbol{\kappa},p)$ of the scattered field has in general components from each mode $(\omega,\boldsymbol{\kappa}',p')$ of the incoming field, the frequency $\omega$ being conserved since we are addressing only stationary processes. The action of these operators is schematically represented in Fig.~\ref{fig:DefRT}.

At this stage, it is interesting to sketch the main steps and assumptions leading to the expression of the radiative heat flux on each body, which can be summarized as follows:
\begin{enumerate}
	\item The fields generated by the fluctuating charges inside each body are identified as the source fields, along with the environmental field in which the system is embedded.
	\item The correlation functions of the individual source fields are deduced from the assumption of local thermal equilibrium.
	\item The total field in each region is explicitly written, in terms of the source fields, as a result of the scattering (reflection and transmission) processes occurring due to the presence of the bodies.
	\item The correlation functions of the total field in each region can be deduced.
	\item These are used for the calculation of the average value of the Poynting vector.
\end{enumerate}
We stress that in point 2 the assumption of local thermal equilibrium is equivalent to stating that the statistical properties of the field emitted by each body are the same we would have if the body was at thermal equilibrium at its own temperature. The details about the derivation of such correlation functions can be found in Ref.~\cite{RMandMA2011b}. This step leads to a source correlation function equivalent to Eq.~\eqref{Eq:FDTdipol} already seen in the case of dipoles, with the difference that in this case the scattering operator, accounting for the geometric and optical properties of the body, will explicitly appear.

The steps described above allow to explicitly write the power absorbed by each body $i$ under the form
\begin{equation}
	\mathcal{P}_{i} = \text{Tr}\Biggl[\hbar\omega \biggl( \sum_{j\neq i} (n_j-n_i) {\mathcal{T}}_{ij} + (n_i - n_b)  \mathcal{T}_{ib} \biggr)\Biggr],
\label{Eq:PowerScatt}
\end{equation}
analogous to Eq.~\eqref{Eq:PowerDipolGeneralb} already encountered in the dipolar case, where the trace operator is defined as
\begin{equation}
  \text{Tr}\mathcal{A} = \sum_p\int\frac{\rd^2\boldsymbol{\kappa}}{(2\pi)^2}\int_0^{+\infty}\!\!\frac{\rd\omega}{2\pi}\,\bra{p,\boldsymbol{\kappa}}\mathcal{A}\ket{p,\boldsymbol{\kappa}}.
\end{equation}
We focus here on the contribution to the heat flux on body 1 associated with the presence of body 2. The corresponding transmission coefficient $\mathcal{T}_{12}$ reads
\begin{equation}
   \mathcal{T}_{12} = U^{(2,1)} \chi_2 U^{(2,1)\dag} \tilde{\chi}_1,
   \label{eq:T12}
\end{equation}
where $U^{(2,1)}=(1-\mathcal{R}^{(2)-}\mathcal{R}^{(1)+})^{-1}$ is the operator decribing the infinite series of reflections inside the \emph{cavity} formed by bodies 1 ans 2 and the generalized susceptibilities are defined as
\begin{align}
    \chi_2 &= f_{-1}(\mathcal{R}^{(2)-})-\mathcal{T}^{(2)-}{P}_{-1}^{\text{(pw)}}\mathcal{T}^{(2)-\dag} \\
    \tilde{\chi}_1 &= f_1(\mathcal{R}^{(1)+})-\mathcal{T}^{(1)-\dag}\mathcal{P}_1^{\text{(pw)}}\mathcal{T}^{(1)-}
\end{align}
by means of the auxiliary functions
\begin{equation}
  f_\alpha(\mathcal{R})=\begin{cases}\mathcal{P}_{-1}^{\text{(pw)}}-\mathcal{R}\mathcal{P}_{-1}^{\text{(pw)}}\mathcal{R}^\dag+\mathcal{R}\mathcal{P}_{-1}^{\text{(ew)}}-\mathcal{P}_{-1}^{\text{(ew)}}\mathcal{R}^\dag\\
 \hspace{4.8cm}\alpha=-1\\
  \mathcal{P}_1^{\text{(pw)}}-\mathcal{R}^\dag\mathcal{P}_1^{\text{(pw)}}\mathcal{R}+\mathcal{R}^\dag\mathcal{P}_1^{\text{(ew)}}s-\mathcal{P}_1^{\text{(ew)}}\mathcal{R}\\
  \hspace{4.8cm}\alpha=1\end{cases}.
\end{equation}
The operators $\mathcal{P}_n^\text{(pw)}$ and $\mathcal{P}_n^\text{(ew)}$ defined (for any integer $n$) as
\begin{equation}
  \bra{p,\boldsymbol{\kappa}}\mathcal{P}_n^\text{(pw/ew)}\ket{p',\boldsymbol{\kappa}'}=k_z^n\bra{p,\boldsymbol{\kappa}}\Pi^\text{(pw/ew)}\ket{p',\boldsymbol{\kappa}'},
  \end{equation}
where $\Pi^\text{(pw)}=\Theta(\omega-ck)$ and $\Pi^\text{(ew)}=\Theta(ck-\omega)$ are the projectors on the propagative and evanescent sector, respectively. 
The transmission coefficient $\mathcal{T}_{12}$ has the same form as in Eq.~(\ref{Eq:Transmission12b}) for two dipolar objects. By choosing the T-operator for dipolar objects or using the plane wave expansion of the T-operators both forms of transmission coefficients can be obtained from the general T-operator expression in \cite{MKetal2012,FHandSAB2019}.

This approach was later generalized to the case of three arbitrary bodies~\cite{RMandMA2014}. The Landauer-like expression \eqref{Eq:PowerScatt} of the power absorbed by each body remains valid, meaning that e.g. the flux on body 1 has contributions coming from bodies 2 and 3, as well as from the environment. It is interesting to investigate here the expression of the transmission coefficient $\mathcal{T}_{12}$ between bodies 1 and 2 in this three-body configuration. It reads
\begin{equation}
\begin{split}
  \mathcal{T}_{12}&=U^{(23,1)}\Bigl(f_{-1}(\mathcal{R}^{(23)-})-\mathcal{T}^{(2)-}U^{(3,2)}f_{-1}(\mathcal{R}^{(3)-})\\
  &\,\times U^{(3,2)\dag}\mathcal{T}^{(2)-\dag}\Bigr)U^{(23,1)\dag} \tilde{\chi}_1,
\end{split}\label{eq:T12mod}
\end{equation}
in which a two-body reflection operator (and the associated multi-reflection operator $U^{(23,1)}$) appears, defined as
\begin{equation}
  \mathcal{R}^{(23)-}=\mathcal{R}^{(2)-}+\mathcal{T}^{(2)-}U^{(3,2)}\mathcal{R}^{(3)-}\mathcal{T}^{(2)+}.
\end{equation}
We immediately see that Eqs.~\eqref{eq:T12} and \eqref{eq:T12mod} are different. The important message behind this comparison is that as for the dipolar case discussed in Sec.~\ref{Sec:NonAddDip} not only does the presence of body 3 introduce an additional source for the energy transfer on body 1, but it modifies the transmission coefficient $\mathcal{T}_{12}$, and consequently the way bodies 1 and 2 exchange heat. In other words, the third body in the system acts both as a source/sink of radiation and as a scatterer (independently of its temperature), modifying the transmission amplitudes of other channels. We conclude that Eq.~\eqref{eq:T12mod} is by itself a proof and a quantitative evaluation of the non-additive nature of RHT, in the simplest possible many-body system made of three bodies.

The same approach described in Sec.~\ref{Sec:HFmany} and applied both to two- and three-body systems has been generalized to some years later to the case of $N$ bodies~\cite{ILetal2017}. In this case, for the sake of simplicity, only planar bodies, i.e. parallel slabs of finite thickness separated by vacuum gaps, have been considered. This assumptions has two main advantages: first, the plane-wave development is particularly convenient for this configuration, since it fully suits its symmetry; moreover the translational invariance along the transverse coordinates makes all the scattering operators diagonal with respect to both $p$ and $\boldsymbol{\kappa}$, significantly simplifying all the expressions. We stress that, since we are dealing here with infinite systems, the power on each body has to be replaced with the heat flux $\Phi$ it receives (power per unit surface).

\subsubsection{Non-additivity in many-body systems}
\label{Sec:NonAddMac}

\begin{figure}
	\begin{center}\includegraphics[width=0.4\textwidth]{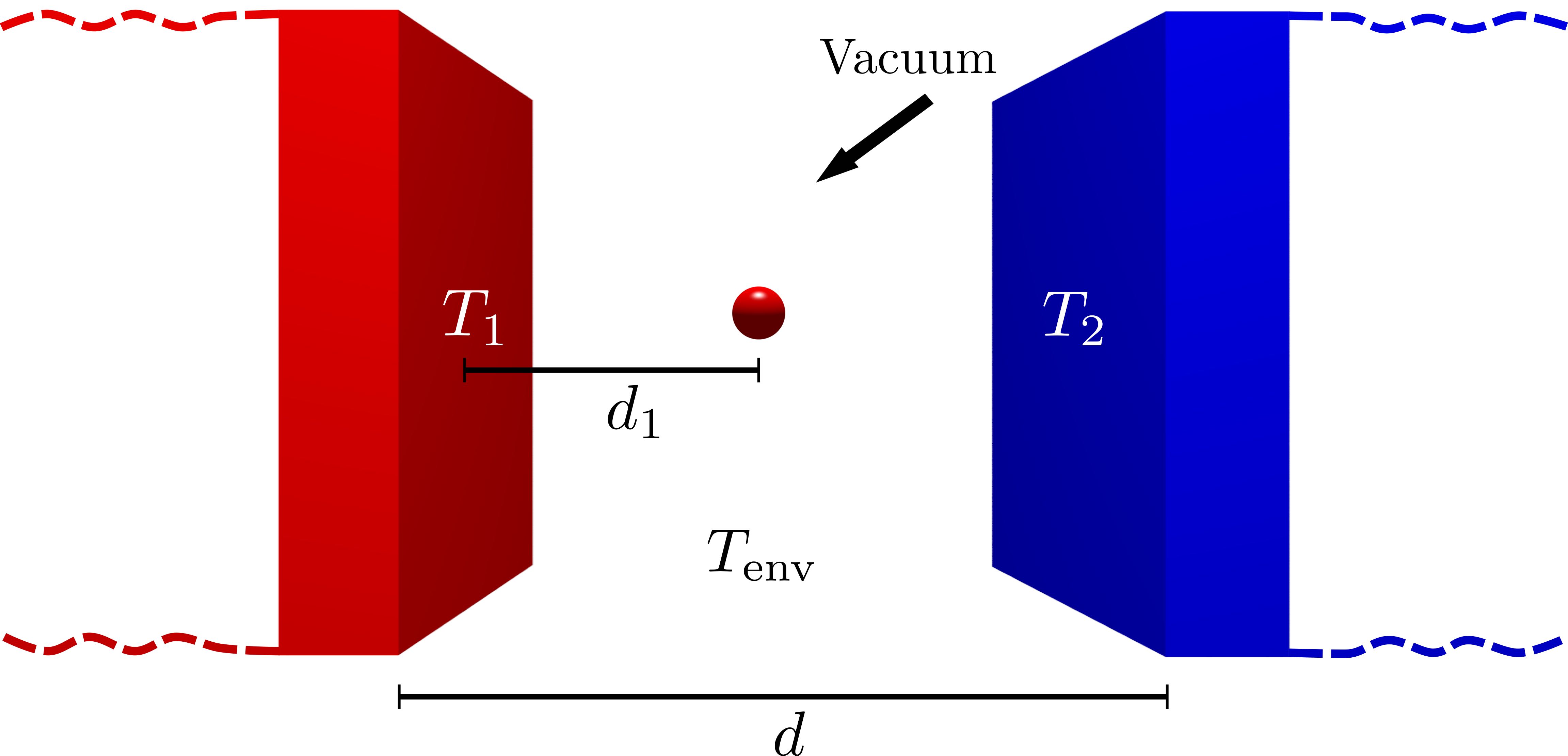}
	\end{center}
	\caption{Two planar slabs (bodies 1 and 2) are placed at distance $d$ and separated by vacuum. A particle of polarizability $\alpha$ is placed at distance $d_1$ from slab 1. From \cite{BMetal2017}.}
	\label{fig:3bodies}
\end{figure}

In the last Section, we have analytically shown the non-additivity of RHT. In the simplest case of three bodies, the appearance of the third one modifies the transmission amplitude $\mathcal{T}_{12}$, namely the way in which bodies 1 and 2 exchange energy. This is shown by the comparison of Eqs.~\eqref{eq:T12} and \eqref{eq:T12mod}. Apart from this formal comparison, it is interesting to address quantitatively the modification to the energy flux between bodies 1 and 2 due to the introduction of a third body in the system. This analysis has been performed analogous to the configuration discussed in \ref{Sec:NonAddDip} by~\cite{BMetal2017}, where the authors generalize the formalism developed in \cite{MKetal2012}, already valid in the scenario of $N$ bodies, to the case of the presence of a nonabsorbing background medium. In this work, the authors apply their formalism to the calculation of RHT between two SiC planar slabs (bodies 1 and 2) separated by a vacuum gap of thickness $d$, when a particle of polarizability $\alpha$ (assumed to be non-dispersive and real) is placed between them, at distance $d_1$ from body 1. The system is depicted in Fig.~\ref{fig:3bodies}.

The heat flux is evaluated after linearizing the general expressions with respect to the particle polarizability, assuming that the scattering contribution is weak. As a result, the heat flux $\Phi$ (power $\mathcal{P}$ per unit area) is conveniently expresses as
\begin{equation}
  \Phi = \Phi_\mathrm{vac} + \Delta \Phi,
\label{eq:CorrAtom}\end{equation}
where $\Phi_\mathrm{vac}$ is the well-known heat flux between two slabs separated by a vacuum gap, and the correction term $\Delta \Phi$ (proportional to $\alpha$ in the linearized approximation) is a direct description of the non-additivity of radiative heat flux.

\begin{figure}
	\begin{center}\includegraphics[width=0.4\textwidth]{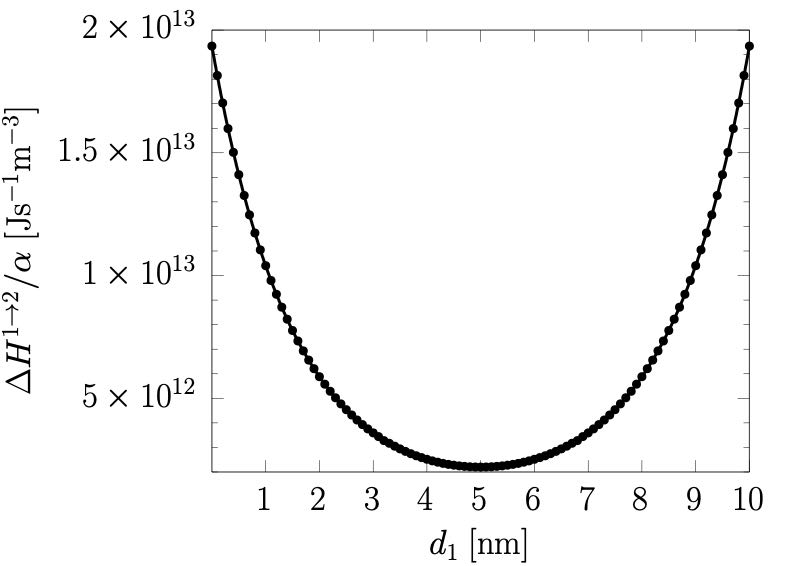}\\
	\includegraphics[width=0.4\textwidth]{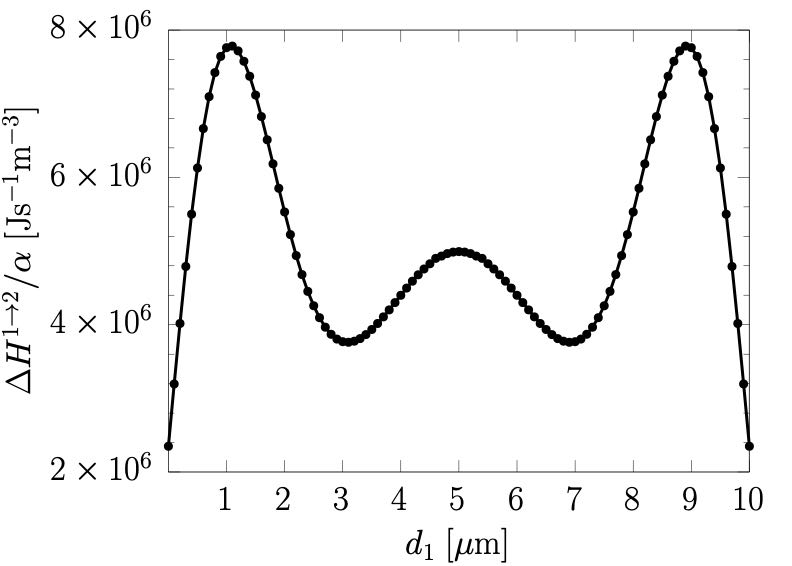}
	\end{center}
	\caption{Non-additive correction to the two-body heat flux $\Delta H^{1\to 2} = \Delta \Phi$ [see Eq.~\eqref{eq:CorrAtom}] in the presence of a particle of polarizability $\alpha$. The upper curve corresponds to the near-field configuration $d=10\,$nm, while the lower one corresponds to the far field ($d=10\,\mu$m). From \cite{BMetal2017}.}
	\label{fig:3bodiesresults}
\end{figure}

The non-additive correction is numerically evaluated for slab temperatures of 301\,K and 300\,K in two difference configurations: for a slab-slab distance $d=10\,$nm (near field) and for $d=10\,\mu$m (far field), as a function of the particle position $d_1$ (see Fig.~\ref{fig:3bodies}). The results are shown in Fig.~\ref{fig:3bodiesresults}. In both configurations we clearly observe the expected symmetry with respect to the central particle position $d_1=\frac{d}{2}$. In the near field, we observe that the effect is maximized when the atom is close to one of the two slabs. This reflects, apart from the symmetry of the system, the typical exponentially decreasing behavior of heat flux in the near field, which is in turn a consequence of the dominating contribution of evanescent waves. The situation is clearly different in the far field. First, not surprisingly, the effect is several orders of magnitude smaller that in the near field, Moreover, the external positions $d=0,d_1$ are now minima of the effect, which oscillates with respect to $d_1$. These oscillations are due to the interferences between propagating waves (dominating in this scenario), reflected between the two plates and scattered by the particle inside the cavity which change the local density of states~\cite{Doro2002,MFetAl2010} at the particle's position which is also known from the context of spontaneous emission of atoms and molecules within a such a configuration~\cite{NDetal2002}.

\begin{figure}
	\begin{center}\includegraphics[width=0.4\textwidth]{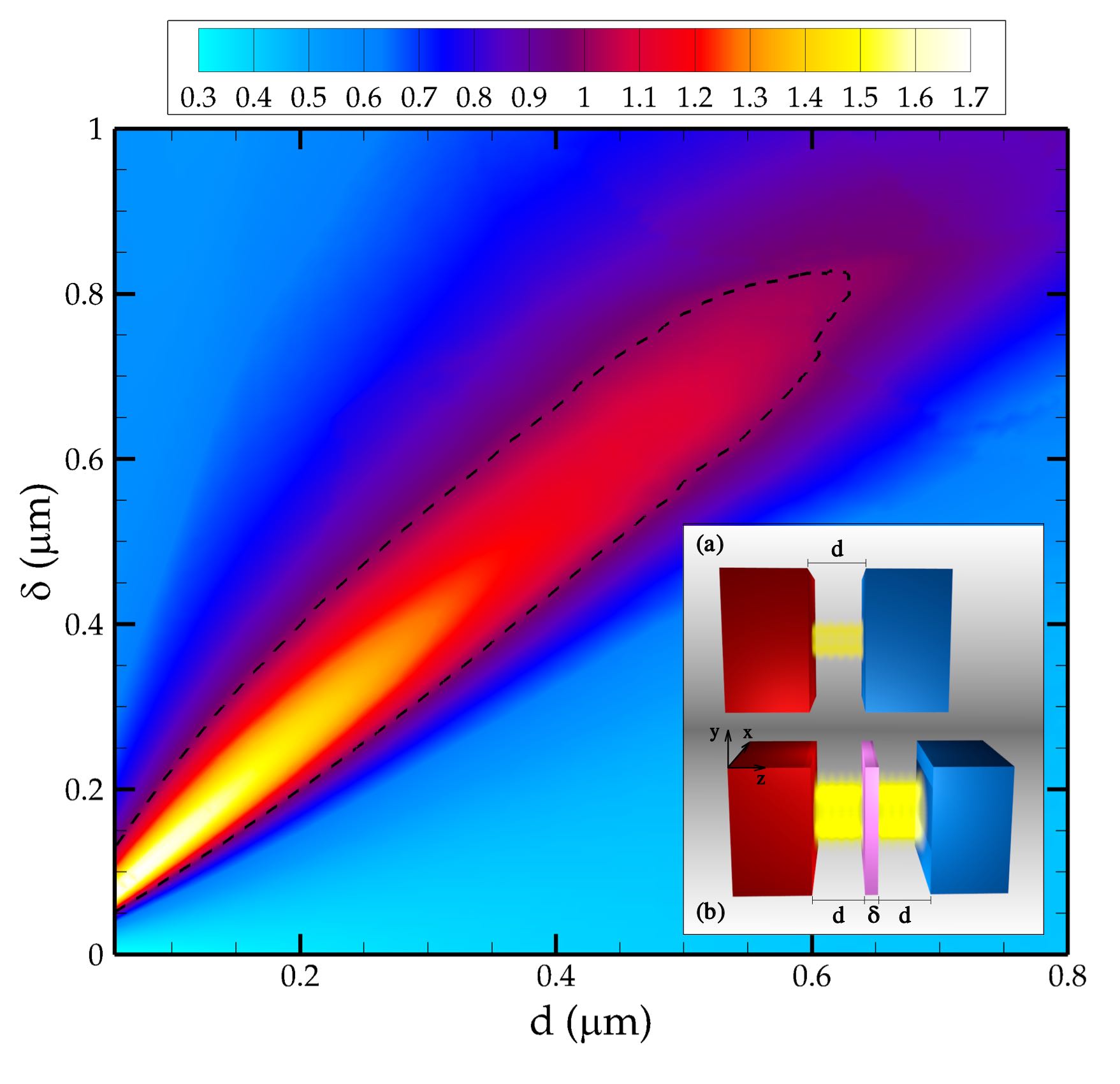}
	\end{center}
	\caption{Heat-flux amplification $\Phi_{\text{3s}}(d,\delta)/\Phi_{\text{2s}}(d)$ in a three-body configuration compared to a two-body configuration shown in inset as a function of distance $d$ and thickness of the intermediate slab $\delta$. The black dashed line corresponds to the constant value $\Phi_{\text{3s}}(d,\delta)/\Phi_{\text{2s}}(d)=1$. From \cite{RMetal2012}.}
	\label{fig:3slabs}
\end{figure}

Another interesting consequence of three-body effects in NFRHT was shown in Ref.~\cite{ZZN2011,RMetal2012}. In these works, the authors considered a system made of three parallel slabs as shown in the inset of Fig.~\ref{fig:3slabs}. The intermediate slab, of thickness $\delta$, is placed at distance $d$ from the external slabs, assumed to have infinite thickness. This configuration is compared to the standard two-body scenario, shown in the inset of Fig.~\ref{fig:3slabs}, where the intermediate slab is removed and $d$ is now the distance between the external slabs. We stress that in both systems the minimum distance between adjacent slabs, very relevant parameter in a near-field configuration, is the same. Moreover, for a chosen couple of temperatures (more specifically, 400 and 300\,K), the temperature of the intermediate slab is taken as the equilibrium one, i.e.\ the one at which the net flux on it vanishes. Based on this assumption, adding the third intermediate slab has no impact on the energy balance of the system, and thus the third body is only acting as a passive relay added to the two-body system. The heat flux amplification, defined as the ratio $\Phi_{\text{3s}}(d,\delta)/\Phi_{\text{2s}}(d)$ between the three- and two-body fluxes, is shown in Fig.~\ref{fig:3slabs}. The figure clearly shows that the flux can be amplified for reasonable values (hundreds of nanometers) of both $d$ and $\delta$, and that this amplification factor goes up to a maximum value around 70\% for small distances. This amplification for $d \approx \delta$ reminiscent of the superlens effect~\cite{JP2000,SABetal2016} which leads to an optimal energy transfer between two atoms which are separated by a superlens if the distance $d$ to the interface of the superlens coincides with the thickness of the superlens $\delta$. Here, it is a purely three-body effect, which is confirmed by the spectral and mode analysis performed in Ref.~\cite{RMetal2012}. More recently patterned intermediate media~\cite{YHKetal2019}, two-dimensional atomic systems~\cite{HS2017} and hyperbolic media~\cite{JLSetal2018} have also been considered to enhance furthermore the transfers. The use of such kind of three-body control of heat flux was proposed to design many-body heat engines~\cite{ILetal2015} with thermodynamic performances better than their two body counterpart and the thermal analog of transistor~\cite{PBAandSAB2014} driven by photons. In the proposed scheme, the combination of many-body effects and the presence of a phase-change material playing the role of the {\it gate/basis} of the transistor, allows to switch, amplify and modulate the heat flux between {\it source/emitter} and {\it drain/collector} (see also Fig.~\ref{Transistor}). 
 
\begin{figure}
\includegraphics[width=0.45\textwidth]{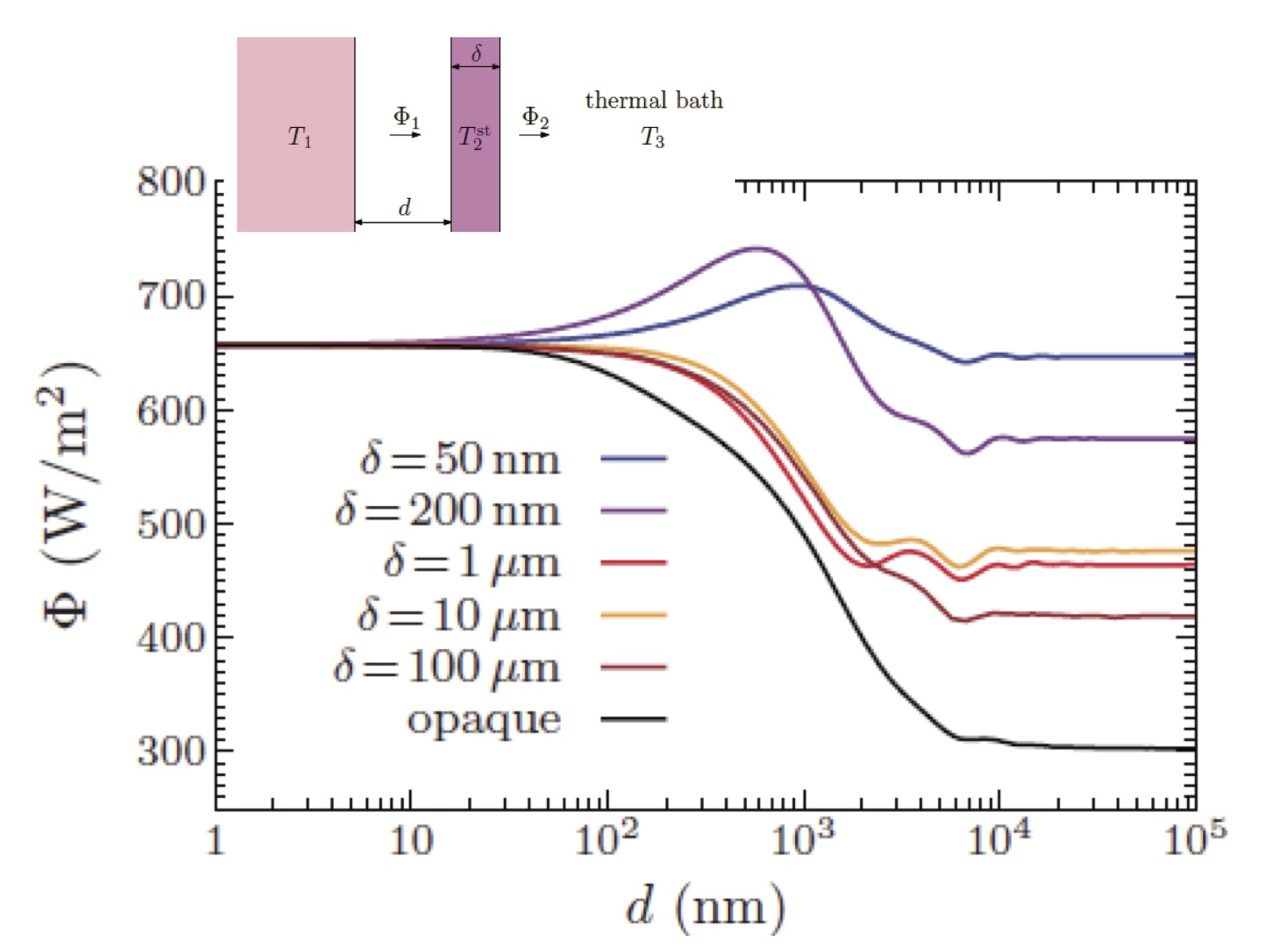}
\caption{Heat flux exchanged between two slabs immersed in a thermal bath with respect to their separation distance $d$. Slab 1 has a fixed temperature of $T_1=400\,$K, while the thermal bath is at $T_3=300\,K$. The second slab, of thickness $\delta$, thermalizes to the equilibrium temperature at which the net flux it receives vanishes. From \cite{ILetal2020}.}
\label{fig:saturationIvan}
\end{figure}

It is interesting to remark that the role of a third thermally interacting body can also be played by a thermal bath, described as a body far from the rest of the system and emitting as a black-body surface at a given temperature. This was recently shown in~\cite{ILetal2020}, where the heat flux between two planar slabs or between a slab and a particle was considered in the presence of a thermal bath. It was shown that, in virtue of many-body interactions taking place in these three-body systems, the flux exchanged between the two slabs (or the slab and the particle) saturates to a constant value when the distance goes to zero even at relatively large separation distance where the non-local optical effects are negligible, as shown e.g. in Fig.~\ref{fig:saturationIvan} in the case of two SiC slabs.

\subsubsection{Steady-state temperatures and multistable states}\label{steady sate}

In arbitrary many-body systems consisting of $N$ objects at temperatures $T_1,\ldots,T_N$ the time evolution reads ($i = 1,\ldots,N$)
\begin{equation}
  I_{i}\frac{\rd T_{i}}{\rd t} = \mathcal{P}_i (T_1,..,T_N;t)\label{Eq:energy_eq},
\end{equation}
where $I_1 = \rho_i C_{i} V_i$ is the termal inertia defined by the heat capacity $C_i$, volume $V_i$, and the mass densite $\rho_i$ of the $i^{th}$ object while $\mathcal{P}_i$ is the net power received by this object. Following expressions in Eq.~(\ref{Eq:PowerScatt}) and (\ref{Eq:PowerDipolGeneralb}) the latter can be broken up into 
\begin{equation}
  \mathcal{P}_i (T_1,..,T_N;t) = \sum_{j\neq i}\mathcal{P}_{ij}(T_1,..,T_N;t) + \mathcal{P}_{ib}(t)
\label{Eq:net_power}
\end{equation}
where $\mathcal{P}_{ij}$ is the power exchanged between the $j^{th}$ and $i^{th}$ object and $\mathcal{P}_{ib}$ is the power exchanged between object $i$ and the background which can also be an external heat bath or thermostat connected to object $i$. If all $\mathcal{P}_i$ are linear functions of the temperatures which is generally the case close to the global equilibrium or non-equilibrium steady state (in the following, for the sake of notation simplicity, we use the abreviation $T^{\rm eq}$ for the steady-state temperatures), i.e. for small temperature differences $\mid T_i-T_j \mid \ll \min(T_1, \ldots, T_N)$), the system of equations can be linearized by introducing the conductances
\begin{equation}
   G_{ij} = \frac{\partial \mathcal{P}_{ij}}{\partial T_j} 
\end{equation}
as done in Eq.~(\ref{Eq:diff2}). For multilayer systems with infinitely large interfaces the above equations can be used as well by simply replacing the quantities by the corresponding quantities normalized to a surface area $A$ so that the thermal inertia becomes the thermal inertial per area $I_i \rightarrow I_i/A$, the dissipated power becomes the heat flux $\mathcal{P}_i \rightarrow \mathcal{P}_i/A \equiv \Phi_i$, and the conductance becomes the heat transfer coefficient $G_{ij} \rightarrow G_{ij}/A \equiv H_{ij}$.

\begin{figure}
  \includegraphics[width=\columnwidth]{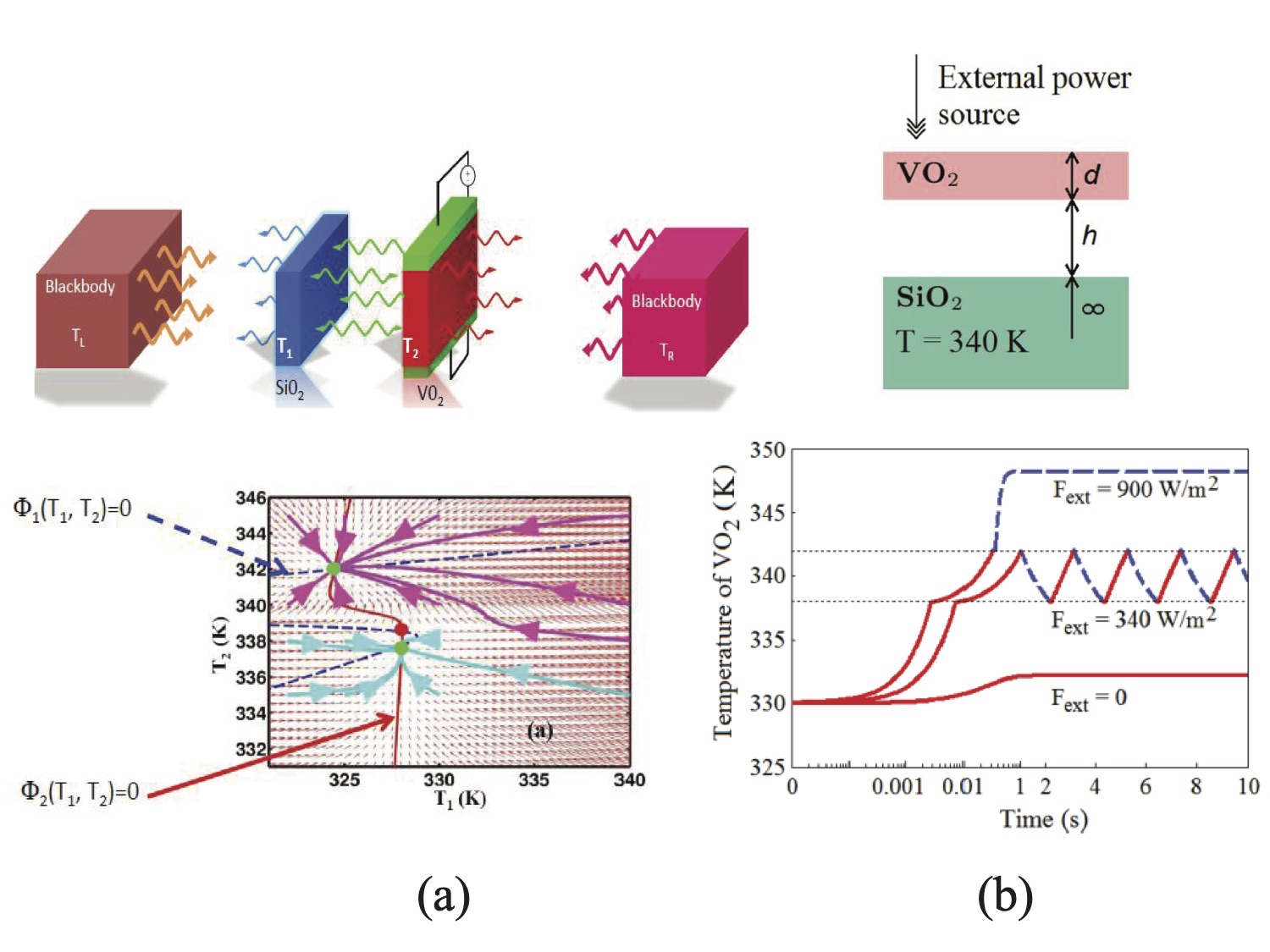}
  \caption{(a) Phase portrait (i.e. trajectories of temperatures) in a bistable system consisting of two membranes of SiO$_2$ and VO$_2$ in interaction with two thermal baths for different initial conditions. The green (red) points denote the stable (unstable) global steady-state temperatures. From~\cite{VKetal2014}.
  (b) Self-oscillation of the temperature of a VO$_2$ membrane in vicinity of a SiO$_2$ substrate when adding a specific external constant power $F_{\rm ext}$. From~\cite{SADetal2015}}.
\label{bistability}
\end{figure}

When assuming that no energy is added or removed from outside of the system, the thermal steady state is a solution of the system of equations ($i = 1,\ldots,N$)
\begin{equation}
  \mathcal{P}_i (T_1,..,T_N) = 0, 
  \label{Eq:steady_eq}.
\end{equation}
The local thermal equilibrium of the $i^{th}$ object is reached when $\mathcal{P}_i(T_1,..,T_N) = 0$. This equation defines a hypersurface in temperature space. The intersection of the hypersurfaces associated to all local equilibria defines the global steady state of the system. In the specific case where the system is composed of two objects the local equilibrium state of each object corresponds to a curve in the two dimensional space of temperatures $(T_1,T_2)$ and the intersection of the two local equilibrium lines defines the global steady-state temperatures.

If all $\mathcal{P}_i$ are linear functions of the temperatures which is generally the case close to the global equilibrium or steady state and when the conductances $G_{ij}$ can be considered as independent of the temperatures, i.e.\ when in particular the material properties can be considered as temperature independent, the system has a unique solution $(T^{\rm eq}_1,...,T^{\rm eq}_N)^t$. On the contrary, when the optical properties of materials are temperature dependent $\mathcal{P}_i$ become nonlinear with respect to the temperatures. In this case, the system of equations (\ref{Eq:steady_eq}) might admit more than one steady-state solution. 
Among these temperature solutions one finds in general stable and unstable solutions. The stability of these temperatures can be assessed by following a perturbative approach. Starting from a steady state $\alpha$ with temperature $(T^{\rm eq}_{1,\alpha},...,T^{\rm eq}_{N,\alpha})^t$ and adding a small perturbation
then the dynamics is described by the following linearized system
\begin{equation}
  \frac{\rd }{\rd t} \begin{pmatrix} \delta T_{1,\alpha} (t) \\ \vdots  \\ \delta T_{N,\alpha}(t)\end{pmatrix} = \mathds{J} \begin{pmatrix} \delta T_{1,\alpha} (t) \\ \vdots \\ \delta T_{N,\alpha}(t)\end{pmatrix} ,
  \label{Eq:diff}
\end{equation}
where $\delta T_{i,\alpha} (t) = T_i - T_{i,\alpha}^{\rm eq}$ ($i = 1,\ldots,N$) is the perturbation from the steady state $\alpha$ and
\begin{equation}
   \mathds{J} = \left(\begin{array}{ccc}
     \frac{\partial \mathcal{P}_1}{\partial T_1}&...& \frac{\partial\mathcal{P}_1}{\partial T_N} \\
      \vdots&&\vdots\\
     \frac{\partial\mathcal{P}_N}{\partial T_1}&...& \frac{\partial\mathcal{P}_N}{\partial T_N} \end{array}\right)
\label{Eq:A0}
\end{equation}
is the Jacobian matrix associated to the dynamical system~(\ref{Eq:energy_eq}). 
As in any linear dynamical system the sign of the eigenvalues of $\mathds{J}$ allows us to conclude on the stability of thermal state.

\begin{figure}
\includegraphics[width=\columnwidth]{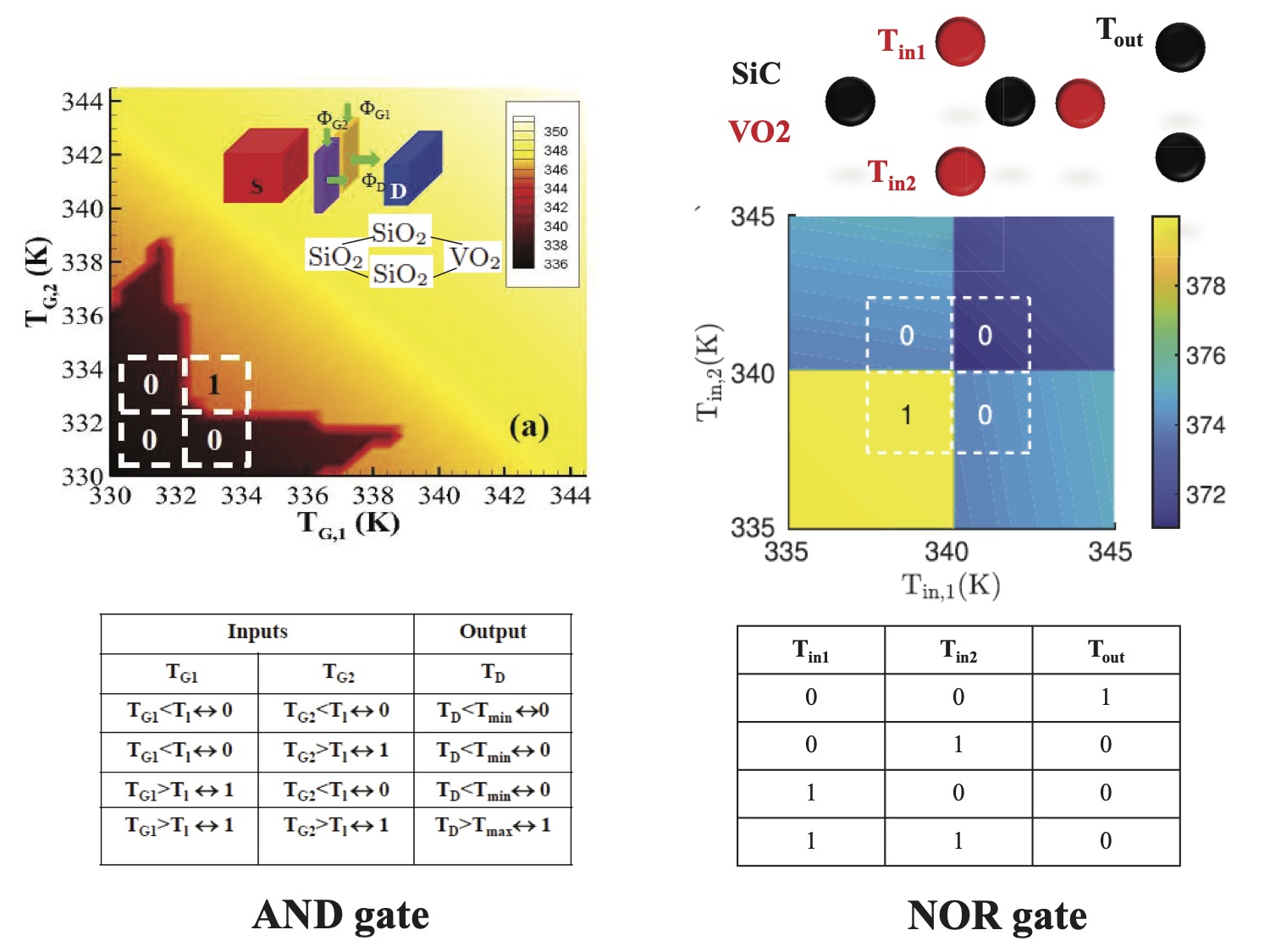}
\caption{(left) AND gate made with two SiO$_2$ membranes (gates) suspended between a thermal SiO$_2$ source and a VO$_2$ drain. The color map represents the output temperature $T_D$ of the drain with respect to the two input temperatures $T_{G1}$ and $T_{G2}$ of the two gates. In the bottom is the thruth table for the AND gate. From~\cite{PBAandSAB2016}. (right) NOR gate designed by coupling of SiC and V0$_2$ nanoparticles. From~\cite{CKetal2020}}.
\label{logic}
\end{figure}

The demonstration of multistable thermal behaviors in many-body systems as shown in Fig.~\ref{bistability} has opened the possibility to design thermal analogs of volatile electronic memories~\cite{VKetal2014,SADetal2015b,PBAandSAB2015,PBAandSAB2017,CKandAWR2017}, logic gates~\cite{PBAandSAB2016,CKetal2020} (see Fig.~\ref{logic}) and self-oscillating systems~\cite{SADetal2015} that allow to switch from one global equilibrium to another and which can be potentially interesting for practical realization of heat engines~\cite{ILetaljap2014,ILetal2015}. 

\subsection{Radiative heat transfer in reciprocal many-body systems}

In the 2000s the first attempts of treating heat transfer in 
$N$-body systemss were made in order to quantify the contribution of plasmonic modes to the thermal conductance in one dimensional arrays of nanoparticles in nanofluids~\cite{PBAapl2006,PBAetal2008}. Inside these simple networks all inner nanoparticles are assumed to be at zero temperature while the two particles at both ends of the chain are connected to two thermostats. In these systems heat carried by photons is simply scattered between the two thermostats. But in contrast to Polder and Van Hove’s theoretical framework, which is based on the FE theory, in these works a kinetic approach has been followed. The main features and limitations of this approach will be discussed in the next section.

\begin{figure}
   \includegraphics[width=0.4\textwidth]{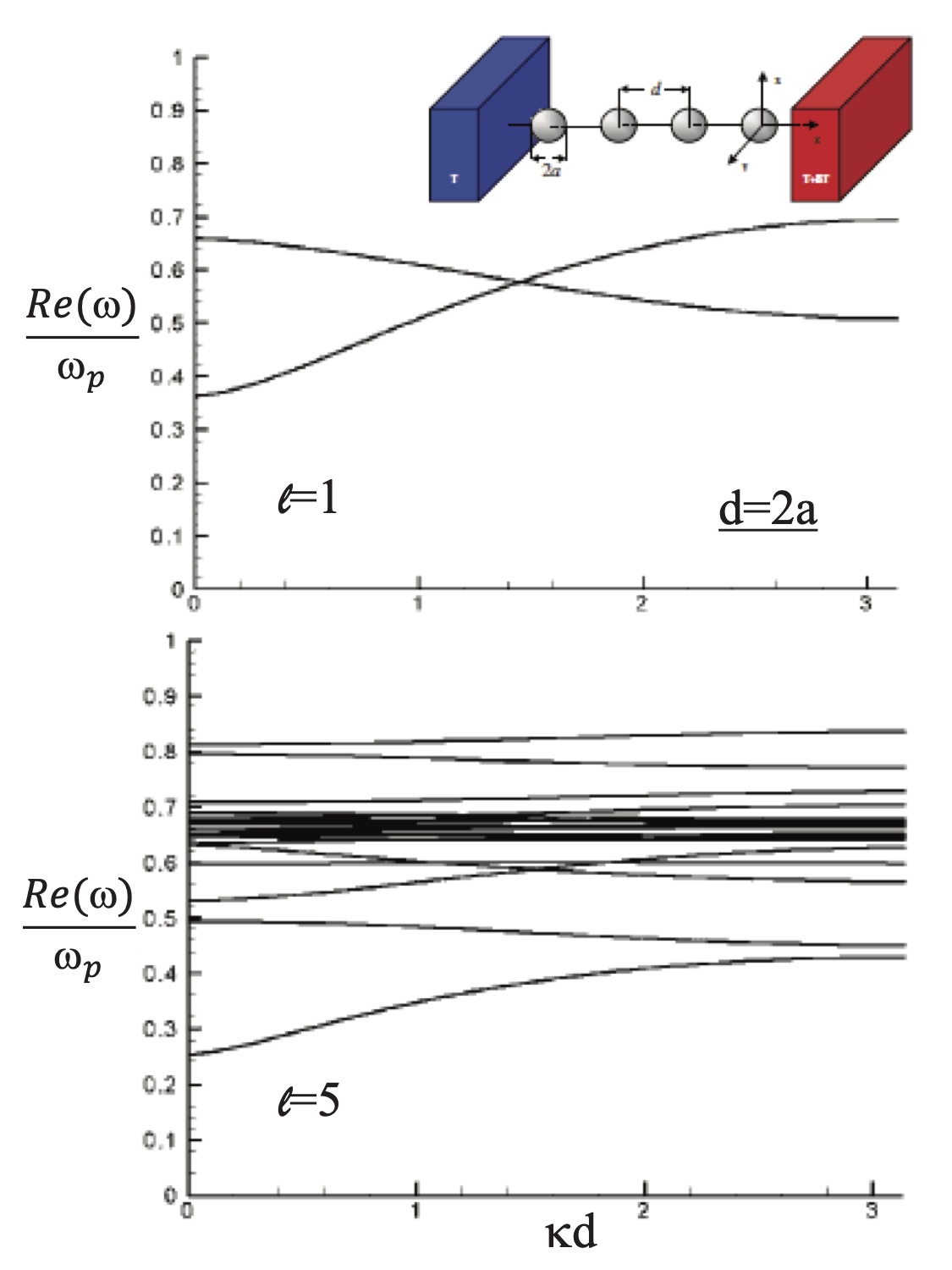}  
   \includegraphics[width=0.4\textwidth]{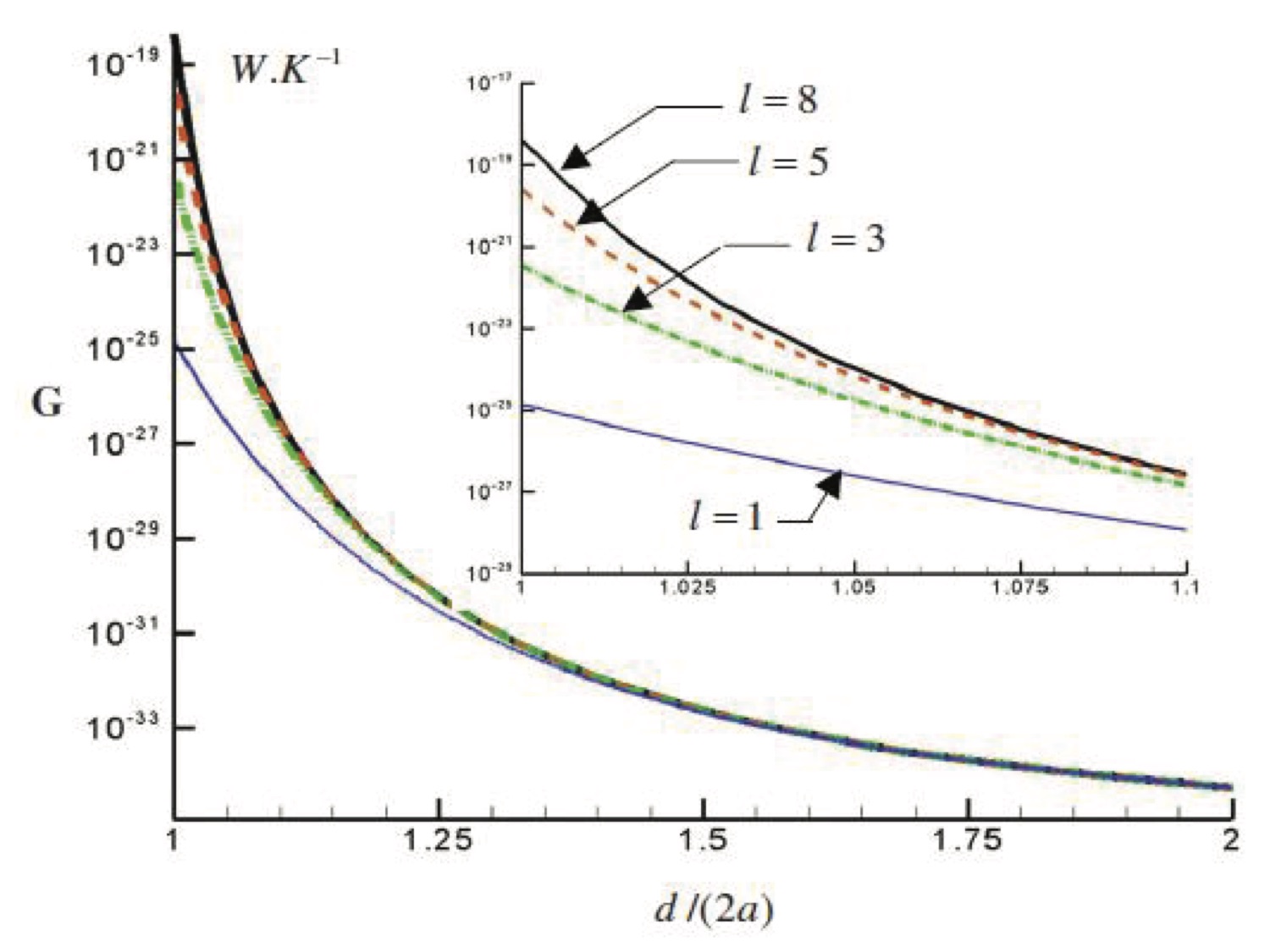}
   \caption{(a) Dispersion curves (real part) of collective plasmonic modes along a chain of copper nanoparticles dispersed in vacuum in the case of dipolar moments ($\ell=1$) and for the multipolar moments of order $\ell = 5$. (b) Thermal conductance $G$ of linear chains of copper particles calculated from the kinetic theory for different multipole orders $\ell$ versus the separation distance $d$ normalized to the particle diameter $2a$. The inset is a zoom on the near-contact region. From~\cite{PBAetal2008}.}
\label{kinetic_multipole}
\end{figure}

\subsubsection{Kinetic approach vs exact calculations}

\begin{figure}
  \includegraphics[width=\columnwidth]{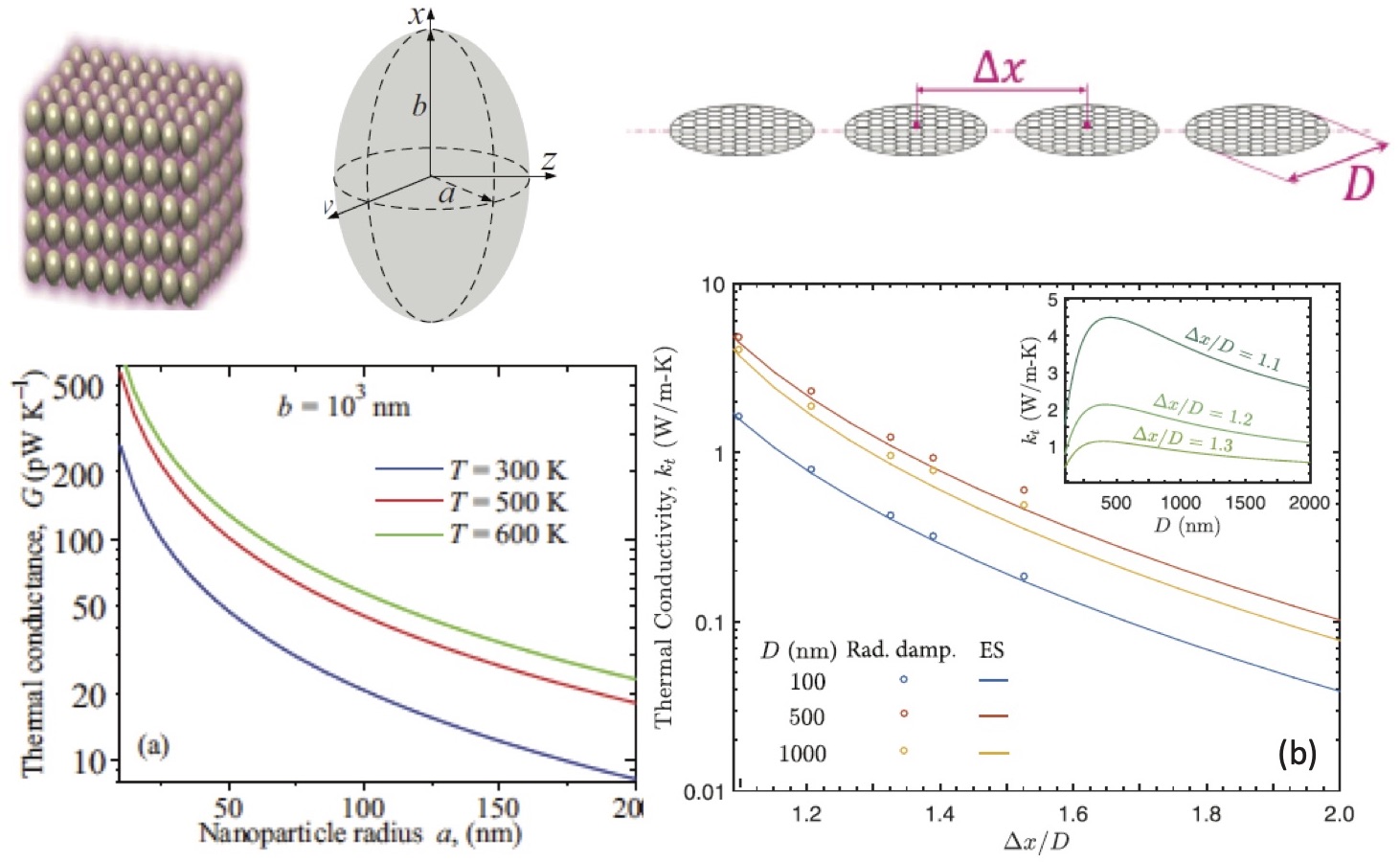}
  \caption{(a) Thermal conductance of the colloidal crystals made up of spheroidal SiC nanoparticles, as a function of their horizontal radius. 
  (b) Thermal conductivity of coplanar disk arrays for different diameters and separations at temperature $T=300\,{\rm K}$. From~\cite{JOetal2016,FVRandAJHM2017}.}
\label{kinetic_ordonez_co}
\end{figure}

This approximate theory is based on the solution of a Boltzmann transport equation
\begin{equation}
 \frac{\partial f}{\partial t}+v_{g}(k)\frac{\partial f}{\partial z}=\biggl[\frac{\partial f}{\partial t}\biggr]_{\rm coll}.
 \label{boltz}
\end{equation}
for the distribution function $f$ of thermal photons inside a given system. Here $v_{g}(k)$ is the group velocity of the mode $k$ and the rhs of this equation stands for the collission term which can be simplified within the relaxation time approximation. When assuming that one thermostat is at temperature $T$ and the other one at zero temperature, then the power $\mathcal{P}$ flowing through this system results from the calculation of first-order moment associated with the photonic equilibrium distribution function $f = n(\omega,T)$~\cite{PBAetal2008} 
\begin{equation}
   \mathcal{P} = \sum_{\ell = 1}^\infty \int_{0}^{\infty} \!\! \frac{\rd k}{2\pi} \hbar \omega_\ell(k) v_{g,\ell}(k) n(\omega_\ell(k), T),
 \label{kinetic}
\end{equation}
where $\omega_\ell(k)$ is the dispersion relation of resonant multipole modes $\ell$ supported by the structure. The conductance is then defined as $G = \partial \mathcal{P} / \partial T$. It is important to note that only the eigenstates of the system are assumed to play a role in the heat transport process. Since these preliminary studies, more complex systems like chains of ellipsoidal polaritonic particles \cite{JOetal2015}, nanoparticle crystals~\cite{JOetal2016,ETetal2016}, nanoresonators inclusions~\cite{ETetal2019} or chains of graphene disks \cite{FVRandAJHM2017} have been investigated (see Fig.~\ref{kinetic_ordonez_co}) as well as multilayer photonic crystals~\cite{WTLetal2008,WTLetal2009} using this kinetic approach. But as shown recently within a full FE calculation based on the $N$-body theory introduced in Sec.~\ref{Sec:NDipol}, the kinetic approach fails in describing heat exchanges in systems where heat is also carried by non-resonant modes over a broad spectral band~\cite{CKetal2018}. This result has been confirmed recently~(\cite{ETetal2020}). Further studies of conductance within two and three dimensional dipolar systems based on the fluctational electrodynamic calculations have been published recently~\cite{ETetal2019b} which opens also to test the validity of the kinetic approach in such systems as studied in \cite{JOetal2016,ETetal2016}. A discussion of the conductance within multilayer photonic crystals within the FE approach discussing the role of surface phonon polaritons can be found in ~\cite{MTetal2012,NarayanaswamyJQSRT2005}.

\subsubsection{Heat transfer in complex networks}
\label{Sec:ComplexNetworks}

\begin{figure}
\includegraphics[width=\columnwidth]{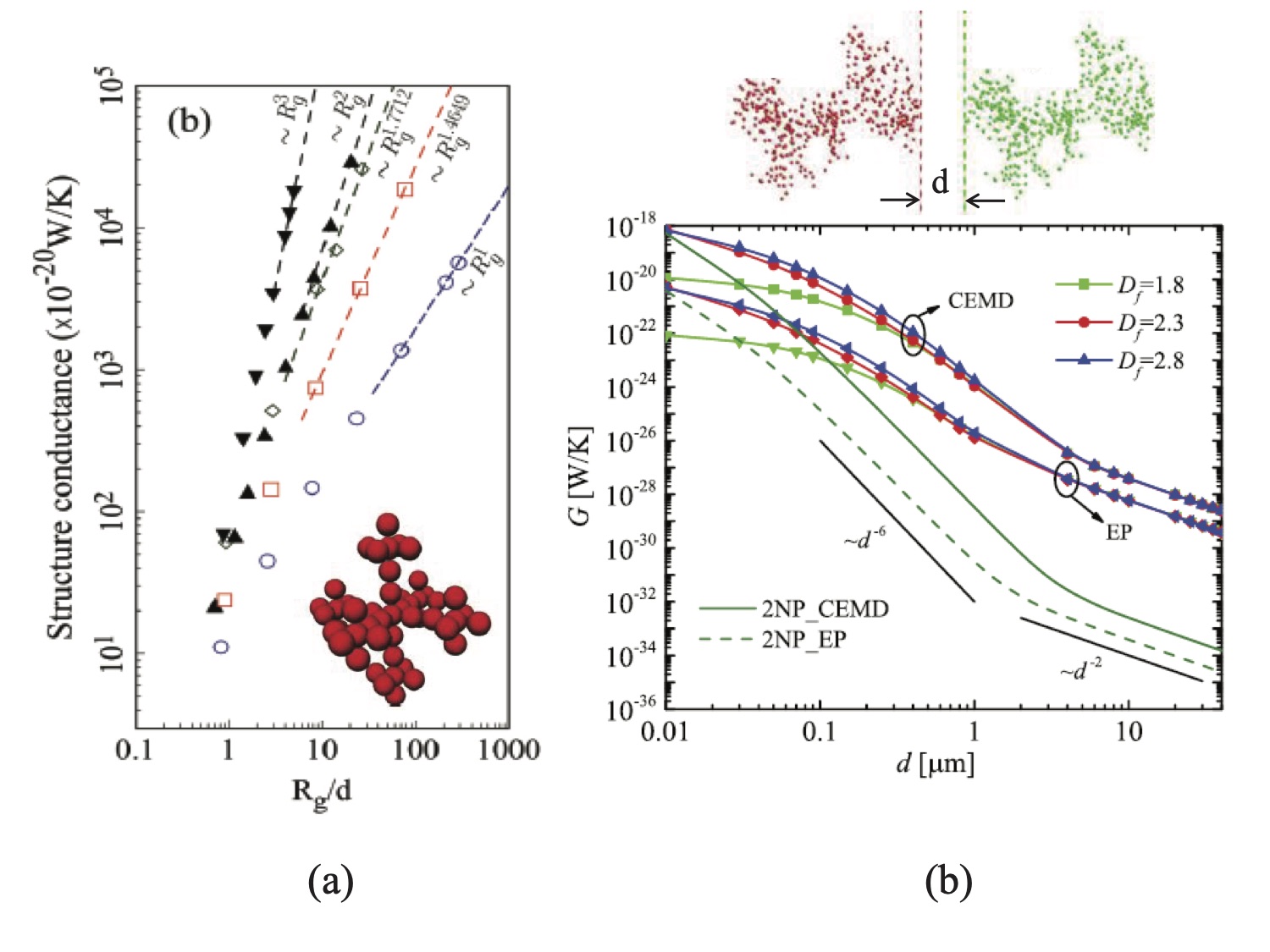}
\caption{(a) Thermal conductance of Vicsek fractal structures
as a function of normalized gyration radius. From \cite{MN2017} (b) Thermal conductance between two Ag nanoparticles
clusters at various fractal dimensions. From \cite{MLetal2019}}
\label{fractal}
\end{figure}

Based on the rigorous FE approach we will now address the heat flux in arbitrary systems. The thermal behavior of fractal structures and the heat exchanges between fractal clusters of nanoparticles has also been theoretically investigated. These studies have revealed ~\cite{MN2017,JDetal2017b} that the (self)conductance increases as $R^{D_f}$ where $R$ is the gyration radius of the structure and $D_f$ its fractal dimension (see Fig.~\ref{fractal}(a)).
When two of these structures interact in near-field the thermal conductance of heat exchange between metallic clusters increases with the fractal dimension as can be seen in Fig.~\ref{fractal}(b). Moreover, in contrast to ordered media, the localization of plasmons or phonon-polaritons in fractal structures could be responsible of a significant reduction of the self-conductance in fractal structures although no clear evidence about this claim has been presented so far. However, a recent study ~\cite{MLetal2019} has revealed that the heat transfer between fractal structures does not depend on their fractality at separation distance larger than the localization lengths, which tends to confirm this statement. 

\begin{figure}
  \includegraphics[width=0.4\textwidth]{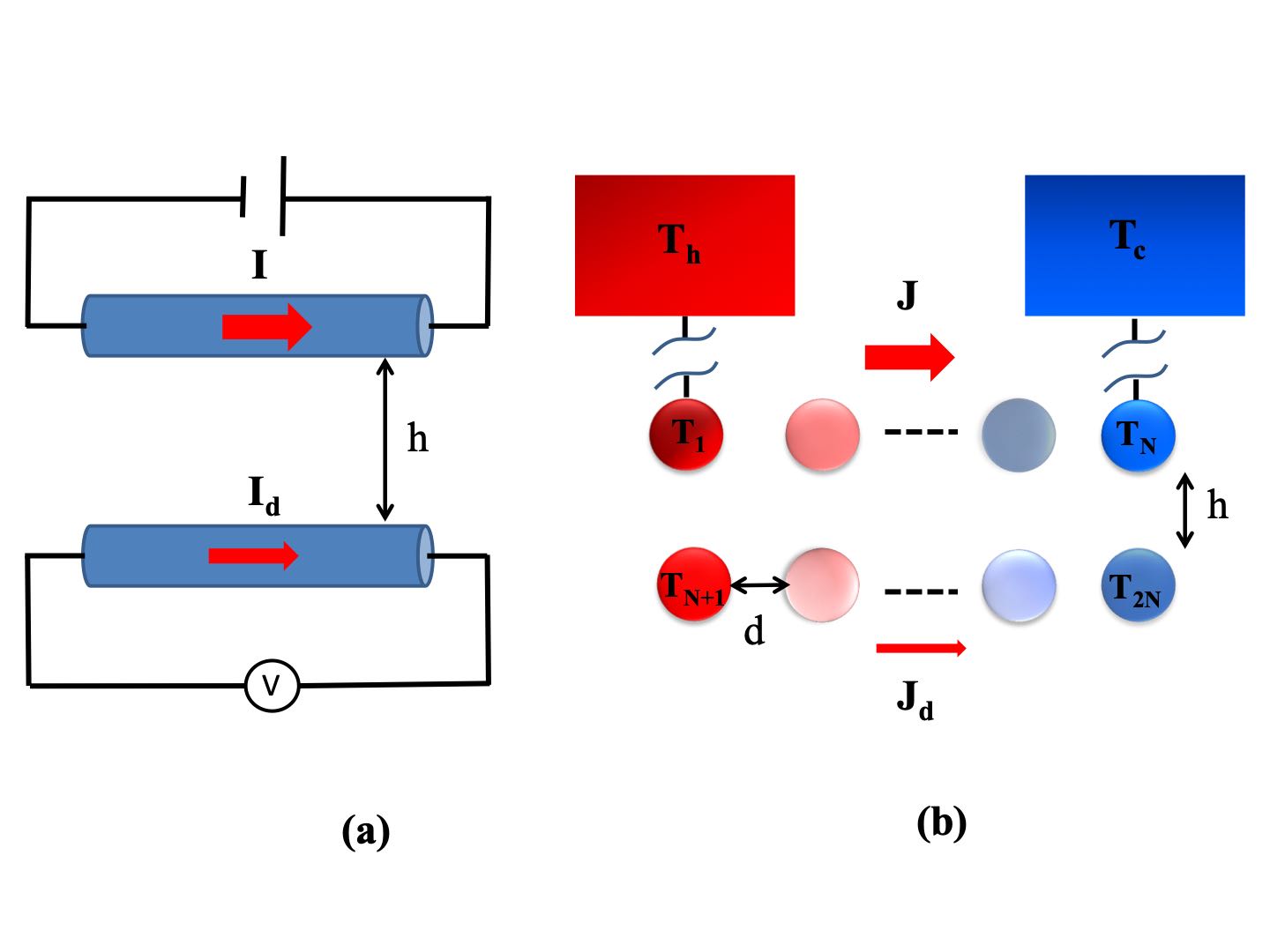}
  \caption{(a) Illustration of the classical Coulomb drag effect. A drag electric current $I_d$ in a passive conducting wire is induced by a primary current $I$ flowing in a driving conductor placed close to it. (b) Radiative drag effect in a many-nody system: a drag heat flux $J_d$ carried by thermal photons between two particles is induced by a heat flux $J$ exchanged between two thermostated objects in a many-body system. From \cite{PBA2019}.}
\label{drag}
\end{figure}

Beside their original thermal properties several physical effect inherent to many-body systems have been highlighted in complex plasmonic structures. 
Among these effects, a thermal analog of Coulomb drag effect in nanoparticle networks has been recently predicted theoretically~\cite{PBA2019}. The configuration is sketched in Fig.\ref{drag}. As in its electric counterpart where interactions at close separation distances (compared to the range of Coulombic interactions) of free charge carriers between two electric conductors gives rise to a drag current in a passive conductor when a bias voltage is applied along the so called drive conductor, 
a radiative heat flux in a many-body systems can be induced in a given region by a primary flux generated by a temperature gradient in another region of the system. 
In the case of two parallel chains of nanoparticles as sketched in Fig.\ref{drag}(b), where the extremities of the first chain are held at fixed temperature with two external thermostats while all other particles can relax to their own local equilibrium temperature, the magnitude and the direction of drag flux can be calculated using the following procedure. 

In the steady state the net power received by each particle vanishes which allows to determine unknown temperatures $(T_2,..,T_{N-1},T_{N+1},...,T_{2N})$ ($T_1$ and $T_N$ are fixed by the thermostats) and the power $\mathcal{P}_{1}$ and $\mathcal{P}_{N}$ coming from the external thermostats in order to keep the temperatures of particle $1$ and $N$ fixed. Then the heat current in the upper chain in Fig.~\ref{drag}(b)
\begin{equation}
  J = \mathcal{P}_{N} - \mathcal{P}_{1} 
\label{Eq:drivePower}.
\end{equation}
as well as the induced heat current in the lower chain in Fig.~\ref{drag}(b)
\begin{equation}
   J_D = \mathcal{P}_{2N} - \mathcal{P}_{N+1}
\label{Eq:inducedPower}
\end{equation}
can be determined. Finally, the thermal drag resistance 
\begin{equation}
  R_D = \frac{T_{N+1}-T_{2N}}{J}.
\label{Eq:drag_resistance}
\end{equation}
quantifies the frictional effect induced by the electromagnetic interactions between the different regions inside the system. In hybrid polar-metal systems 
this friction can be negative~\cite{PBA2019} proving the existence of regions within these systems where heat can locally flow in an opposite direction to 
the applied temperature gradient. 

\begin{figure}
  \includegraphics[width=\columnwidth]{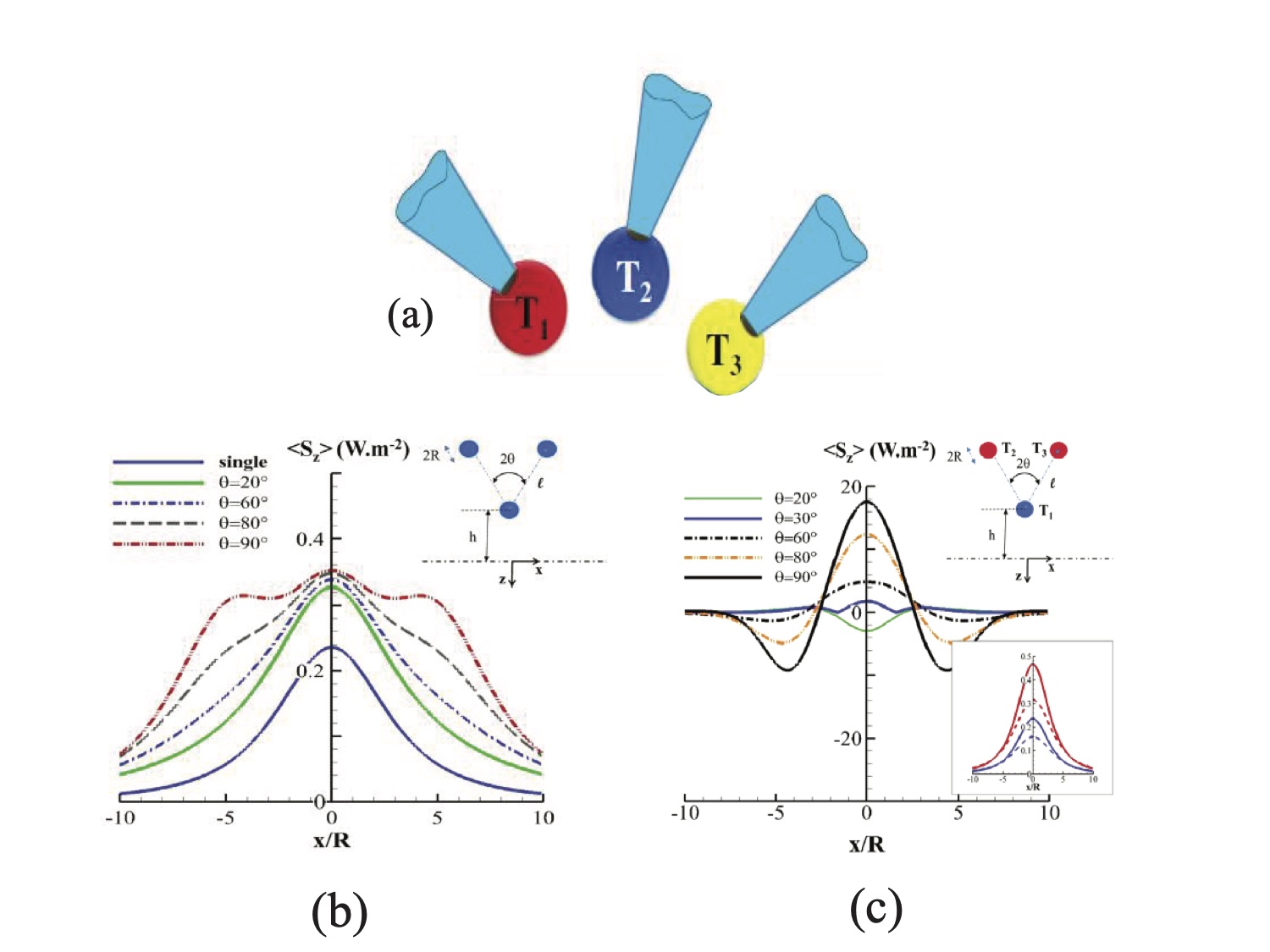}
  \includegraphics[width=\columnwidth]{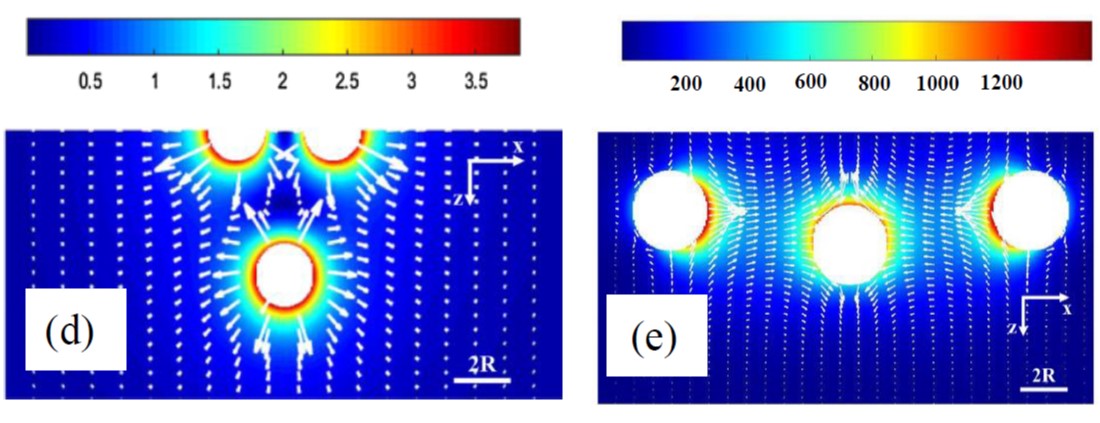} 
  \caption{(a) Schematic of a multi-tip SThM platform with three tips. Nano-spheres (thermal emitters) are grafted on single scanning probe tips and held close to a substrate. Their temperatures and positions are individually controlled. (b) Normal component $\langle S_z \rangle$ of the Poynting vector radiated through the substrate surface at $z=0$ by a three-tip SThM setup with glass nano-emitters at $T=300\:{\rm K}$. (c) As in (b) but with $T_2 = T_3=350 {\rm K}$ (red) and $T_1=300\:{\rm K}$ (blue). The inset shows the flux at $z=0$ for a single particle at $T=300\:{\rm K}$. (d)-(e) Magnitude of Poynting vector field in the $(x, z)$ plane radiated by a multi-tip setup in the case (d) for an angular opening of $\theta=20^\mathrm{o}$ and in the case (e) for an angular opening $\theta=80^\mathrm{o}$. From~\cite{PRL2019}.}
\label{focusing}
\end{figure}

Beside this generation of heat flux by frictional effect in many-body systems the temperature of the particles in particle networks can be individually addressed with a subwavelength accuracy~\cite{VYandNVV2013} using external excitations such as chirped pulses and can be controlled by adaptive optimization techniques at the time scale of thermal relaxation processes. The interplay between nano-objects can also be used to focus and even pump heat~\cite{PRL2019} outside of the system itself. The heat flux radiated through an oriented surface by a collection of emitters held at different temperature $T_i$ ($i = 1,\ldots,N)$ can be calculated from Eq.~(\ref{Eq:PVDipolgeneralNoBG}). By tuning the temperature of three thermal emitters in vicinity of a substrate as shown in Fig.\ref{focusing}, for instance, the heat flux can be locally focused and even amplified in tiny regions which are much smaller than the diffraction limit and even smaller than the regions heated with a single emitter~\cite{PRL2019}. This control of flux lines by a collection of nano-sources can be used to tailor the heat flux at the nanoscale or to analyze and change at this scale the local temperature of solid surfaces. 

\subsubsection{Long range heat transport and amplification of heat flux}
\label{Sec:Guiding}

\begin{figure}
  \includegraphics[width=0.4\textwidth]{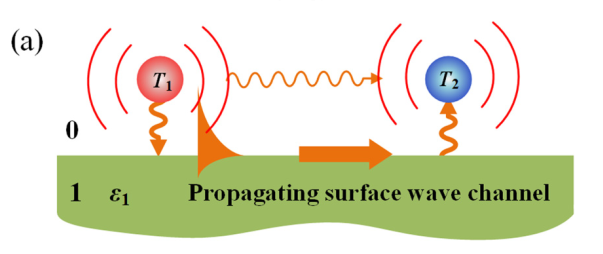}
  \includegraphics[width=0.4\textwidth]{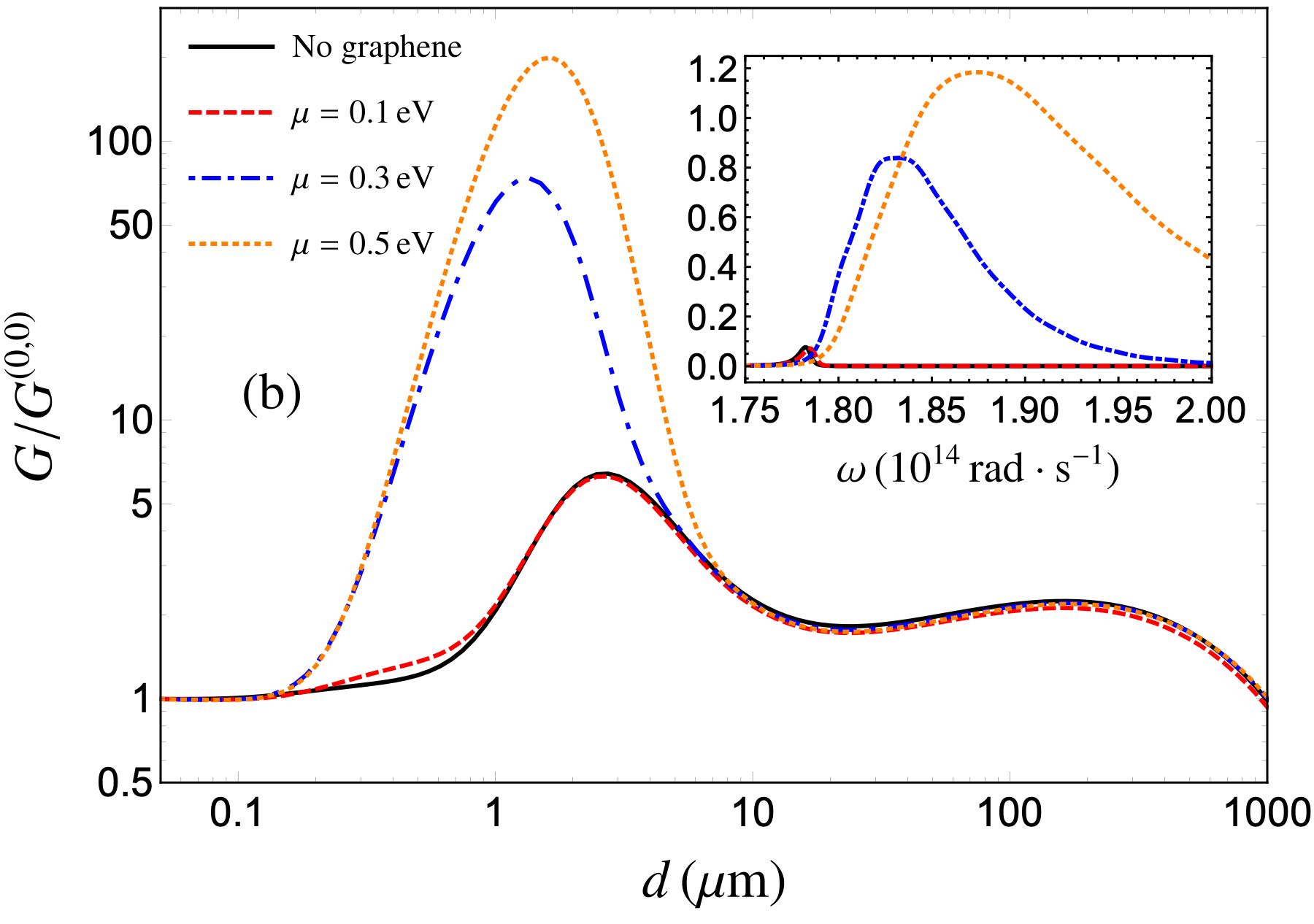}
  \caption{(a) Heat flux between two nanoparticles at inter-particle distance $d$ by coupling via the surfacce modes of an inferface. From Ref.~\cite{JDetal2018}. (b) conductance ratio $G/G^{(0,0)}$ ($G$ conductance with interface and $G^{(0,0)}$ without interface) as a function of $d$ between two Au nanoparticles placed at distance $z=150\,$nm from a SiC substrate. The four lines correspond to the absence of graphene (black solid line), and to configurations with graphene having $\mu=0.1\,$eV (red dashed line), 0.3\,eV (blue dot-dashed line) and 0.5\,eV (orange dotted line). The inset shows the spectral conductance associated with the four same configurations. From Ref.~\cite{RMetal2018}. \label{NanoparticleInterface}}
\end{figure}

Instead of enhancing the heat flux between two nanoparticles or two slabs by introducing an intermediate nanoparticle or slab as discussed in Sec.~\ref{Sec:NonAddDip} and \ref{Sec:NonAddMac} it is also possible to guide the radiative heat flux over a long distance by exploiting the properties of specific modes such as surface or hyperbolic modes supported by some structures. This guiding can for example be done by bringing two nanoparticles close to a planar interface as sketched in Fig~\ref{NanoparticleInterface}(a) which supports a surface polariton in the infrared. Then the hot nanoparticle can directly couple to this surface mode and subsequently transfer its heat to the second (cold) particle over relatively long distances. 

Such a transport has been first investigated in Ref.~\cite{KSetal2014} between polar nanoparticles above single polaritonic surfaces and inside cavities formed of two mirrors or made with slabs supporting surface modes. This study and more recent studies~\cite{JDetal2018,RMetal2018,KAandMK2018} have shown that the heat current between dipoles placed in a cavity can be enhanced by several orders of magnitude as compared to the free-space heat current with a similar interparticle distance. In particular, in Ref.~\cite{RMetal2018} it has been shown that a similar enhancement and long range heat transport can be also observed between metallic particles when a graphene sheet cover a SiC interface. In this case the heat-flux can be enhanced by several orders of magnitude at interparticle distance of about 1-10$\mu {\rm m}$ as shown in Fig.~\ref{NanoparticleInterface}(b) suggesting that the near-field enhanced thermal radiation can be brought to distances which are comparable to the thermal wavelength. Similar enhancement effects were reported for the heat flux along chains of nanoparticles close to a phonon-polaritonic interface~\cite{JDetal2018}, between two nanoparticles mediated by an intermediate macroscopic phonon polaritonic sphere~\cite{KAetal2017}, by an anisotropic meta-surface made of graphene stripes~\cite{YZetal2019} or a stack of graphene sheets~\cite{HeEtAl2019}. As shown in ~\cite{Ott2020} the distance at which the maximum heat flux enhancement occurs is connected to the propagation length of surface modes~\cite{Ott2020}. Hence, the enhancement mechanism for the heat flux is reminiscent of the enhancement of F\"{o}rster resonance energy transfer between atoms, molecules, or quantum dots which are brought in close vicinity to a plasmonic interface where also a maximal enhancement is found at distances coinciding with the propagation length of the surface modes involved in the energy transport~\cite{KAVTVS2012,SABGSA2013,DBEtAl2016,APEtAl2016} allowing for a long-range energy transfer. 

\begin{figure}
  \includegraphics[width=0.4\textwidth]{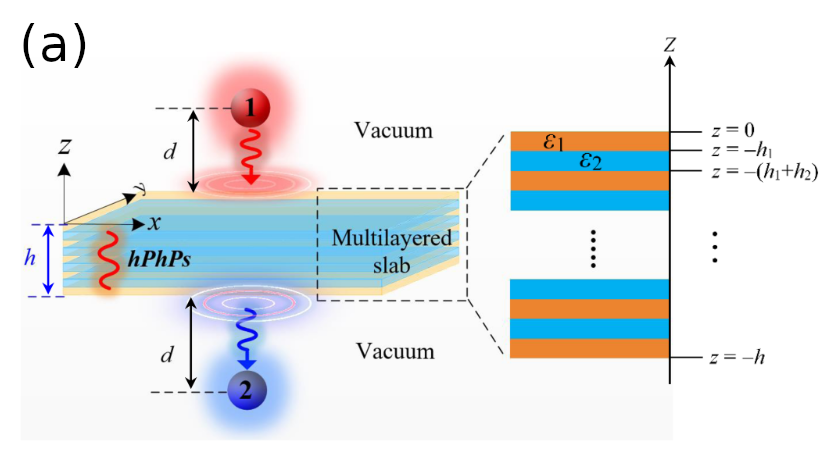}
  \includegraphics[width=0.35\textwidth]{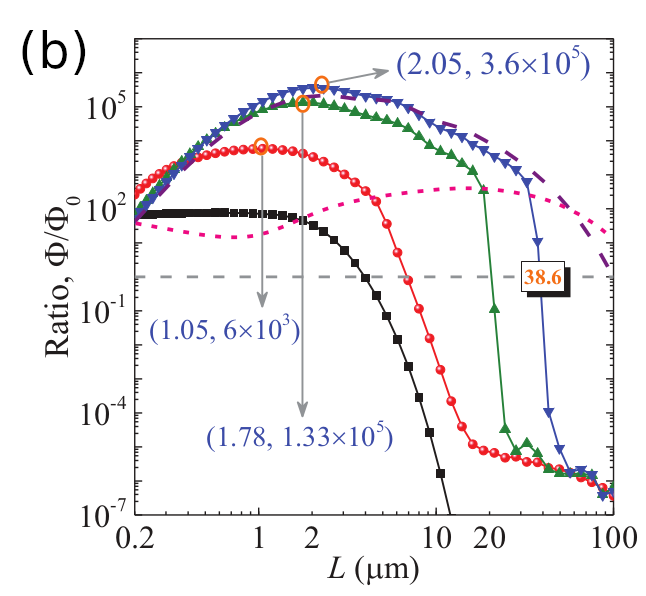}
  \caption{(a) Sketch of heat flux between two nanoparticles through a hyperbolic multilayer meta-material. (b) Exchanged power $\Phi$ as function of interparticle distance $L$ normalized to the exchanged power $\Phi_0$ where the hyperbolic multilayer meta-material has been replaced by vacuum. From Ref.~\cite{RYZetal2019}.}
\label{NanoparticleHMM}
\end{figure}

 Motivated by the very promising properties of hyperbolic metamaterials for long-range F\"{o}rster energy transfer~\cite{SABetal2016,RDetal2018,WDNetal2018} another strategy has been explored to transport the near-field heat flux over long distances using such hyperbolic guides. Hence it has been shown that the large wavevector surface waves supported by polaritonic materials can be converted into propagating hyperbolic modes inside these media so that the usual ultra-small penetration depth of near-field heat flux~\cite{SBZM2009} can become very large~\cite{SLetal2014,MTEtAl2015,SABandPBA2017} in these guides as well as the amount of heat they can transport~\cite{JLEM2015,SABetal2015}. Since the hyperbolic media can support hyperbolic modes over a broad spectral band the flux they can transport can be very high. It seems even possible to achieve with hyperbolic metamaterials a radiative thermal conductivity which can in principle be comparable to the phononic conductivity~\cite{JLEM2015,SABetal2015}. Recently, first experimental steps have been made to verify this claim~\cite{HSetal2019}, but the experimental results are not yet convincing. In a more detailed study it could be demonstrated that the near-field heat flux between two slabs can be guided through a hyperbolic waveguide over distances larger than the thermal wavelength so that larger heat fluxes than the black-body value are achievable for far-field distances~\cite{RMetal2016}. On the other hand, it could also be shown that the guiding performance highly depends on the dissipative properties of the waveguide material and that for long-distance guiding also low-loss infrared materials like Ge, for instance, would already have very good long-range guiding properties~\cite{RMetal2016}. The long-range guiding effect has also been verified for the heat flux between two nano-particles through a hyperbolic multilayer structures~\cite{RYZetal2019} as shown in Fig.~\ref{NanoparticleHMM}. 

Even though the enhancement of the heat flux due to coupling to the surface modes of the phonon-polaritonic or plasmonic structures can be several orders of magnitude it has to be kept in mind that the mentioned studies consider the steady-state heat flux between the nanoparticles and that the enhancement is relativ to the case where the interface is removed. Hence, even by increasing the heat flux by several orders of magnitude at a distance of 1 micron the absolute value of the heat flux is still small, because the heat flux between the nanoparticles follows the $1/d^6$ law in the near-field regime~\cite{AIV2001}. Furthermore, it should be kept in mind that by bringing the nanoparticles in close vicinity of an interface not only the heat flux between the particles increases, but also the thermal emission of the hot particle into the substrate so that the hot particle will rather tend to cool by thermal emission into the substrate then by heating the cooler nanoparticle. However, a first thermal relaxation study shows~\cite{Ott2020} that by choosing wisely the distances between the nanoparticles and between the nanoparticles and the interface, a substantial heating of the cold nanoparticle can be observed. Similar considerations also hold for the heat flux though a structure. Hence, it is very useful to focus in future studies on heat fluxes and the thermal relaxation or actual heating/cooling performance as well.

\subsubsection{Relaxation dynamics}
\label{Sec:RelaxDyn}

The temporal dynamics of any many-body system in interaction with an external environment or with local thermostats is simply driven by the competition between its thermal inertia and the strength of the thermal link with the external environment and these thermostats. Close to the thermal equilibrium, the time evolution of temperatures $\mathbf{T}=(T_{1},...,T_{N})$ in Eq.~(\ref{Eq:net_power}) is driven by the linear dynamical system 
\begin{equation}
  \mathds{I}\frac{\rd \mathbf{T}}{\rd t} = -\mathds{C}\mathbf{T}(t)+\mathds{C}_b\mathbf{T}_b.
\label{Eq:diff2}
\end{equation}
where $\mathds{I}={\rm diag}(I_1,...,I_N)$ is the diagonal inertia matrix which depends on the mass density, heat capacity and size of each element, $\mathbf{T}_b=(T_{b1},...,T_{bN})$ is the temperature of external bath and reservoirs with which each elements interact, $\mathds{C}_b= {\rm diag}(G_{1b},...,G_{Nb})$, $G_{ib}$ being the thermal conductance between the $i^{th}$ element and the bath or a thermostat while $\mathds{C}$ is the general conductance matrix with components
\begin{equation}
  \mathds{C}_{ij}= \biggl(\sum_{k\neq i} G_{ik}+G_{ib}\biggr)\delta_{ij}- (1-\delta_{ij})G_{ij}.
\label{mat_cond1}
\end{equation}
with $G_{ij}$ the conductance between the $i^{th}$ and $j^{th}$ element defined as follows
\begin{equation}
	G_{ij} =  3\int_0^\infty\!\!\frac{\rd \omega}{2 \pi}\, \hbar \omega \frac{\partial n}{\partial T} \biggr|_{T = T_j} \mathcal{T}_{ij}(\omega). 
	\label{Eq:ConductanceDipoles}
\end{equation}
A corresponding definition can also be used for slabs. Note that this definition is only valid in the absence of temperature dependence of optical properties of the materials involved. When the conductance matrix is independent of time the thermal state of the system reads
\begin{equation}
\begin{split}
  \mathbf{T}(t) &= \exp[-\mathds{I}^{-1}\mathds{C}\:t]\mathbf{T}(0)\\
                &\qquad +\int_{0}^{t}\!\! \exp[-\mathds{I}^{-1}\mathds{C}(t-\tau)]\mathds{I}^{-1}\mathds{C}_b\mathbf{T}_b(\tau)d\tau   
\end{split}
 \label{Eq:temp_time}
\end{equation}
 $\mathbf{T}(0)$ being the initial state. Hence, it is clear that the relaxation dynamic is driven by the set $\{\Gamma_i\}$ of eigenvalues of the matrix $\mathds{I}^{-1}\mathds{C}$, the dominant relaxation time is given by $\tau = 1/\min(\Gamma_i)$ ($\mathds{C}$ being a strictly diagonally dominant matrix with positive diagonal elements).

 \begin{figure}
  \includegraphics[width=0.45\textwidth]{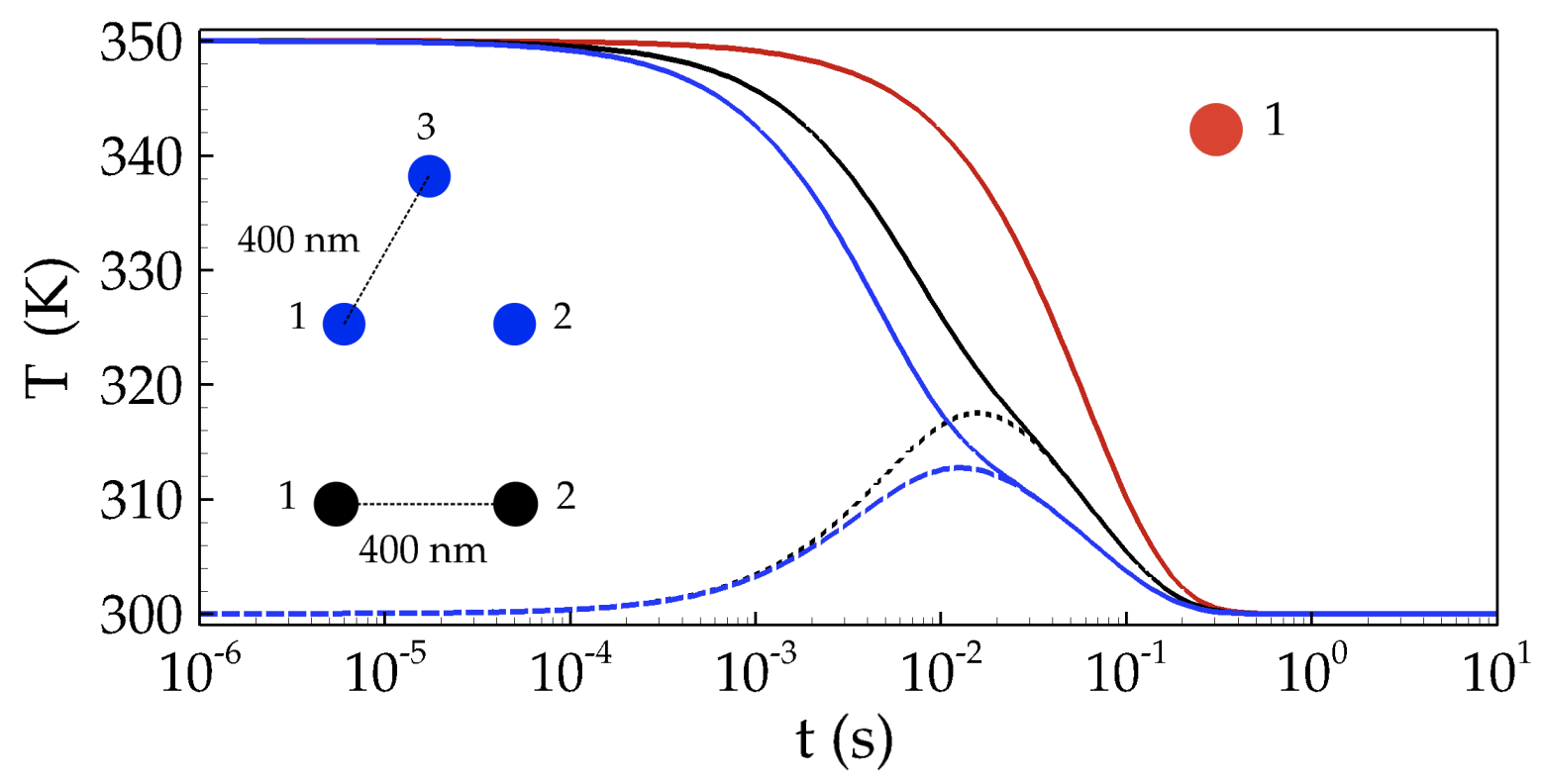}
   \caption{Time evolution of thermal state in a single (red), two-body (black) and a three-body system (blue) in a bath at temperature $T=300\:{\rm K}$. The distance between particles 1 and 2 is 400 nm, while the distances (solid line for dipole 1, dashed line for dipole 2, and dot-dashed line for dipole 3). From~\cite{RMetal2013}.}
\label{relax}
\end{figure}

Generally speaking the relaxation process takes place at different scales~\cite{RMetal2013}. When the separation distance between the different elements is subwavelength they are first thermalized in near-field regime at the same temperature. This generally happens in few millisecond~\cite{YWandJW2016} for objects of nanometric size (Fig.\ref{relax}) and even in hundreds microseconds for two dimensional nanosystems~\cite{ZundelPRB2020}. In a second step each element and therefore the whole system thermalizes in far-field toward the bath temperature. 

This difference in the time scales for the relaxation dynamics can also be studied in a simpler system when considering a single nanoparticle at temperature $T_1$ close to a sample with a fixed background temperature $T_b$ then the dynamical equation in (\ref{Eq:diff2}) reduces to
\begin{equation}
  \frac{\rd T_1}{ \rd t} =  \frac{G_{1b}}{I_1} (T_b - T_1)
\end{equation}
or equivalently
\begin{equation}
	\frac{\rd \Delta T}{ \rd t} = - \Gamma \Delta T 
\end{equation}
where $I_1 = \rho C_{\rm p} V$ is the thermal inertia of the nanoparticle and
$\Delta T = T_1 - T_b$ and the relaxation rate $\Gamma = G_{1b}/I_1$. The solution to this differential equation is simply $\Delta T(t) = \Delta T (0) \exp(- \Gamma t)$ or $T_1(t) = \bigl(T_1(0) - T_b\bigr) \exp(- \Gamma t) + T_b$. Hence, the relaxation time in this case is the inverse of the relaxation rate $\tau = \Gamma^{-1}$ which is itself determined by the thermal inertia and the heat conductance between the nanoparticle and the sample. 

The heat conductance for this configuration has been studied for spherical dielectric and metallic nanoparticles close to a sample with a flat surface~\cite{IAD1998,JPMetal2001,AIV2002,DedkovKyasov2007,POC2008}, between a spherical dielectric nanoparticle and a structured or rough surface~\cite{SABetal2008,FRetal2012,AKetal2008,SABJJG2010} and between dielectric and metallic ellipsoidal particles and a flat or structured surfaces~\cite{OHetal2010,SABetal2010}. Here we focus on a spherical nanoparticle with radius $R$ in a distance $d$ over a planar interface. For $d \gg R$ it can be shown~\cite{IAD1998,JPMetal2001,AIV2002,DedkovKyasov2007,POC2008} that $G_{1b}$ is proportional to the electric (magnetic) photonic local density of states $D^{\rm E}(\omega,d)$ ($D^{\rm H}$) as defined in \cite{GSA1975b,WE1982} for dielectric (magnetic) nanoparticles above a dielectric (magnetic) substrate. Hence, when disregarding mixed cases as considered in \cite{AM2012,JDetal2017} the relaxation rate can be written as~\cite{MTetal2012b}
\begin{equation}
   \Gamma = \frac{1}{I_1} \sum_{i = E, H}\intop_{0}^{\infty} \rd\omega\, 2 \hbar \omega^2 \Im(\alpha^i) D^i (\omega,d)\frac{\rd n}{\rd T}\biggr|_{T_b}
\label{Eq:Lifetime1}
\end{equation} 
where $\alpha^E$ is its electric and $\alpha^H$ its magnetic polarizability. The latter takes the magnetic moments due to eddy currents into account which play an important role for thermal emission of metallic nanoparticles~\cite{YVM2005,PMT2006,DedkovKyasov2007,POC2008}. Hence, we find that in comparison to the spontaneous emission of an atom or molecule above a substrate~\cite{Novotny} where the emission rate is proportional to the local density of states for the transition frequency, the thermal emission rate is given by a spectral average of the local density of states with respect to $\hbar \omega \rd n /\rd T$. Hence, the thermal relaxation rate ressembles the spontaneous emission rate if the nanoparticles have a narrowband emission spectrum. 

\begin{figure}
  \includegraphics[width=0.4\textwidth]{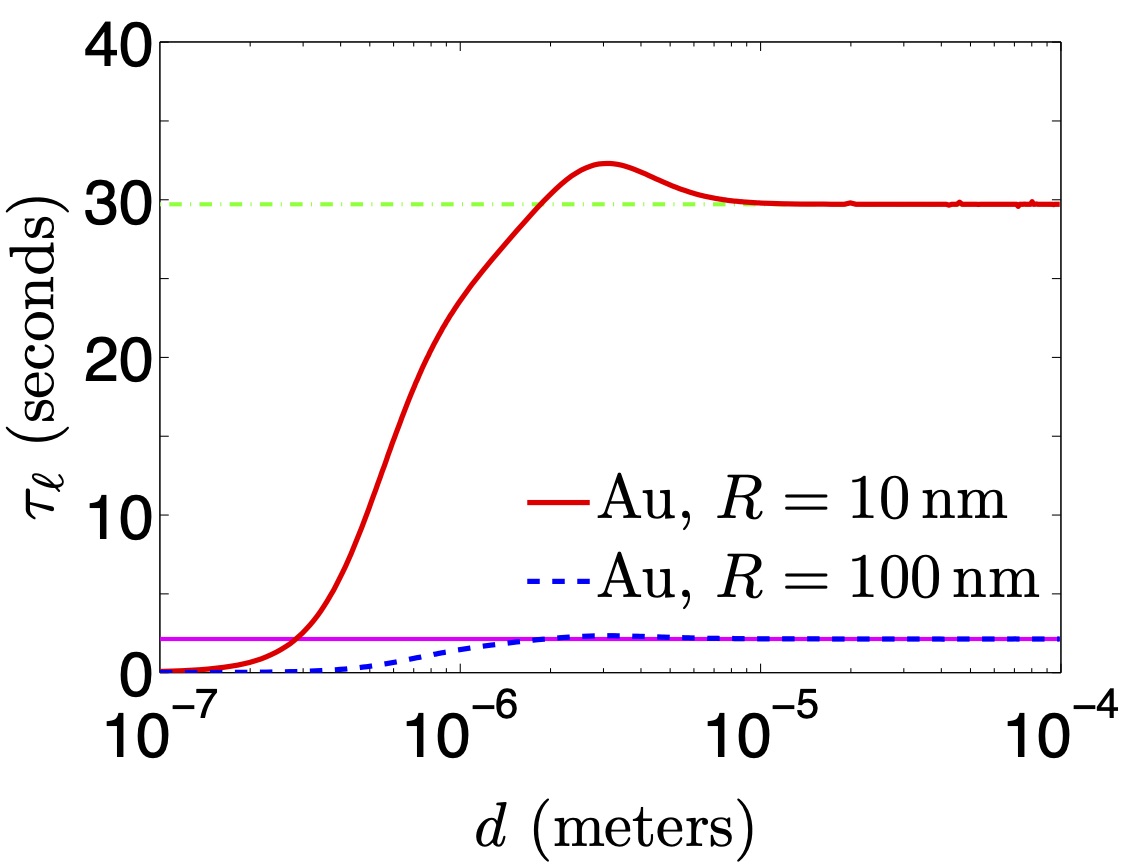}
  \includegraphics[width=0.4\textwidth]{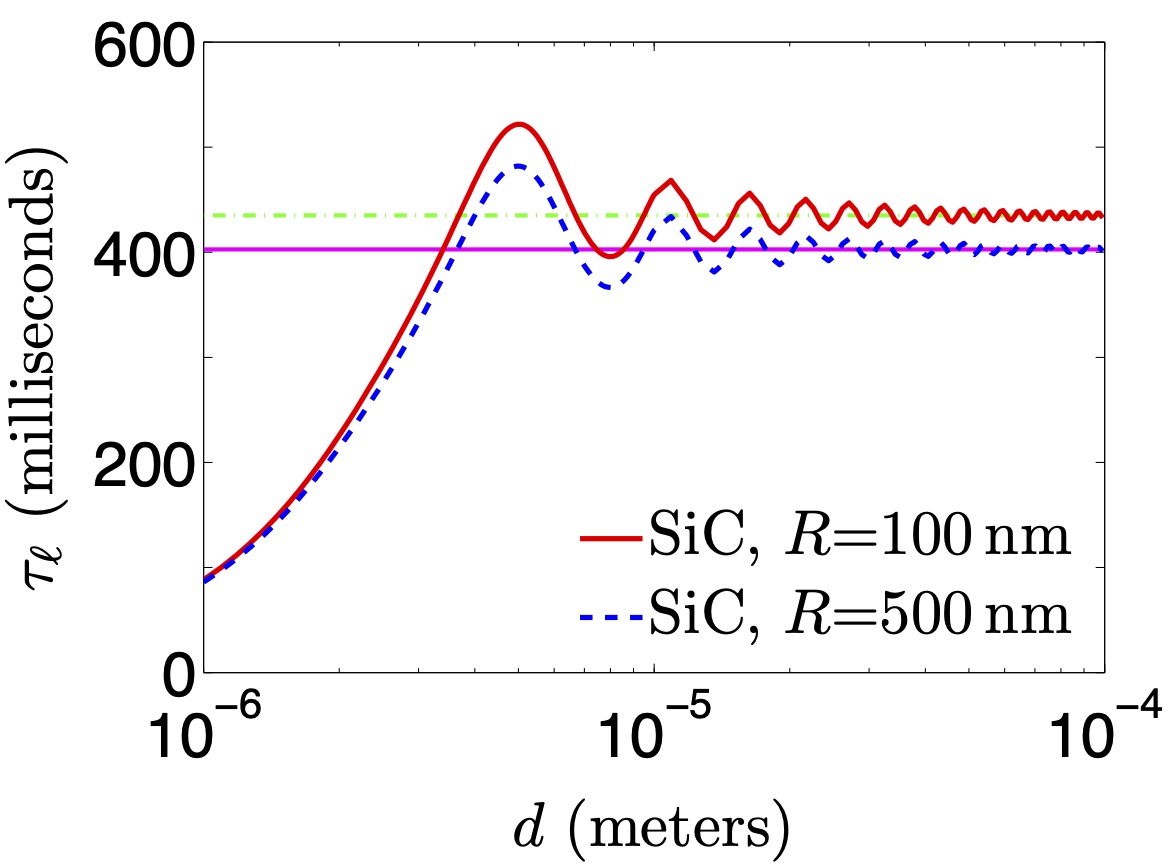}
	\caption{Distance dependence of the relaxation time $\tau = \Gamma^{-1}$ of a nanoparticle above a substrate with temperature $T_b = 300\,{\rm K}$
           (a) for a gold nanoparticle above a gold surface, (b) a SiC nanoparticle above a SiC surface. 
           We use $\rho^{\rm Au} C_{\rm p}^{\rm Au} = 2.404\cdot10^6 \,{\rm J} {\rm m}^{-3} {\rm K}^{-1}$ and 
	   $\rho^{\rm SiC} C_{\rm p}^{\rm SiC} = 2.212\cdot10^6 \,{\rm J} {\rm m}^{-3} {\rm K}^{-1}$. From~\cite{MTetal2012b}.}
\label{Fig:ThermalRelaxationHalfspace}
\end{figure}

In Fig.~\ref{Fig:ThermalRelaxationHalfspace} it can be nicely seen that the thermal relaxation time changes by orders of magnitude when going from the far-field into the near-field regime which is due to the strong increase in $G_{1b}$, i.e.\ the local density of states, in the near-field regime~\cite{Doro2011,KJetal2003}. There is also a large difference for metallic and dielectric nanoparticles due to the fact that thermal radiation is more efficient for dielectric than for metals. Furthermore, it can be seen that for SiC oscillations in the transition region between near-field and far-field regime which can be interpreted as the photonic counterpart of the Friedel oscillations~\cite{KJetal2003}. These oscillations are due to the oscillations in the local density of states which average out for the gold nanoparticle (broad band thermal emission spectrum) but remain for the SiC nanoparticle (narrow band thermal emission spectrum). A detailed discussion can be found in \cite{MTetal2012b}.

\subsubsection{Dynamical control}

\begin{figure}
   \includegraphics[width=0.4\textwidth]{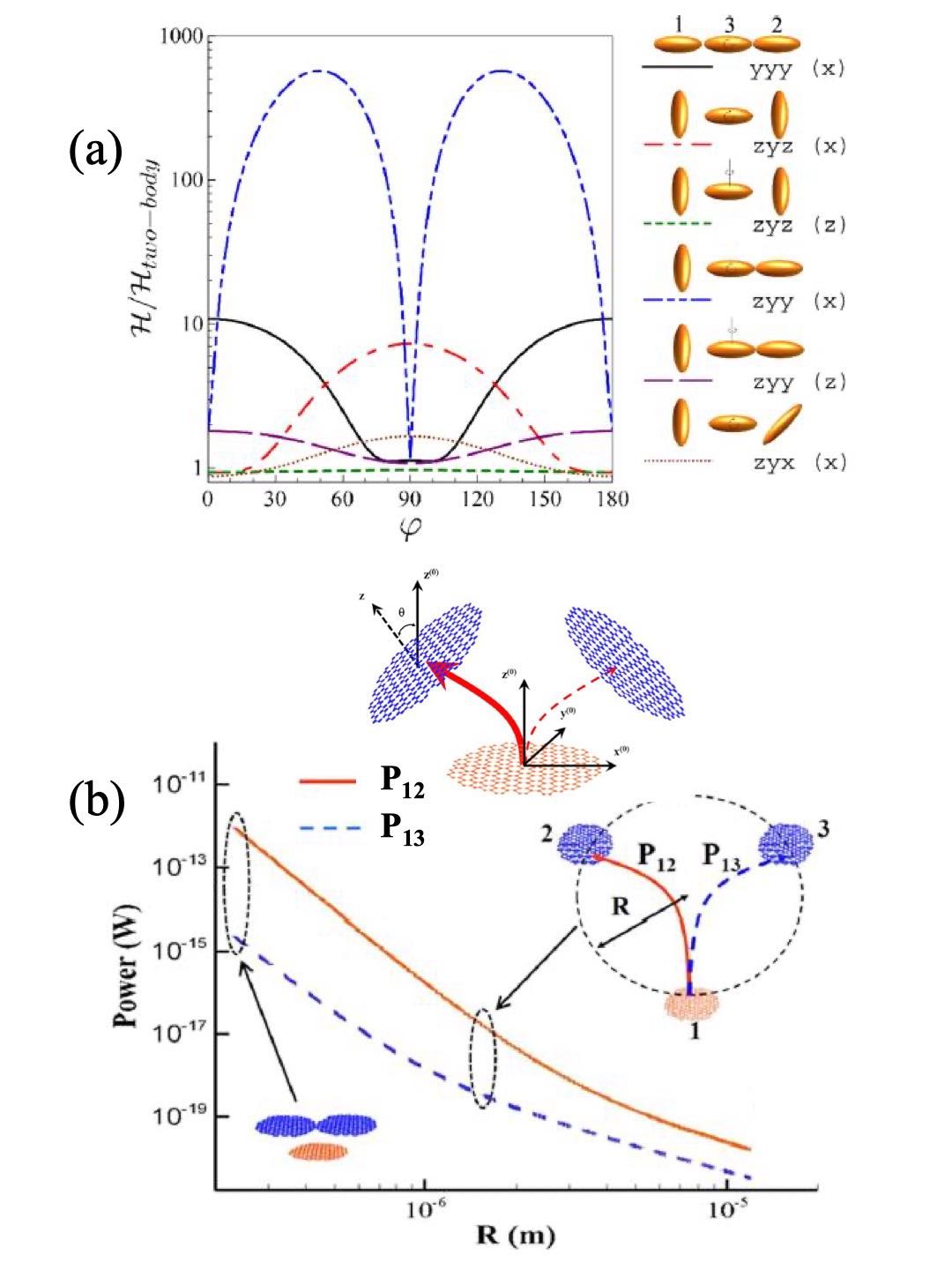}
   \caption{(a) Normalized heat flux between two spheroidal nanoparticles with respect to the orientation of a third particle placed in between. From~\cite{MN2014}. (b) Graphene-based heat flux splitter. Three graphene disks with different Fermi levels controlled by external gating exchange thermal energy in the near-field through many-body interactions. The magnitude of heat flow from 1 to 2 and 1 to 3 can be controlled by an appropriate tuning of the Fermi level of the graphene disks 2 and 3. The thermal power exchanged in the near-field between graphene disks of $100\,{\rm nm}$ radius versus the separation distance in a three body system. From \cite{PBAetal2015}.
   \label{Fig:Switch}}
\end{figure}

A control of the magnitude of heat flux has been highlighted in layered many-body systems~\cite{MJHetal2017b} coated by graphene sheets simply by tuning the doping level of graphene.
Beyond this control several principles have been introduced during the last decade to dynamically control both the magnitude and the direction of heat flux at nanoscale with many-body systems. For example, by changing the shape and orientation of elements~\cite{MN2014} the heat flux can be modulated by several orders of magnitude with anisotropic particles as shown in Fig.~\ref{Fig:Switch}(a). Another example for a dynamical modulation which can by realized by electrical gating is the heat flux splitter as sketched in Fig.~\ref{Fig:Switch}(b). It enables to control the direction of the heat flux in the near-field regime. The design is based on a network of tunable graphene palets~\cite{PBAetal2015} which allow us to control spatially the near-field interactions and therewith the direction of heat flux by dynamically tuning the graphene plasmons. A similar control also has been performed with polar particles covered by graphene~\cite{JSetal2019}. 

\begin{figure}
  \includegraphics[width=0.35\textwidth]{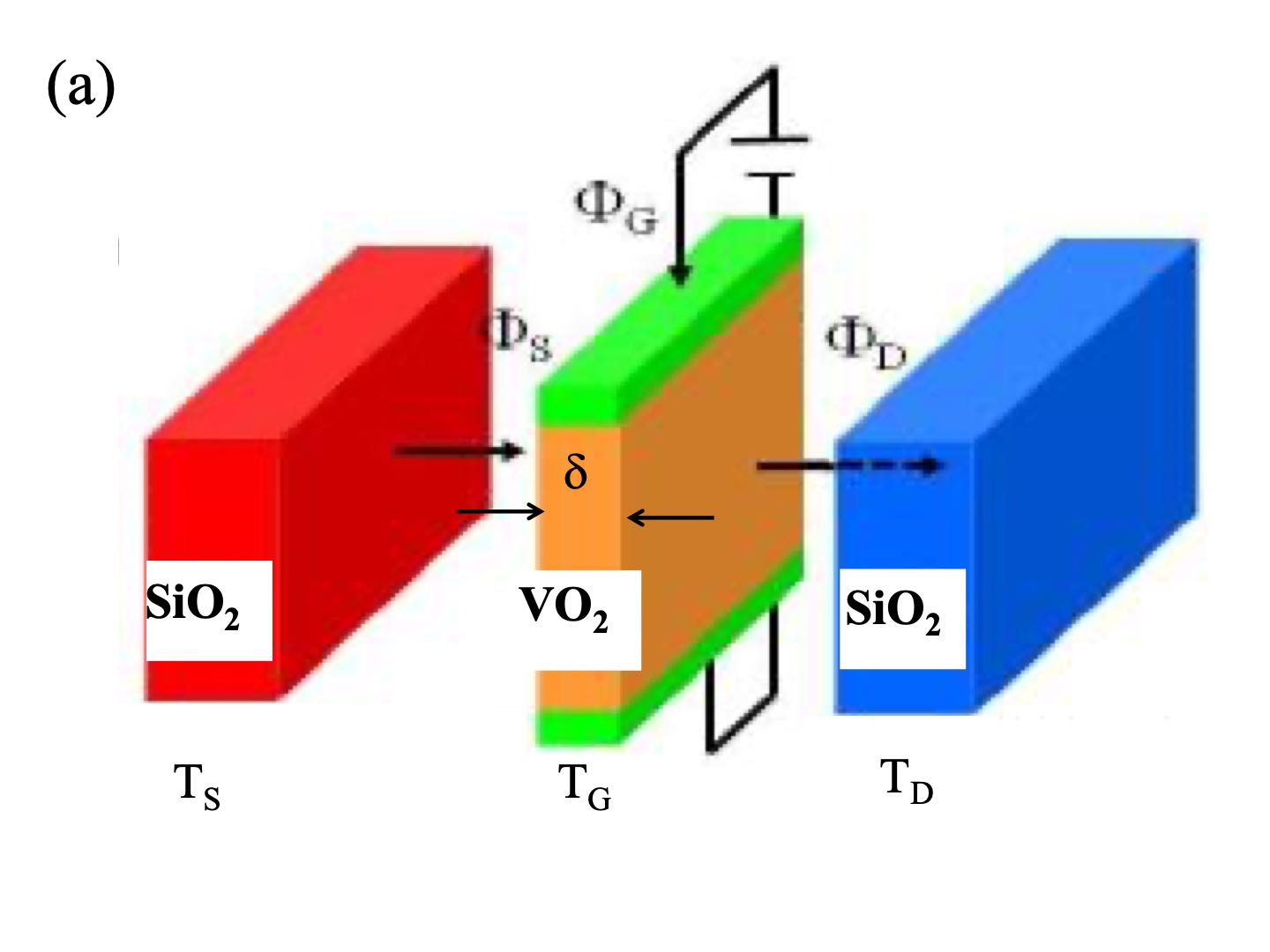}
  \includegraphics[width=0.45\textwidth]{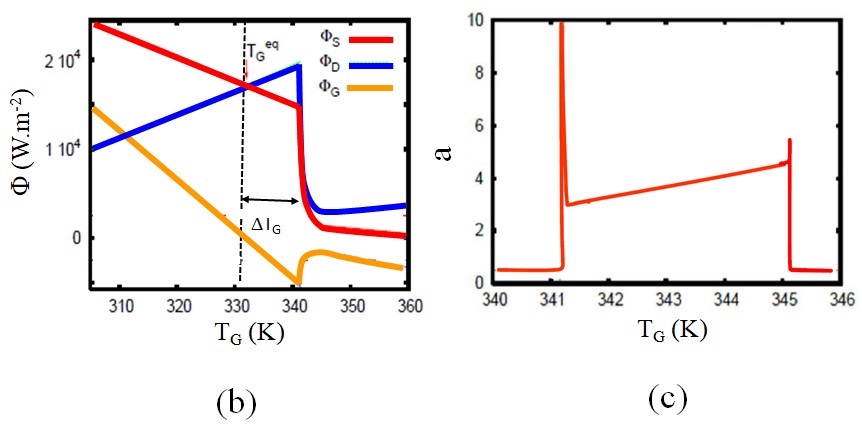}
	\caption{(a) Radiative thermal transistor made of a three-terminal system composed of a SiO$_2$ source, a VO$_2$ gate and a SiO$_2$ drain. The gate is a layer based on a phase-change material and its temperature can be actively controlled around its local equilibrium value $T^{eq}_G$ by an external thermostat while the temperature $T_S = 360\,{\rm K}$ and $T_D = {\rm 300}$ of source and drain are fixed so that $T_S > T_D$. (b) Radiative fluxes $\Phi_S, \Phi_D$, and $\Phi_G$ exchanged between the different parts inside the transistor. (c) Amplification factor with respect to the gate temperature. From~\cite{PBAandSAB2014}.}
\label{Transistor}
\end{figure}

Recently it could be demonstrated that the flux exchanged between two solids can even be amplified through a transistor effect~\cite{PBAandSAB2014} by using a phase-change material like VO$_2$ for an intermediate relay also called gate between two SiO$_2$ slabs functioning as source and drain at temperatures $T_S = 360\,{\rm K}$ and $T_D = {\rm 300}$ as illustrated in Fig.~\ref{Transistor}. Since this configuration corresponds to two oppositly connected heat radiation diodes~\cite{PBASAB2013,YYetal2013,KIetal2014,WGetal2015,AFetal2018} this transistor corresponds to a bipolar transistor so that the terminology emitter, base, and collector would be more appropriate but this has no impact on the physics involved. In the region of the phase-transition around its critical temperature $T_c \approx 340\,{\rm K}$ even though the temperature difference between the gate and the drain is increased a drastic reduction of flux $\Phi_D$ received by the drain takes place due to the strong change in the optical properties of the VO$_2$ gate from a dielectric to a metallic response shielding the heat flux from the source towards the drain as can be seen in Fig.~\ref{Transistor}(b). This variation corresponds to the presence of a negative differential thermal conductance or resistance~\cite{BLIetal2006} $R_{D} = (\frac{\partial \Phi_{D}}{\partial T_G})^{-1}$ induced by the phase transition. In the transition region, the amplification factor
\begin{equation}
  a= \biggl| \frac{\partial \Phi_D}{\partial \Phi_G}\biggr|
\end{equation}
 of the flux received by the drain $\Phi_D$ compared to the heat flux $\Phi_G$ removed or added to the gate can be defined. It can also be recast in terms of the thermal resistances of the source and the drain as
\begin{equation}
  a= \biggl| \frac{R_S}{R_S + R_D} \biggr|
\end{equation}
with the postive resistance $R_{S} = - (\frac{\partial \Phi_{S}}{\partial T_G})^{-1}$. This expression clearly shows that the amplification factor can only become larger than one if $R_D$ is negative so that a negative thermal resistance is a necessary condition for obtaining an amplification. For the thermal transistor the amplification factor is clearly larger than one in the phase-change temperature region as can be seen Fig.~\ref{Transistor}(c). Note that the peaks at the edges of the phase transition are an artefact of the effective medium model used to model the transition of the optical properties of VO$_2$ in this region. Investigations of the same effect in the far-field regime, the impact of the hysteresis of the transistor can be found in~\cite{KJetal2015,HPetal2016,HPetal2018} while the dynamical response of transistors can be found in ~\cite{ILetal2019}.

\begin{figure}
   \includegraphics[width=0.35\textwidth]{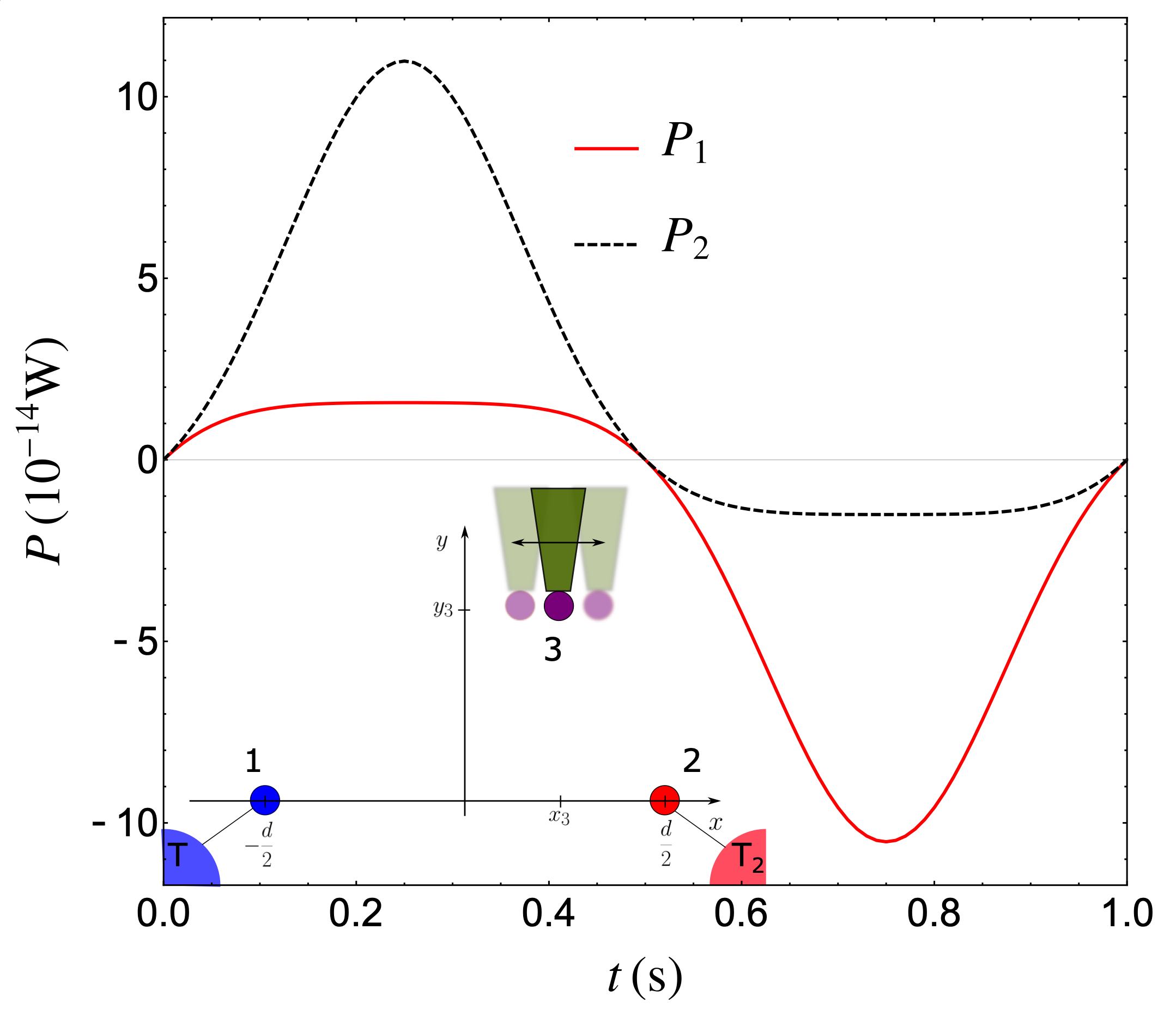}
   \caption{Inset: geometrical configuration of a three-particle system, where the position of particle 3 is periodically modulated. Main part of the figure: radiative heat pumping by modulation of control parameters in a three-particles system sketched in the inset. The three particles are made of SiC. In this specific case, particles 1 and 2 are thermostated at temperature $T_1=T_2$ while the temperature $T_3$ and the $x_3$ coordinate of particle 3 can be modulated with respect to time. Powers $\mathcal{P}_1$ and $\mathcal{P}_2$ absorbed by particles 1 (solid red line) and 2 (dashed black line) as a function of time for the a periodic variation of the coordinate and temperature of particle 3 of frequency $\omega = 2\pi\,$s$^{-1}$ and amplitudes $\Delta x=100\,$nm and $\Delta T=5\,$K around $x_3=0$ and $T_3=300\,$K. We have $d=600\,$nm and $y_3=300\,$nm, and the radius of the particle is $R=50\,$nm. From \cite{RMandPBA2020}.}
\label{fig:pumping}
\end{figure}

The principle of negative thermal resistance plays further an important role for the so-called radiative heat shuttling which has been proposed~\cite{ILetal2018}, recently. In a system consisting of only two parallel slabs, it has been shown that the periodic modulation of the temperature and/or chemical potential of the two bodies can be exploited to control the heat flux between them. More specifically, it has been proven that in order to thermally insulate them a negative thermal differential resistance is required. A further step in this direction has been done in~\cite{RMandPBA2020}, where the heat flux between two particles is tailored by periodically modulating the temperature $T_3$ and the position $x_3$ of a third particle in a three-particle system as sketched in the inset of Fig.~\ref{fig:pumping}. This many-body configuration allows for controlling the direction and amplitude of the heat exchanged between the two particles $1$ and $2$, even when they are kept at the same temperature and (differently from the shuttling effect mentioned above) in the absence of a negative thermal differential resistance~\cite{RMandPBA2020}. This possibility can be anticipated already by performing a Taylor expansion up to second order, around the middle position $x_3=0$ and the equilibrium temperature $T_3=T_{3,\text{eq}}$ of particle 3. This gives
\begin{equation}\begin{split}
   \mathcal{P}_1 &\simeq \mathcal{P}_1(0,T_{3,\text{eq}}) + \frac{\partial \mathcal{P}_1}{\partial x_3}x_3 + \frac{\partial \mathcal{P}_1}{\partial T_3}(T_3 - T_{3,\text{eq}})\\
 &\,+ \frac{1}{2} \frac{\partial^2 \mathcal{P}_1}{\partial x_3^2}x_3^2 + \frac{1}{2} \frac{\partial^2 \mathcal{P}_1}{\partial T_3^2}(T_3 - T_{3,\text{eq}})^2\\
 &\,+ \frac{\partial^2 \mathcal{P}_1}{\partial x_3\partial T_3}x_3(T_3 - T_{3,\text{eq}}).
\end{split}\end{equation}
For a time variation of the form $T_3(t)=T_{3,\rm{eq}} + \Delta T\sin(\omega t)$ and $x_3(t)=\Delta x\sin(\omega t+\phi)$, and in the specific case $T_1=T_2=T_{3,\rm{eq}}$, the time average over a period reads
\begin{equation}
  \langle \mathcal{P}_1\rangle_t \simeq \frac{\Delta T}{2}\Bigl(\Delta x\frac{\partial^2 \mathcal{P}_1}{\partial x_3\partial T_3}\cos\phi + \frac{\Delta T}{2}\frac{\partial^2 \mathcal{P}_1}{\partial T_3^2}\Bigr),
\end{equation}
This equality clearly shows that magnitude of the first term can be easily modulated simply by changing the dephasing $\phi$ between $x_3$ and $T_3$, paving the way to an active heat pumping mechanism. More intringuing , the sign can be changed as well so that the heat can flow from cooler to warmer regions.
A numerical example of this modulation for a vanishing dephasing $\phi=0$ is shown in Fig.~\ref{fig:pumping}, where the average over a period of the powers $P_1$ and $P_2$ absorbed by particles 1 and 2 (having temperatures $T_1=T_2=300\,$K) are positive and negative, respectively.

\subsubsection{Heat transport regimes}\label{Sec:regimes}

\begin{figure}
 \includegraphics[width=0.4\textwidth]{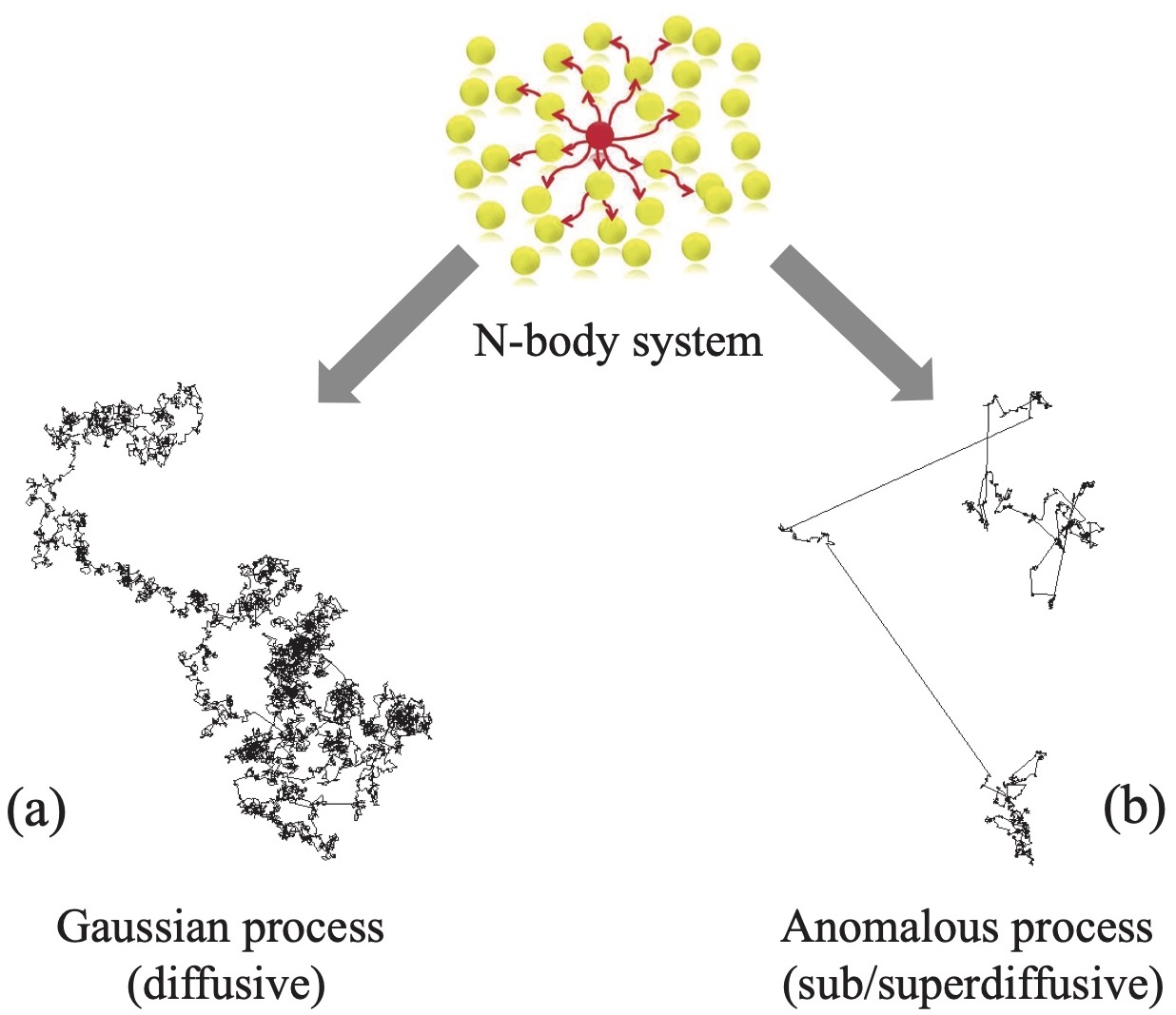}
  \caption{Types of heat transport regimes in a $N$-body system. When an element (red) is heated up its heat spread out through the system either by (a) a classical (Gaussian) diffusion process or (b) an anomalous process. The trajectories correspond to random walks with a Gaussian and a non-Gaussian probability distribution function, respectively. Here, the non-Gaussian process is a Levy flight with an algebraic jpg.}
\label{regimes}
\end{figure}

It is commonly admitted that heat conduction inside a bulk solid is governed by a normal diffusion process. Heat carriers that are electrons or phonons move through the atomic lattice following a usual random walk which is driven by a Gaussian distribution function as in Fig.~\ref{regimes}(a). In this section we discuss how heat carried by thermal photons is transported in many-body systems. We demonstrate the existence of anomalous regimes of transport as in Fig.~\ref{regimes}(b). In dilute systems we show that heat can spread out following a superdiffusive process \cite{Levy,Schlesinger} while in dense systems it can be ballistically transported.
 
To start this analysis, let us consider a network of small objects at temperature $T_{i}$ which are distributed inside an background or environment at temperature $T_b$. When the separation distance between two arbitrary objects in this network is much larger than their characteristic size and that their size is small enough compared with the thermal wavelengths $\lambda_{T_{i}} = c\hbar/(\kb T_{i})$ then this network can be modelled as a set of simple dipoles located at positions $\mathbf{r_{i}}$ in mutual interaction and in interaction with the surrounding bath. In near-field regime the power exchanged with the bath is negligible as discussed in Sec.~\ref{Sec:RelaxDyn} compared to the internal exchanges. Then the time evolution of objects temperature is governed by Eq.~(\ref{Eq:diff2}) neglecting the heat exchange with the background yielding
\begin{equation}
  I_{i}\frac{\rd T_{i}}{\rd t}= \sum_{j\neq i} G_{ij}(T_j-T_i)
  \label{Eq:heat_eq},
\end{equation}
where $I_{i}$ represents the thermal inertia of object $i$ while $G_{ij}$ stands for the thermal conductance between the $j^{th}$ and the $i^{th}$ dipole as defined in Eq.~(\ref{Eq:ConductanceDipoles}) which depends only on the distance between the dipoles
\begin{equation}
  G_{ij}\equiv G(\mid\mathbf{r}_{i}-\mathbf{r}_{j}\mid).
   \label{Eq:conductance}.
\end{equation}

\begin{figure}
\includegraphics[width=0.35\textwidth]{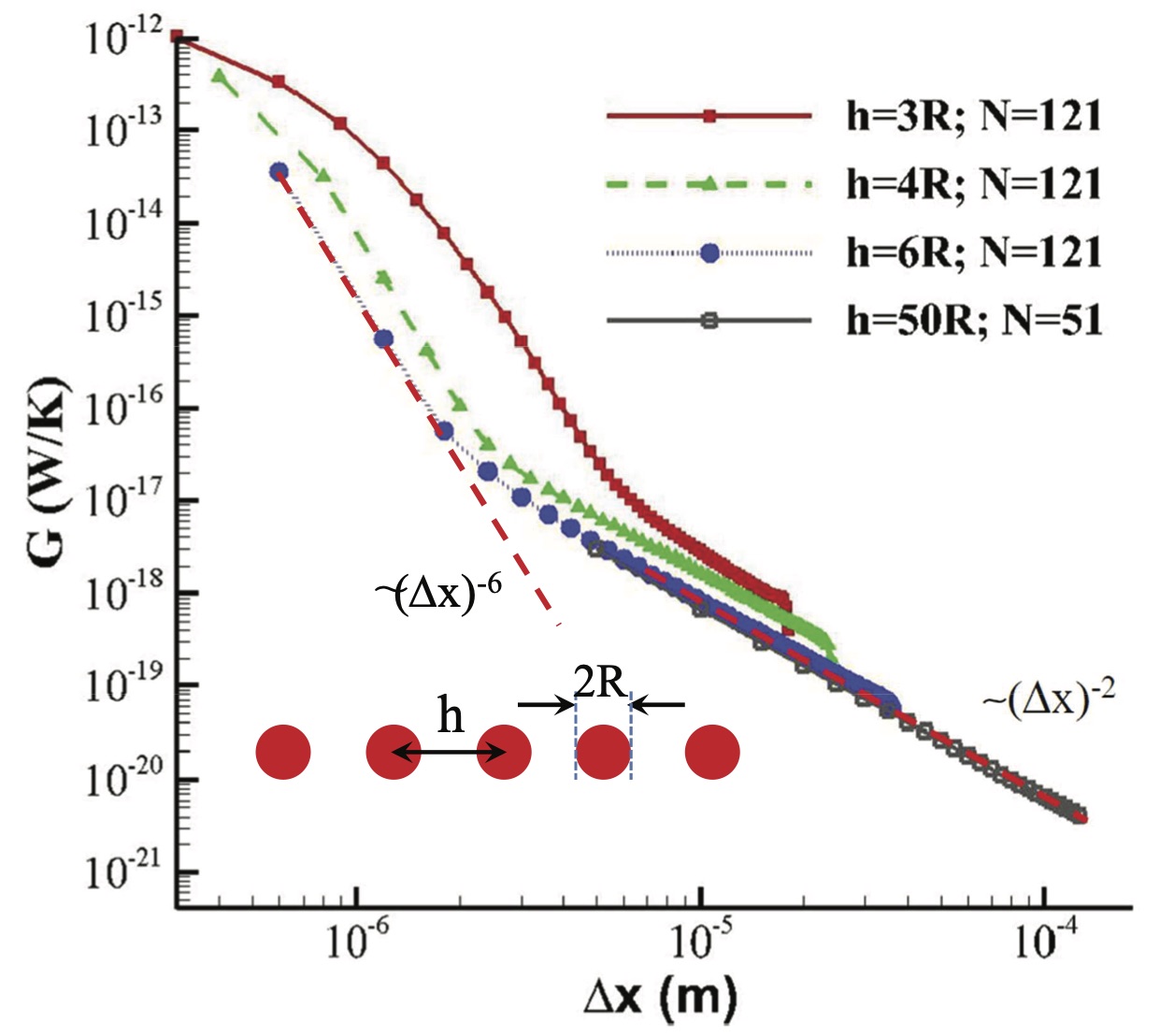}
\includegraphics[width=0.4\textwidth]{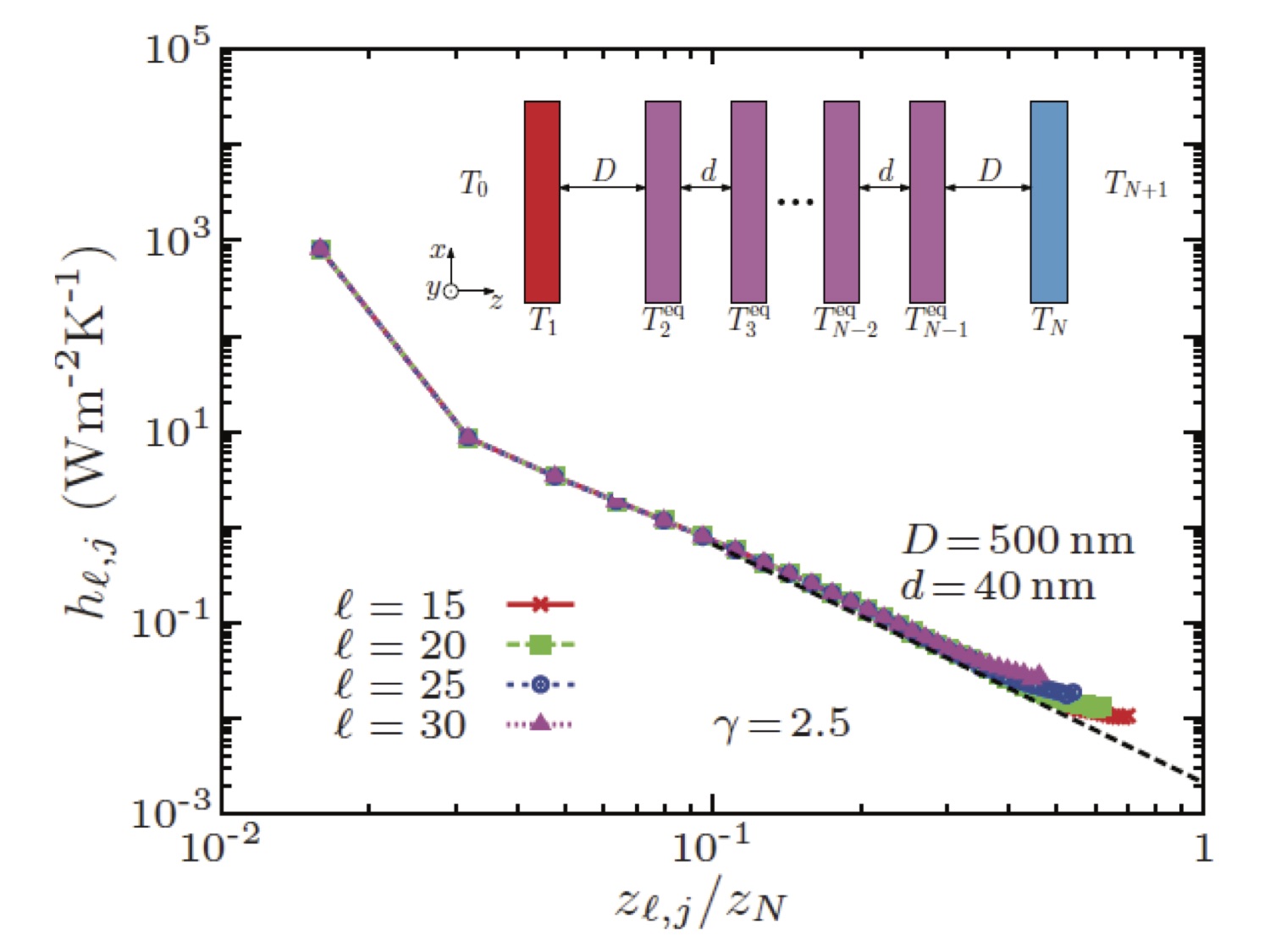}
   \caption{(a)Thermal conductance $G$ in log-log scale along a chain of SiC spherical particles 100 nm radius with different inter-particle distances $h$ and different particle numbers $N$ as a function of the separation distance $\Delta x= | \mathbf{r} - \mathbf{r}'|$ at temperature $T=300\,$K. From \cite{PBAetal2013}.
   (b) Heat-transfer coefficients $h_{\ell,j}$ with respect to the normalized separation $z_{\ell,j}/z_N$ in a dilute multilayer system made with SiC layers 200 nm thick separated by a distance d=40 nm at $T=300\,$K. From \cite{ILetal2018b}.}
\label{superdiffusion}
\end{figure}

In the continuous limit the energy balance equation (\ref{Eq:heat_eq}) can be recast as~\cite{PBAetal2013}
\begin{equation}
  \frac{\partial T_{i}}{\partial t}= \int_{R^{d}} \!\! \rd\mathbf{r} \, p(\mathbf{r}_{i},\mathbf{r})\frac{T(\mathbf{r},t)}{\tau(\mathbf{r})} - \frac {T(\mathbf{r}_{i},t)}{\tau(\mathbf{r_{i}})} 
  \label{Eq:chapman1},
\end{equation}
where the integration is done over the whole space of dimension $d$. 
This equation is formally analog to a Chapman-Kolmogorov master equation which drives a generalized Markov process. The temperature field $T(\mathbf{r},t)$ is a passive scalar which evolves by following a generalized random walk of probability distribution function (jpg) 
\begin{equation}
  p(\mathbf{r},\mathbf{r}')= \frac{G(|\mathbf{r}-\mathbf{r}'|)} { \int_{R^{d}} \rd \mathbf{r}' G(|\mathbf{r}-\mathbf{r}'|)}
  \label{Eq:chapman3}
\end{equation}
and the rate of jumps between two collission events
\begin{equation}
  \tau(\mathbf{r})= \biggl(\int_{R^{d}} \!\!\! \rd\mathbf{r}' \, G(|\mathbf{r}-\mathbf{r}'|) \biggr)^{-1}
  \label{Eq:chapman4}.
\end{equation}
Hence, by analyzing the spatial variation of the jpg and therefore of the conductance as well between two points inside the system we can identify the regime of heat transport. If the asymptotic behavior of the jpg $p(x)$ (where we have set $x=\mid\mathbf{r}-\mathbf{r^{'}}\mid$) is Gaussian, all its moments $M^{(n)} = \int x^{n}p(x)dx$ are finite so that the regime of transport is diffusive. On the other hand if it decays algebraically, i.e.\ $p(x)=O(1/x^\gamma)$ and hence $G(x) =O(1/x^\gamma)$, then there is a given order $\overset{\sim}{n}$ beyond which $M^{(n)}$ diverges for any $n>\overset{\sim}{n}$. In this case, the heat transport regime becomes superdiffusive (see right trajectory on Fig.\ref{regimes}). In this specific case the (continuous) energy balance equation takes the form \cite{PBAetal2013}.
\begin{equation}
  I \frac{\partial T}{\partial t}=-\kappa(-\Delta)^{(\gamma-d)/2}T(\mathbf{r}),
  \label{Eq:heat_eq2}
\end{equation}
where $\kappa$ is a parameter which depend on the dimension $d$ and $(-\Delta)^{\alpha/2}$ denotes the fractional Laplacian~\cite{Schlesinger}
\begin{equation}
  (-\Delta)^{\alpha/2}T(\mathbf{r})=c_{d;\alpha}\,\text{PV}\int_{R^{d}} \!\! \rd\mathbf{r}' \, \frac{T(\mathbf{r})-T(\mathbf{r}')}{|\mathbf{r}-\mathbf{r}'|^{d+\alpha}}
\end{equation}
with $c_{d;\alpha}=\frac{2^{-\alpha} \pi^{1+d/2}}{\Gamma(1+\alpha/2)\Gamma(\frac{d+\alpha}{2})\sin(\alpha \pi/2)}$. It is worthwhile to note that Eq.(\ref{Eq:heat_eq2}) is general and can be applied for describing the energy balance in arbitrary dipolar or macroscopic systems. When $\gamma\rightarrow d+2$ the fractional Laplacian degenerates into its classical form, i.e.\ $(-\Delta)^{\alpha/2})=(-\Delta)$, and the transport regime is diffusive. On the other hand when $\gamma \rightarrow d$ the fractional Laplacian approaches the identity operator and the transport becomes ballistic. Finally when $d<\gamma<d+2$ the regime is superdiffusive.

\begin{figure}
  \includegraphics[width=0.4\textwidth]{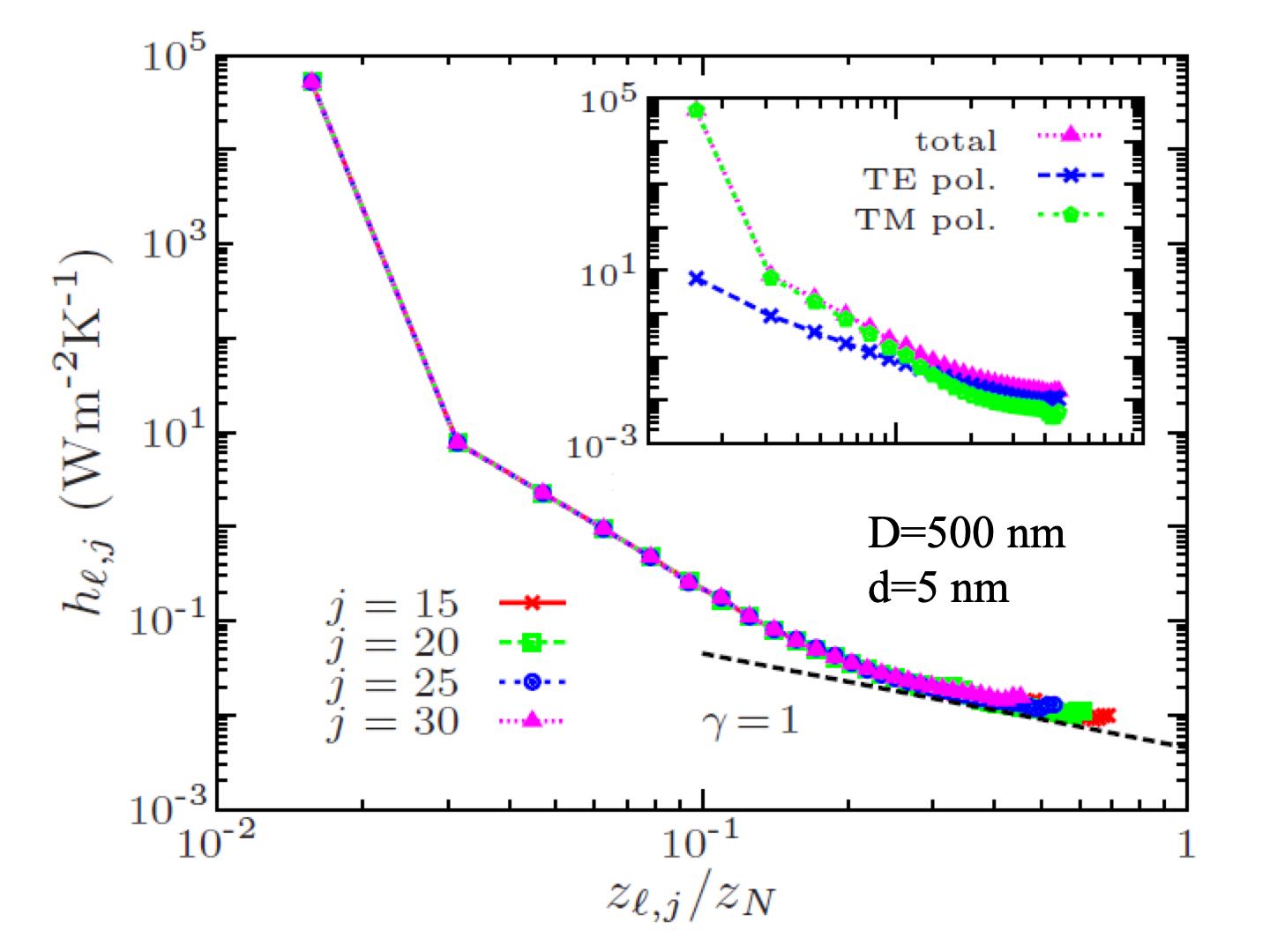}
\caption{Heat transfer coefficient $h_{\ell,j}$ with respect to the normalized separation $z_{l,j}/z_N$ in a dense multilayer system made with SiC layers 200 nm thick separated by a distance  d=5 nm at $T=300\,$K The inset decomposes $h_{\ell,j}$ into TE and TM polarization contributions. From~\cite{ILetal2018b}.}
\label{ballistic}
\end{figure}

\begin{figure}
  \includegraphics[width=0.4\textwidth]{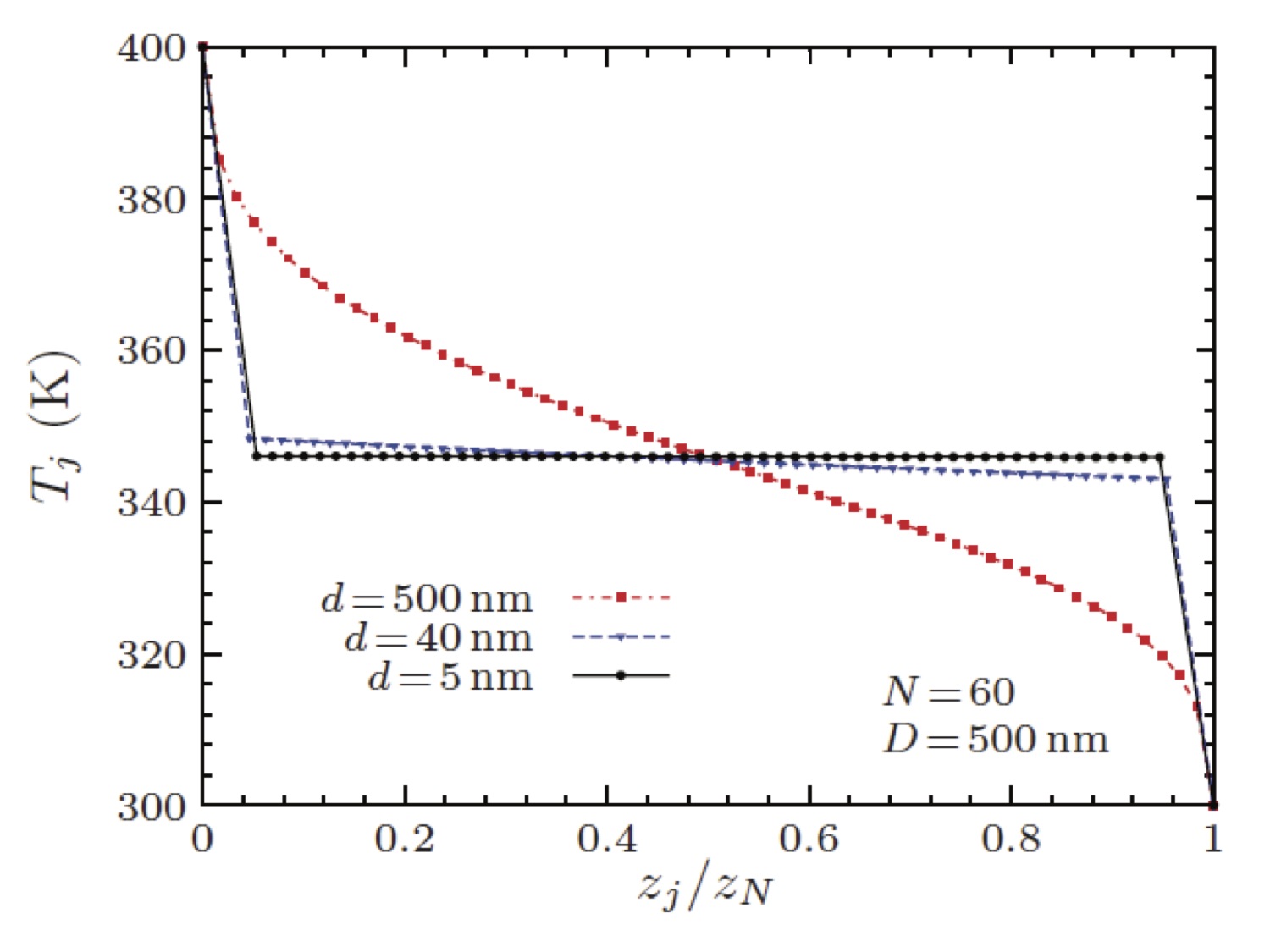}
  \caption{Temperature profile as a function of the normalized position $z_j/z_N$ along a multilayer made with $N=60$ SiC layers 200\,nm thick for different separation distances $d$ and at fixed distance $D=500\,$nm from the thermostats. From~\cite{ILetal2018b}.}
\label{temp_ballistic}
\end{figure}

In Figs.\ref{superdiffusion} and \ref{ballistic} we show the existence of those regimes in two simple many body systems: (1) linear chains of nanoparticles periodically dispersed in vacuum and (2) multilayer periodic systems. In the first system (see Fig.~\ref{superdiffusion}(a)), the thermal conductance $G(\Delta x)$ between a central particle and another particle at a distance $\Delta x$, is calculated for different filling factors ($2R / h$). For any filling factor, we see that $G$ decays asymptotically at long separation distance as $ 1 / \Delta x^2$, i.e.\ $\gamma = 2$, showing according to our previous discussion that the regime of heat transport is superdiffusive. In the example plotted in Fig.~\ref{superdiffusion}(a) the long range interactions which give rise to this anomalous regime comes from the presence of collective electromagnetic modes supported by the whole structure. In the case of a chain made with silicon carbide (SiC) particles these modes result from the coupling of surface phonon-polaritons localized on each particle~\cite{PBAetal2013,JOetal2015,CKetal2018,ETetal2020}. 

A similar superdiffusive regime is observed in dilute multilayer systems (see Fig.~\ref{superdiffusion}(b)) where the heat transfer coefficient $h_{l,j}$ between layers $l$ and $j$ decays algebraically and scales as $1/z^{2.5}_{l,j}$ where $z_{l,j}$ is the distance between layers $l$ and $j$ so that $\gamma = 2.5$. On the other hand, in a dense multilayer system as considered in Fig.~\ref{superdiffusion}(b) a transition occurs between this superdiffusive regime and a ballistic regime~\cite{ILetal2018b,MJHetal2019}. In this case we see that $h_{l,j}$ scales as $1/z_{l,j}$ meaning that the transport becomes clearly ballistic and the temperature profile inside the structure submitted to a temperature gradient is constant as can be seen in Fig.~\ref{temp_ballistic} having a value $T^*$ which is close to the Casimir temperature $T_C=\frac{T_1+T_N}{2}$. This regime of heat transport seems to be inconsistent with the previous arguments about the collective modes supported by the structure, but it occurs due to the fact that the coupling of the inner dense multilayers is much stronger than the coupling to the two outer baths when $d \ll D$. For $D = d$ on the other hand the temperature profile in Fig.~\ref{temp_ballistic} is reminiscent of a quasi-ballistic temperature distribution. Although the transition mechanism remains today partially elusive it has been shown in~\cite{ILetal2018b} that it is related to a change of channel for heat exchanges in dense systems from TM dominated to TE dominated heat transfer (see inset of Fig.~\ref{ballistic}). For this TE polarization state the slabs do not support anymore surface waves.



\subsection{Non-reciprocal systems}
\label{Sec:NonReciprocal}

In electromagnetics, a nonreciprocal system is defined as a system that exhibits different received-transmitted field ratios when a source and a detector are interchanged. This concept is also closely related to a time reversal symmetry breaking of Maxwell’s equation. In this case the classical Lorentz’s reciprocity is violated~\cite{CCetal2018}. Here below we discuss first the general formulation of radiative power exchange between non-reciprocal objects and then show how RHT is taking place in non-reciprocal many-body systems made for sets of simple non-reciprocal nano-particles.

\subsubsection{General discussion}

As a first step, let us consider only two objects $1$ and $2$ having temperatures $T_1$ and $T_2$, respectively, which are immersed into a background or environment having another temperature $T_{\rm b}$. Under the assumption that the objects and the environment can be considered to be in local thermal equilibrium, the power absorbed by object $1$ can be determined with the conventional FE approach analogous to Eq.~(\ref{Eq:PowerDipolGeneral}) as~\cite{ILandPBA2017,FHandSAB2019}
\begin{equation} 
  \mathcal{P}_1= 3 \int_{0}^\infty \!\!\frac{\rd \omega}{2 \pi}\, \hbar \omega \bigl[ (n_1 - n_{\rm b}) \mathcal{T}_{11} + (n_2 - n_{\rm b}) \mathcal{T}_{12} \bigr],
\label{Eq:PowerTwoParticlesEnvironment}
\end{equation}
where $n_{1/2} = n(T_{1/2})$ and $n_{\rm b} = n(T_{\rm b})$. The transmission coefficients $\mathcal{T}_{\alpha/\beta}$ are, for example, explicitly given in terms of the $T$ operators of the objects in~\cite{FHandSAB2019} where they were derived within the scattering approach~\cite{MKetal2012}. Here we only give explicitly the expression for $\mathcal{T}_{21}$ which is given by~\cite{FHandSAB2019}
\begin{equation}
  \mathcal{T}_{12} = \frac{4}{3} {\rm Tr}\bigl[\mathds{D}^{-1} \mathds{G} \chi_2  (\mathds{D}^{-1} \mathds{G})^\dagger \tilde{\chi}_1 \bigr], 
 \label{Eq:T2alpha2} 
\end{equation}
where the trace is the operator trace, $\mathds{G}$ is the operator for the Green's function, $\mathds{D} = (\mathds{1} - \mathds{G} \mathds{T}_2 \mathds{G} \mathds{T}_1)$ written in terms of the T-operators $\mathds{T}_{1/2}$ of both objects and the generalized suszeptibilities are defined as
\begin{align}
  {\chi}_{2} &= \frac{\mathds{T}_{2} - \mathds{T}_{2}^\dagger}{2 \ri} - \mathds{T}_{2} \frac{\mathds{G} - \mathds{G}^\dagger}{2 \ri} \mathds{T}_{2}^\dagger, \\
  {\tilde{\chi}}_{1} &= \frac{\mathds{T}_{1} - \mathds{T}_{1}^\dagger}{2 \ri} - \mathds{T}_{1}^\dagger \frac{\mathds{G} - \mathds{G}^\dagger}{2 \ri} \mathds{T}_{1}.
\end{align}
Note that this expression is formally equivalent to the expressions in Eqs.~(\ref{Eq:Transmission12b}) and (\ref{eq:T12}). 
Analogous expressions can also be found in the work \cite{LZetal2018} and more explicitly for spherical nanoparticles in~\cite{Ott2020}. The corresponding expression for the absorbed power $\mathcal{P}_2$ in object $2$ can be obtained by exchanging $1 \leftrightarrow 2$ in the above expression. First of all, it can now easily be seen in Eq.~(\ref{Eq:PowerTwoParticlesEnvironment}) that in global thermal equilibrium the overall absorbed power is zero. Secondly, when setting $T_{1} = T_{\rm b}$ then the expression in Eq.~(\ref{Eq:PowerTwoParticlesEnvironment}) can only describe the absorbed power in object $1$ due to the heat flow coming from or going towards object $2$. Thus, $\mathcal{T}_{12}$ can be identified as the transmission coefficient describing the heat flow from object $2$ to $1$. Thirdly, when assuming that $T_{2} = T_{\rm b}$ then Eq.~(\ref{Eq:PowerTwoParticlesEnvironment}) describes the heat flow from object $1$ to the environment and to object $2$ or vice versa. Therefore we can identify $\mathcal{T}_{11}$ as the transmission coefficient standing for the so called ``self emission'' of object $1$~\cite{MKetal2012}. Finally, when taking $T_1 = T_2$ then Eq.~(\ref{Eq:PowerTwoParticlesEnvironment}) describes merely the power flowing from the environment towards object $1$ either directly or via object $2$. Therefore $ \mathcal{T}_{11} + \mathcal{T}_{12}$ equals the transmission coefficient $\mathcal{T}_{\rm 1b}$ as discussed also for $N$ dipolar objects when deriving Eq.~(\ref{Eq:PowerDipolGeneralb}). These observations allows us to rewrite Eq.~(\ref{Eq:PowerTwoParticlesEnvironment}) as
\begin{equation}
\begin{split}
   \mathcal{P}_1 &= 3 \int_{0}^\infty \!\!\frac{\rd \omega}{2 \pi}\, \hbar \omega \bigl[ n_1 \mathcal{T}_{11} + n_2 \mathcal{T}_{12} - n_{\rm b} \mathcal{T}_{\rm 1b}  \bigr] \\
                 &\equiv \mathcal{P}_{1 \rightarrow 1}(T_1) +  \mathcal{P}_{2 \rightarrow 1}(T_2) + \mathcal{P}_{\rm b \rightarrow 1}(T_{\rm b})
\end{split}
\label{Eq:PowerTwoParticlesEnvironmentSplit}
\end{equation}
introducing
\begin{align}
    \mathcal{P}_{1 \rightarrow 1}(T_1) &= + 3 \int_{0}^\infty \!\!\frac{\rd \omega}{2 \pi}\, \hbar \omega n_1 \mathcal{T}_{11}, \label{P11}\\
    \mathcal{P}_{2 \rightarrow 1}(T_2) &= + 3 \int_{0}^\infty \!\!\frac{\rd \omega}{2 \pi}\, \hbar \omega n_2 \mathcal{T}_{12}, \label{Eq:PowerOneTwo}\\
    \mathcal{P}_{\rm b \rightarrow 1}(T_{\rm b}) &= - 3 \int_{0}^\infty \!\!\frac{\rd \omega}{2 \pi}\, \hbar \omega  n_{\rm b} \mathcal{T}_{\rm 1b}.
\end{align}
where the first term stands for the self-emission of body 1, the second is the emission toward body 2 and the last term is the power coming from the bath. Notice that when the two bodies are set at the same temperature we can make a connection between the transmission coefficient $\mathcal{T}_{\rm b \rightarrow 1}$ and the thermal emissivity $\epsilon=\sigma_{\rm abs}/S$~\cite{SABandPBA-prb2016}. More specifically, $\mathcal{T}_{\rm b \rightarrow 1}$ can be expressed as a function of its absorption-cross section (\ref{Eq:cross_section_def}), its geometrical cross section $S$ and the absorbed power as follows
\begin{equation}
  \mathcal{T}_{\rm b \rightarrow 1}(\omega)=\frac{A}{2\pi}\epsilon(\omega)  \frac{\omega^2}{c^2}=\frac{A}{2\pi}\frac{\sigma_{\rm abs}(\omega)}{S}  \frac{\omega^2}{c^2}, \label{Eq:emissivity_transmission}
\end{equation}
where $A$ is the surface of the object (assumed convex).

The self-emission term $\mathcal{P}_{1 \rightarrow 1}$ appearing in Eq.~\eqref{P11} must balance the energy flow from the other object $2$ and the environment described by $ \mathcal{P}_{2 \rightarrow 1}$ and $\mathcal{P}_{\rm b \rightarrow 1}$ to establish global equilibrium so that this term describes the power needed to keep the temperature of object $1$ constant in thermal equilibrium. Hence, when taking $T_1 = T_2 = T_{\rm b}$ we have $\mathcal{P}_1 = 0$ and therefore
\begin{equation}
   \mathcal{P}_{1 \rightarrow 1}(T_{\rm b}) = -\mathcal{P}_{2 \rightarrow 1}(T_{\rm b}) - \mathcal{P}_{\rm b \rightarrow 1}(T_{\rm b}).
\end{equation}
This equation relates $\mathcal{P}_{1 \rightarrow 1}$ to $\mathcal{P}_{2 \rightarrow 1}$ and $\mathcal{P}_{\rm b \rightarrow 1}$ and therefore allows us to eliminate the background term $\mathcal{P}_{\rm b \rightarrow 1}(T_{\rm b})$ from the expression for the overall absorbed power giving \cite{MKetal2012}
\begin{equation}
\begin{split}
   \mathcal{P}_1 &= \mathcal{P}_{1 \rightarrow 1}(T_1) -  \mathcal{P}_{1 \rightarrow 1}(T_{\rm b}) \\
                 &\quad +  \mathcal{P}_{2 \rightarrow 1}(T_2) - \mathcal{P}_{2 \rightarrow 1}(T_{\rm b}).
\end{split}
\end{equation}
This elimination of the background term is of course clear from the above definitions showing that $\mathcal{T}_{1 \rightarrow \rm b}$ can also be expressed by $\mathcal{T}_{1 \rightarrow 1}$ and $\mathcal{T}_{2 \rightarrow 1}$ and obviously the implementation of the equilibrium condition brought us back to Eq.~(\ref{Eq:PowerTwoParticlesEnvironment}). As described in \cite{MKetal2012} this expression for $ \mathcal{P}_1$ can now be generalized to the case of $N$ objects in a given environment. In this case ($i = 1,\ldots,N$)
\begin{equation}
   \mathcal{P}_i = \sum_{j = 1}^N \bigl[ \mathcal{P}_{j \rightarrow i}(T_j) - \mathcal{P}_{j \rightarrow i}(T_{\rm b}) \bigr].
\end{equation}
This is the general $N$-body formula for the power absorbed by object $i$ of which Eq.~(\ref{Eq:PowerDipolGeneral}) can be considered as a special case for dipolar objects. For an explicit calculation of the absorbed power it is, of course, necessary to determine the transmission coefficients for the studied configuration. Before focusing on the heat flow in some specific cases, we want to discuss in the next section the impact of the non-reciprocity in a similarly general way.

\subsubsection{General impact of non-reciprocity}
\label{non-reciprocal framework}

For Lorentz-reciprocal objects and their environment the coresponding response functions, i.e.\ the permittivity tensor, the polarizability tensor, T-operator, Green's function etc.\ are symmetric~\cite{CCetal2018}. Consequently, in this case we have symmetric transmission coefficients $\mathcal{T}_{12} = \mathcal{T}_{21}$ or more generally for $N$ objects $\mathcal{T}_{ij} = \mathcal{T}_{ji}$ ($i \neq j$). This means that we have detailed balance for the heat flux between any two objects~\cite{MKetal2012}. In contrast, for configurations where the objects or the environment do not fulfill the conditions for Lorentz reciprocity it has been explicitly proven in \cite{FHandSAB2019} that in general $\mathcal{T}_{12} \neq \mathcal{T}_{21}$. More precisely, $\mathcal{T}_{12} = \mathcal{T}_{21}$ if and only if the objects and their environment are both reciprocal. Therefore non-reciprocity introduces in general a directionality for the heat flow. 

One of the astonishing consequences is that in non-reciprocal systems one has $\mathcal{P}_{12} \neq \mathcal{P}_{21}$ in general so that the heat flux related expressions for the reciprocal case fulfilling detailed balance need to be generalized to the non-reciprocal case where detailed balance is broken~\cite{LZandSF2014} like, for instance, the Green-Kubo relation for heat radiation~\cite{GolykEtAl2013,FHandSAB2019}. More surprisingly, this asymmetry in the heat flow from object $1$ to object $2$ and from $2$ to $1$ even exists in global thermal equilibrium suggesting that there might be a net heat flow even though there is no temperature difference. However, by looking at Eq.~(\ref{Eq:PowerTwoParticlesEnvironment}) it is clear that although the heat flux from object $1$ towards object $2$ is different from the heat flux from object $2$ to $1$ there is no net heat flow in global equilibrium because $\mathcal{P}_1 = \mathcal{P}_2 = 0$ in that case. The same is also true for $N$ objects where due to non-reciprocity one has in general $\mathcal{P}_{i \rightarrow j} \neq \mathcal{P}_{j \rightarrow i}$ ($i \neq j$). As we discuss below in more detail, this can result in a so-called {\itshape persistent heat current} in a $N$-body configuration in global thermal equilibrium~\cite{LZandSF2016,LZetal2018,AOetal2019b}.
%

In many works on the radiative heat exchange between two objects the contribution of the environmental field is neglected. In that case, as pointed out in \cite{ILandPBA2017}, the global equilibrium can only be achieved if the transmission coefficients fulfill the condition 
\begin{equation}
  \sum_{j \neq i} \bigl[\mathcal{T}_{ij} - \mathcal{T}_{ji} \bigr] = 0.
\label{Eq:ConditionTij}
\end{equation}
In particular, this implies that when having only two objects $\mathcal{T}_{12} = \mathcal{T}_{21}$. Hence, for two isolated objects the non-reciprocity has no impact and therefore at least three objects are necessary to observe for example a broken detailed balance. From this very general findings it can be understood that in \cite{LZandSF2014} it was necessary to consider three non-reciprocal thermal emitters to show that detailed balance can be broken for thermal radiation and in \cite{LZandSF2016} it was necessary to consider three non-reciprocal nanoparticles to observe the persistent heat current. On the other hand, it is also clear that the heat exchange between two non-reciprocal halfspaces will not show any rectification effect~\cite{MoncadaEtAl2015,FanEtAl2020}. On the other hand, when considering two objects with an environment, then the environment can be regarded as a third object. This explains why in general for only two objects in a given environment the transmission coefficients can be asymetric ($\mathcal{T}_{1 \rightarrow 2} \neq \mathcal{T}_{2 \rightarrow 1}$) so that we have here no contradiction to the discussion at the beginning of this paragraph.

\subsubsection{Magneto-optical nano-particles}
\label{MONnano}

In the following we will review the results obtained for the RHT in many-body systems consisting of subwavelength nano-particles. Most of the works are neglecting the coupling to the background which can be justified in steady-state situations when the distance between the objects is much smaller than the thermal wavelength so that the near-field coupling dominates over the coupling to the environment~\cite{RMetal2013}. Therefore, we will work with expression~(\ref{Eq:PowerDipolGeneralNoBG}) together with the transmission coefficients $\mathcal{T}_{ij}$ as defined in Eq.~(\ref{Eq:Transmissionij}). Neglecting the radiation correction it can also be
written as
\begin{equation}
  \mathcal{T}_{ij}(\omega) = \frac{4}{3}{\rm  Tr}\left[\alphamat^{-1}\boldsymbol{T}^{-1}_{ij}\frac{\alphamat-\alphamat^\dagger}{2 \ri}(\alphamat^{-1}\boldsymbol{T}^{-1}_{ij})^\dagger\frac{\alphamat-\alphamat^\dagger}{2 \ri}\right].
\label{Eq:Tij}
\end{equation}
assuming that all particles have the same polarizability $\alphamat$ defined for spherical nano-particles by means of the permittivity in Eq.~(\ref{Eq:quasistaticPolarizability}) with the volume $V = 4 \pi R^3/3$~\cite{ALetal1991}
\begin{equation}
  \alphamat = 4\pi R^3(\underline{\underline{\epsilon}}-\mathds{1})(\underline{\underline{\epsilon}}+2\mathds{1})^{-1}.
\label{Eq:AlphaSphere}
\end{equation}
The transmission coefficients in Eq.~(\ref{Eq:Tij}) are equal to the expressions given in~\cite{PBAetal2011,RMAEetal2017} for spherical nano-particles within the so called weak-coupled dipole limit~\cite{AL1992} where the radiation correction can be neglected~\cite{SAetal2010}. They can also be derived from the general $T$-operator expressions obtained within the scattering approach for the reciprocal~\cite{MKetal2012} and for the non-reciprocal case~\cite{LZetal2018,FHandSAB2019}. 

As already done in Sec.~\ref{sec-MO1} we consider again InSb as magneto-optical material for which the permittivity tensor becomes asymmetric $\uuline{\epsilon}^t \neq \uuline{\epsilon}$, i.e.\ the material properties are non-reciprocal, when a magnetic field is applied. As a consequence, the polarizability tensor then has the same asymmetry $\alphamat \neq \alphamat^t$. Furthermore, due to the applied field the three-fold degeneracy of the dipolar localized plasmon resonances, solution of the transcendental equation $det(\underline{\underline{\epsilon}}+2\mathds{1})=0$, with magnetic quantum number $m = 0,\pm 1$ is lifted~\cite{GWandDW2011,FPetal2013}. In particular, there is a red-shift of the resonance with $m = +1$ and a blue-shift of the resonance with $m = -1$. The size of the splitting is proportional to the cyclotron frequency $\omega_c = e B/m^*$, $m^*$ being the effective mass of electrons~\cite{GWandDW2011,FPetal2013}. To be more precise, in the regime where the dissipation can be neglected we find the resonances~\cite{AOetal2018} 
\begin{equation}
\begin{split}
  \omega_{m = \mp1} &= \sqrt{\left(\frac{\epsilon_\infty\omega_{\rm p}^2}{\epsilon_\infty+2}+\frac{\omega_{\rm c}^2}{4}\right)} \pm \frac{\omega_{\rm c}}{2}, \\
  \omega_{m = 0} &= \sqrt{\frac{\epsilon_\infty\omega_{\rm p}^2}{\epsilon_\infty+2}},
\end{split}\end{equation}
which are deterimined by the poles of the polarizability tensor. Therefore, for small magnetic fields the two circular resonances with $m = \pm 1$ are shifted by $\mp \omega_c/2$ with respect to the unaffected resonance for $m = 0$.

\subsubsection{Giant magneto-resistance}
\label{magnetoresistance}

\begin{figure}
	\includegraphics[width=\columnwidth]{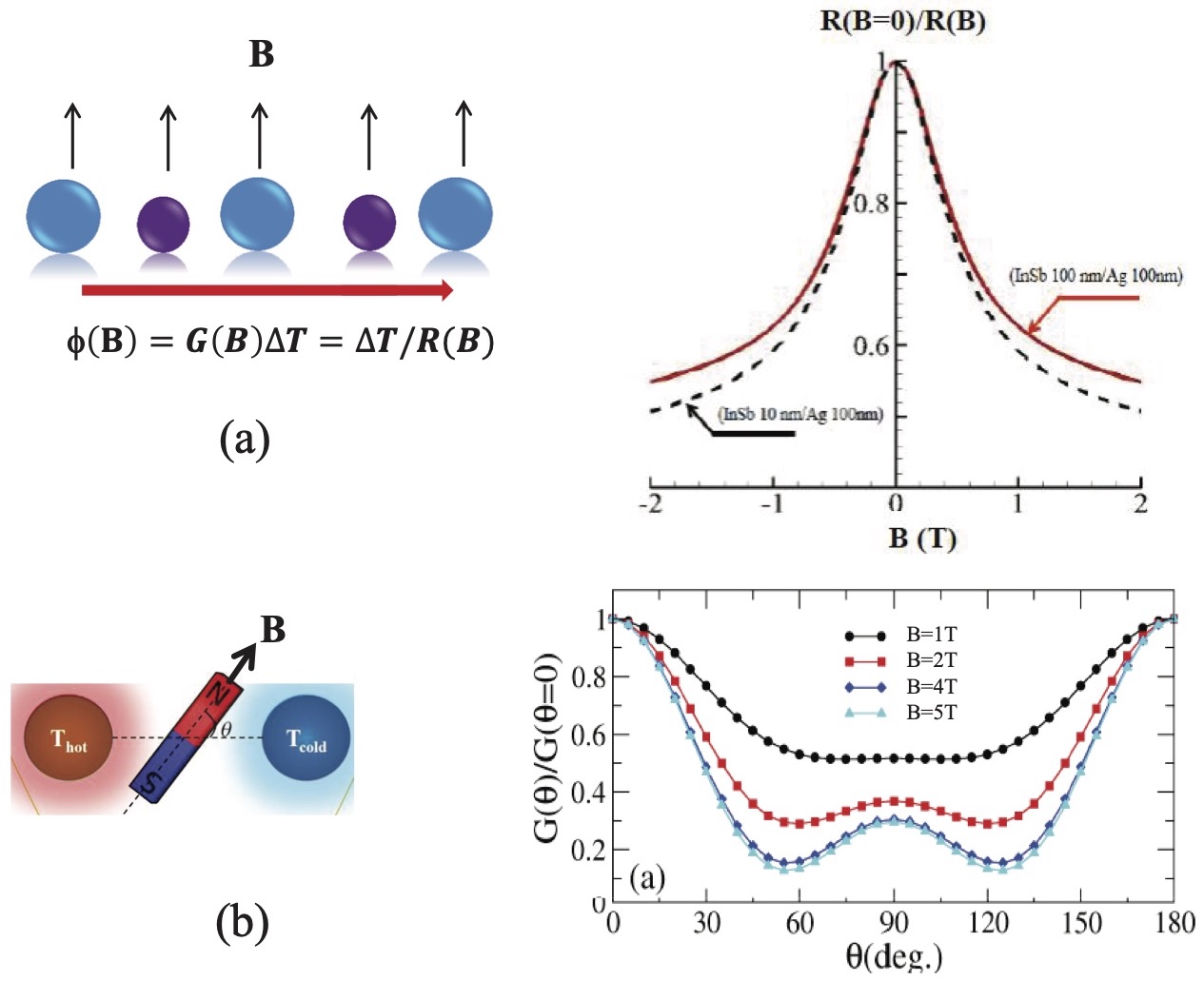}
    \caption{(a) Giant thermal magneto-resistance along linear chains of InSb and InSb/Ag nanoparticles at $T=300\,$K as a function of the strength of an external magnetic field $B$ applied in the direction orthogonal to the chain axis. From~\cite{ILandPBA2017}.
    (b) Anisotropic magneto-resitance between two InSb nanoparticles with respect to the orientation of magnetic field. From ~\cite{RMAEetal2018}.}
	\label{Magnetoresistance}
\end{figure}

Due to the strong dependence of dipolar resonances of particles on the 
magnetic field the heat flux emitted by a magneto-optical particle can drastically change by tuning this field \cite{ILandPBA2017,RMAEetal2018}. It turns out that the thermal magneto-resistance 
\begin{equation}
  R_{ij}(\mathbf{B})=\left(3 \int_{0}^{\infty}\frac{\rd\omega}{2\pi} \hbar \omega \frac{\partial n}{\partial T}\mathcal{T}_{ij}(\omega,\mathbf{B})\right)^{-1}\label{Eq:neq-resistance}
\end{equation}
between two particles in a many-body system is strongly dependent on the magnitude of applied magnetic field as it can be seen in Fig.~\ref{Magnetoresistance}(a). Variations of about 50\% along nanoparticle chains has been highlighted with magnetic fields of magnitude of about $500\,$ mT~\cite{ILandPBA2017}. This sensitivity to the magnetic field is of the same order of magnitude than the giant electric magneto-resistance reported in ferromagnetic/normal metal multilayers~\cite{Fert1988}. This resistance can also be tuned by changing the direction of applied magnetic field~\cite{RMAEetal2018}. In this case we speak of an anisotropic magneto-resistance. As shown in Fig.~\ref{Magnetoresistance}(b), for certain orientations of the magnetic field the heat flux can drop by more than 90\%. These effects open up the opportunity to control or modulate the amplitude of the heat flux between nano-particles by external means. A more detailed discussion can be found in ~\cite{ILandPBA2017,RMAEetal2018,AOetal2019b}.

\subsubsection{Persistent heat flux, angular momentum, spin and heat current}
\label{supercurent and spin}

As shown in~\cite{AOetal2018} the circular plasmonic resonances for $m = \pm 1$ of a single particle responsible for magnetic circular dichroism~\cite{FPetal2013} and ``inverse Faraday effect''~\cite{YGandKGK2010} are connected with a circular mean heat flux 
\begin{equation}
  \langle \mathbf{S} \rangle = \langle \mathbf{E}\times\mathbf{H} \rangle    
\end{equation}
emitted by the nano-particle in planes perpendicular to the applied magnetic field. This results in a certain spectral angular momentum density $\langle \mathbf{J}_\omega \rangle = \langle \mathbf{L} \rangle_\omega + \langle {\mathbf{S}_d} \rangle_\omega$ which can be divided in an orbital $\langle \mathbf{L}\rangle_\omega $ and spin angular momentum density $\langle {\mathbf{S}_{\rm d}} \rangle_\omega$ defined as~\cite{BliokhNori2015}
\begin{align}
     \langle \mathbf{L} \rangle_\omega &= \mathbf{r}\times\langle \mathbf{P} \rangle_\omega, \\
     \langle\mathbf{S}_d \rangle_\omega &= \frac{g}{2}{\rm Im} \left(\langle\mathbf{E}^*\times\mathbf{E} \rangle + \frac{\mu_0}{\epsilon_0} \langle \mathbf{H}^*\times\mathbf{H} \rangle\right),
\label{spin}
\end{align}
with $g = \epsilon_0 / \omega$ and the canonical spectral momentum density is given by
\begin{equation}
   \langle \mathbf{P} \rangle_\omega = \frac{g}{2}{\rm Im}[ \langle \mathbf{E}^*(\nabla)\mathbf{E} \rangle +\frac{\mu_0}{\epsilon_0} \langle \mathbf{H}^*(\nabla)\mathbf{H} \rangle], 
\label{kanimpuls}
\end{equation}
adopting the notations from~\cite{BliokhNori2015} that $\vec{X}(\vec{Y})\vec{Z} = \sum_{i}X_i\vec{Y}Z_i$. Using these definitions together with FE the persistent angular momentum close to the walls of a cavity was first evaluated and discussed in~\cite{MGS2017} and the angular momentum and spin for a thermally emitting nanoparticle by~\cite{AOetal2018}. A more detailed study of the angular momentum and spin close to a planar interface has been published recently~\cite{Khandekar2019}.

That there is a finite angular momentum and spin of the thermally emitted radiation is not surprising, because the Lorentz force constrains the electrons in the nanoparticles on a circular orbit so that the dipolar resonance is rotating in the plane perpendicular to the magnetic field which is the microscopic origin of the circular heat flux and the total angular momentum. The right-hand rule determines the direction of the circular heat flux in the near-field regime~\cite{AOetal2018}. It is interesting to note that the angular momentum of the $m = +1$ ($m = -1$) resonance is oriented in the (opposite) direction of the magnetic field as one would expect, whereas the spin of the $m = -1$ ($m = +1$) is oriented in the (opposite) direction of the magnetic field in the near-field regime. From this perspective the splitting of the $m = \pm 1$ resonances can also be understood as a Zeeman splitting, where $m = -1$ ($m = +1$) is blue-shifted (red-shifted) because the near-field direction of the spin is in direction (opposite) to the magnetic field, but of course the correct quantity determining the Zeeman splitting is the magnetic momentum of the dipolar resonance itself~\cite{YGandKGK2010}. The presence of a finite spin means that the thermal emission of the non-reciprocal nanoparticle will be circularly polarized in general as is well known for solid matter within a magnetic field like semi-conductors~\cite{Kollyukh2005} but also white dwarfs~\cite{Kemp1970,KempEtAl1970}, for instance. More recently, circularly polarized thermal emitters based on chiral meta-surfaces~\cite{DyakovEtAl2018} and nano-antennas~\cite{CKandZJ2018} have been proposed.

\begin{figure}
	\centering
	\includegraphics[width=0.35\textwidth]{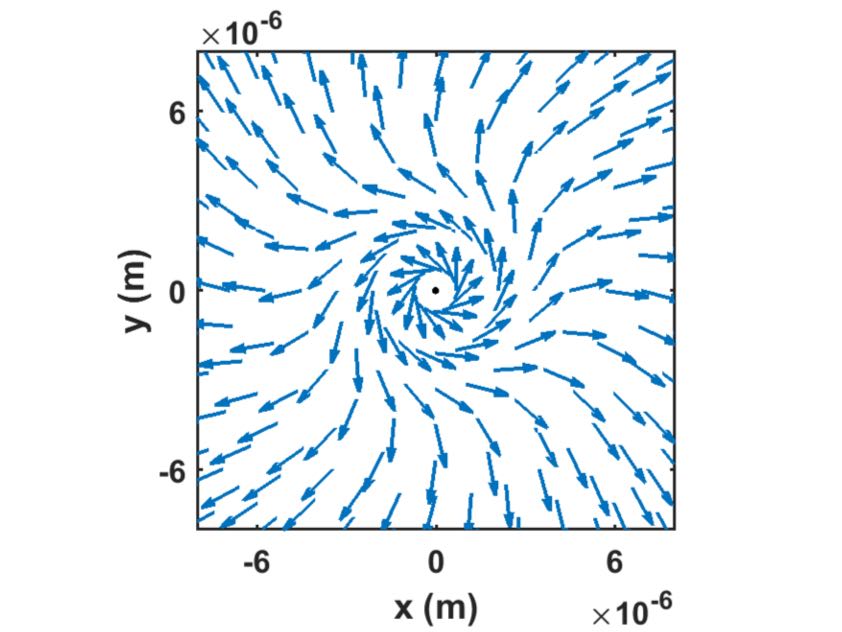}
	\caption{Normalized mean Poynting vector $\langle \mathbf{S} \rangle$ of thermal radiation emitted by an InSb nanoparticle at the origin of the coordinate system with radius of 300nm at a temperature of $300\,{\rm K}$ into a cold environment (vacuum) at $T_b = 0\,{\rm K}$ when a magnetic field is applied in $z$ direction. This circular heat flux persists in global thermal equilibrium. From~\cite{AOetal2018} }
	\label{CircularHeatFlux}
\end{figure}

Interestingly, it turns out that these three quantities, mean heat flux, orbital angular momentum, and spin persist in global equilibrium if $\uuline{\alpha} \neq \uuline{\alpha}^t$ and therefore is a direct consequence of the non-reciprocity of the permittivity or polarizability. Even though it might seem strange to have a non-zero mean heat flux in global equilibrium circulating around the nanoparticle, this does not pose any problem from the thermodynamical point of view, since it can be shown that $\nabla \cdot \langle \mathbf{S}_{\rm pers} \rangle = 0$, which means that there is no heat flux through any closed surface including the nanoparticle~\cite{AOetal2018}. In other words, no heat is finally emitted. Similar conclusions have been made for the thermal radiation field of the non-reciprocal surface modes on planar interfaces~\cite{MGS2017,Khandekar2019}.

\begin{figure}
	\centering
	\includegraphics[width=0.35\textwidth]{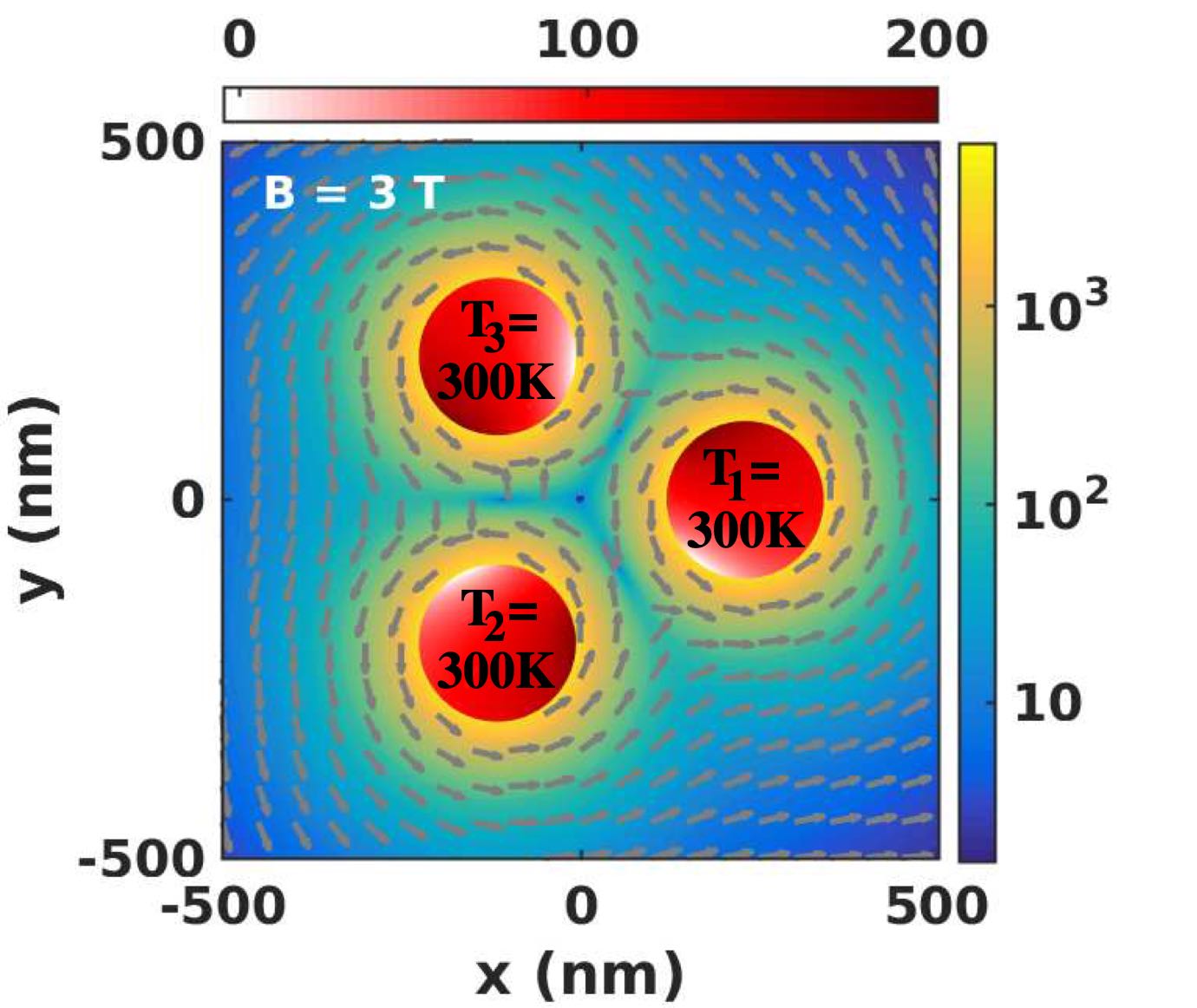}
	\caption{Normalized mean Poynting vector $\langle \mathbf{S} \rangle$ and its magnitude (${\rm W} {\rm m}^{-2}$ in color scale) of thermal radiation emitted by three InSb nanoparticles with a radius of 300\,nm having the same temperatures $T_1 = T_2 = T_3 = 300\,{\rm K}$ when a magnetic field is applied in $z$ direction. From~\cite{AOetal2019}.}
	\label{CircularHeatFlux3p}
\end{figure}

\begin{figure}
	\includegraphics[width=0.4\textwidth]{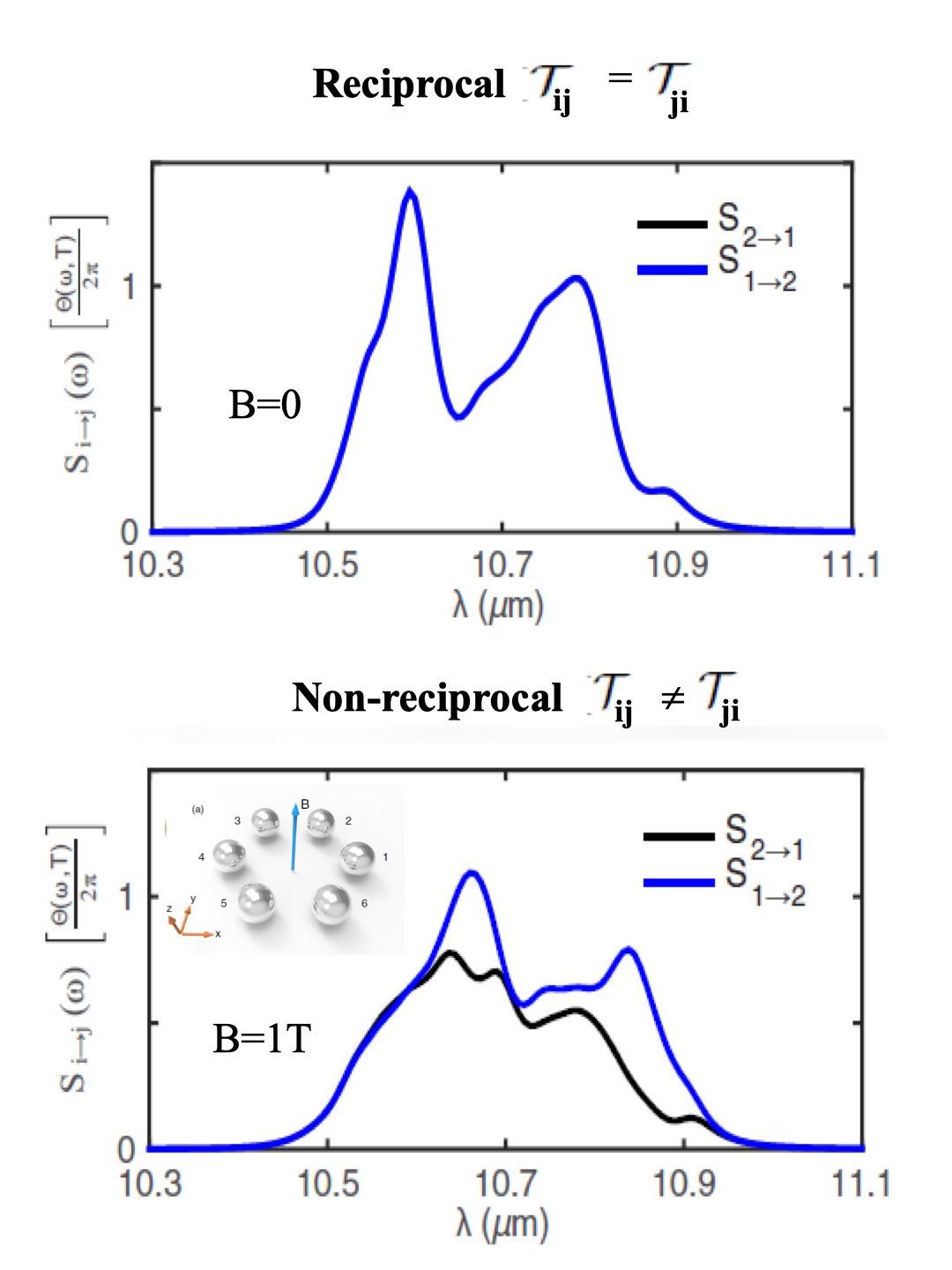}	
	\caption{Heat transfer spectra in a many-body system consisting of six InSb nanospheres placed at the vertices of a regular hexagon on $x$-$y$ plane without (reciprocal) and with (non-reciprocal) an externally applied magnetic field in the $z$-direction. From ~\cite{LZandSF2016,LZetal2018}}
	\label{supercurrent}
\end{figure}

Instead of a persistent heat flux, i.e.\ a non-zero heat flux in global thermal equilibrium, as observed from the mean Poynting vector around a non-reciprocal nanoparticle or in the vicinity of a planar interface of a non-reciprocal sample, there can also be a persistent heat current as first discussed in ~\cite{LZandSF2016} for the thermal radiation exchanged by three nanoparticles, but it of course exists also for more than three particles~\cite{LZetal2018}. As clear from the above discussion, when neglecting the contribution of the environment of the nanoparticles, then it follows from the constraint in Eq.~(\ref{Eq:ConditionTij}) that for only two nanoparticles $\mathcal{T}_{12} = \mathcal{T}_{21}$ and consequently $\mathcal{P}_{1\rightarrow2} = \mathcal{P}_{2\rightarrow 1}$ if $T_1 = T_2$. Therefore it is necessary to have at least three nanoparticles to have $\mathcal{T}_{12} \neq \mathcal{T}_{21}$. For three particles as in Fig.~\ref{CircularHeatFlux3p} the constraint in Eq.~(\ref{Eq:ConditionTij}) demands $\mathcal{T}_{12} = \mathcal{T}_{23} = \mathcal{T}_{31}$ and $\mathcal{T}_{13} = \mathcal{T}_{32} = \mathcal{T}_{21}$ due to the C$_3$ symmetry. If the three nanoparticles are now non-reciprocal then it can be shown from the definition of the transmission coefficient in Eq.~(\ref{Eq:Tij}) that $\mathcal{T}_{12} \neq \mathcal{T}_{21}$ and hence 
\begin{equation}
  \mathcal{P}_{1 \rightarrow 2} = \mathcal{P}_{2 \rightarrow 3} = \mathcal{P}_{3 \rightarrow 1} \neq  \mathcal{P}_{1 \rightarrow 3} = \mathcal{P}_{3 \rightarrow 2} = \mathcal{P}_{2 \rightarrow 1}.
\end{equation}
This means there is a clockwise heat flow exchanged by the nanoparticles which is different from the heat flow in counter-clockwise direction even if $T_1 = T_2 = T_3$ and therefore there is a persistent heat current in clockwise or counterclockwise direction depending on which of the two heat flows is larger. This persistent heat flow or better heat current~\cite{LZandSF2016} is the many-body analogue of the persistent heat flux, which of course also exists in the three-body configuration. 
It is worthwhile to note from relation (\ref{Eq:ConditionTij}) that, in a non-reciprocal system at temperature $T$, the body $i$ and $j$ still exchange a power~\cite{ILandPBA2017}
\begin{equation}
\begin{split}
   \mathcal{P}^{eq}_{i \leftrightarrow j} &= \mathcal{P}^{eq}_{j \rightarrow i} - \mathcal{P}^{eq}_{i \rightarrow j} \\
   &= \int_{0}^{\infty}\frac{\rd\omega}{2\pi}\,\hbar\omega\, n(\omega,T)[\mathcal{T}_{ij}-\mathcal{T}_{ji}]
   \label{Eq:equilibriumHeatFlux},
\end{split}
\end{equation}
although the net power $\mathcal{P}^{eq}_j=\sum_{i \neq j}\mathcal{P}^{eq}_{i \leftrightarrow j}$ vanishes so that the persistent heat flux does not lead to any heating or cooling. Hence the magnitude of asymmetry of transmission coefficients spectra (Fig.~\ref{supercurrent}) and the value of the equilibrium temperature are directly responsible for the value of persistent current. Today, the measurement of this current is still a challenging problem. 
Recently, a setup has been proposed in Ref.~\cite{Khandekar2019} which might be able to access it in the vicinity of a magneto-optical planar sample.

\subsubsection{Hall effect for thermal radiation}

The asymmetry in the exchanged heat flux in many-body configurations observed in global equilibrium, i.e.\ the persistent heat current, has directly measurable consequences when driving the system out of global equilibrium. An astonishing consequence is the Righi-Leduc or Hall-effect for thermal radiation~\cite{PBA2016}. Classically, the Righi-Leduc effect~\cite{Righi,Leduc} is just the thermal analogue of the Hall-effect~\cite{Hall}. When applying a temperature difference in a metallic sample together with a magnetic field the heat current by the electrons will be deflected due to the Lorentz-force acting on the electrons such that a temperature difference perpendicular to the initially applied temperature difference will build up in steady state. Such an effect has also been highlighted for other heat carriers in solids like magnons and spinons~\cite{Fujimoto2009,KatsuraEtAl2010,OnoseEtAl2010} or even phonons~\cite{Inyushkin2007,StrohmEtAl2005}. 

\begin{figure}
	\includegraphics[width=0.35\textwidth]{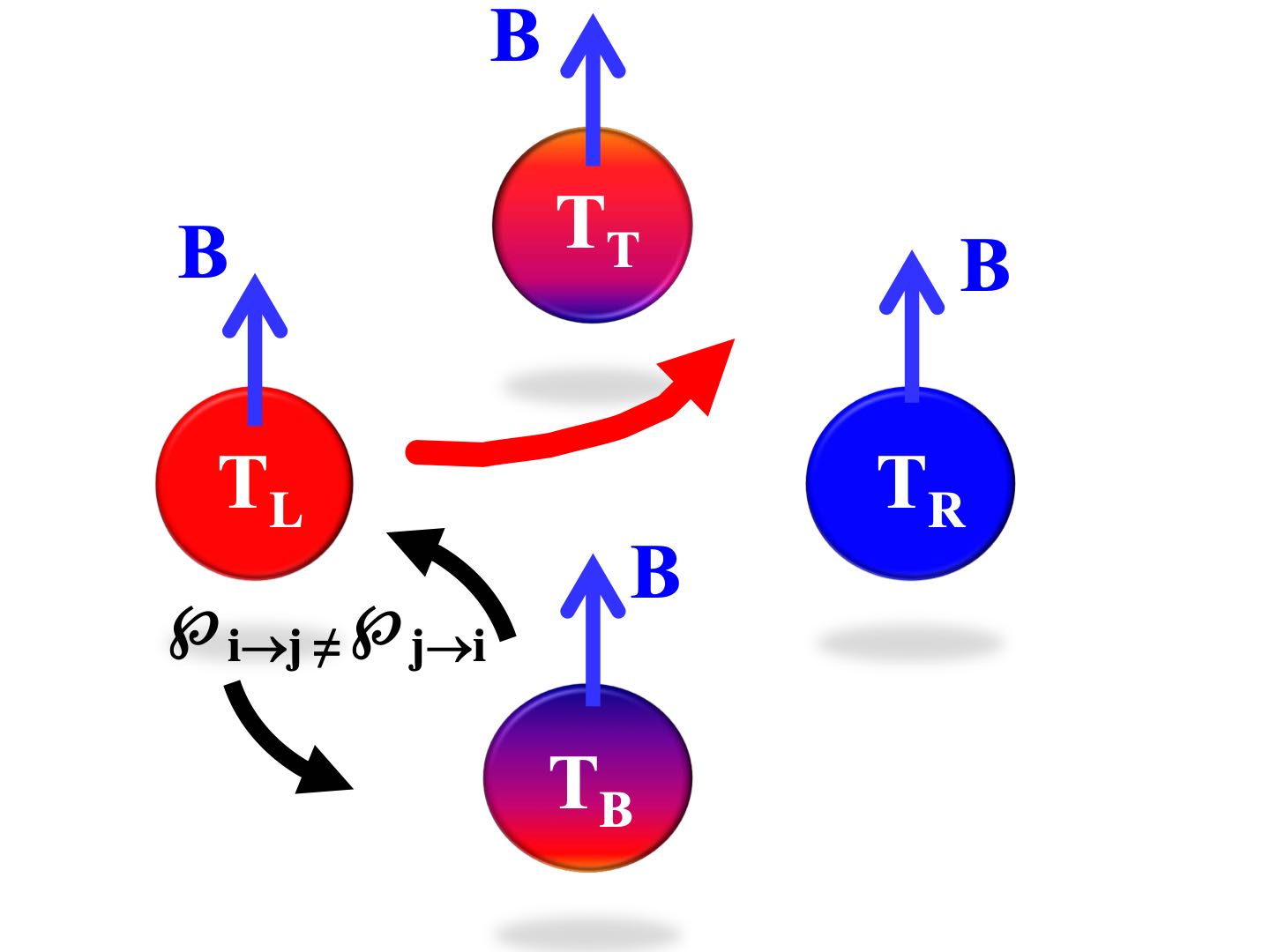}
     \caption{Photon thermal Hall effect: a four terminal junction with magneto-optical particles forming a square with C$_4$ symmetry is submitted to an external magnetic field $\mathbf{B}$ in the direction orthogonal particle plane. When a temperature gradient $\Delta T=T_L - T_R$ is applied between the particles $L$ and $R$, a Hall flux transfers heat transversally between particles $B$ and $T$, thus bending the overall flux (red arrow) towards the top or the bottom. In this case the heat $\mathcal{P}_{i\rightarrow j}$ and $\mathcal{P}_{j\rightarrow i}$ exchanged between two particles $i$ and $j$ is not symmetric.}
	\label{Halleffectconf}
\end{figure}

\begin{figure}
	\centering
	\includegraphics[width=0.35\textwidth]{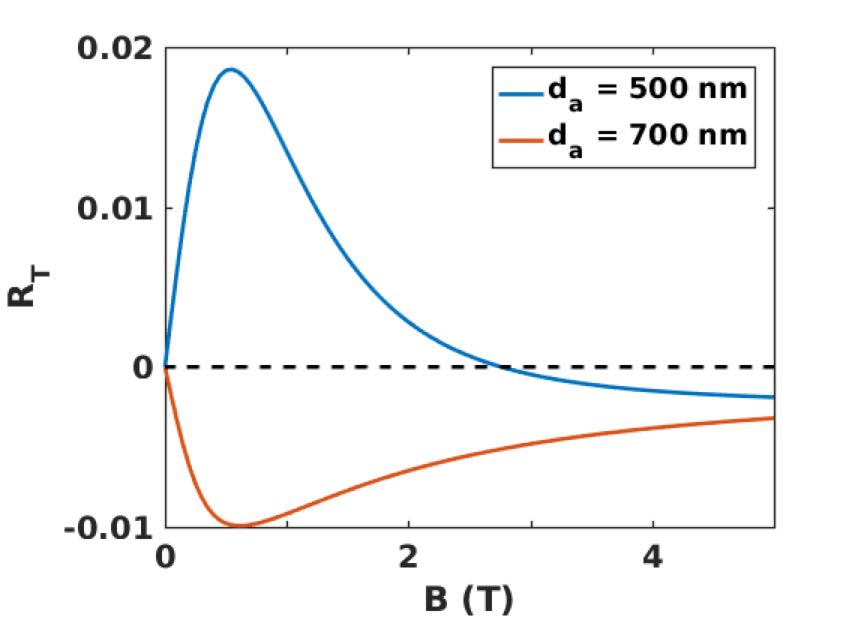}
	\caption{Magnetic field strength dependence of the Righi-Leduc-like coefficient defined in Eq.~(\ref{Eq:RighiLeducCoeff}) for four spherical InSb nanoparticles with a radius of 100nm in a C$_4$-symmetric configuration as depicted in Fig.~\ref{Halleffectconf} choosing an interparticle distance of opposite particles of $d_a = 500\,{\rm nm}$ and $d_a = 700\,{\rm nm}$. From~\cite{AOetal2019}.}
	\label{RighiLeducCoeff}
\end{figure}

Now, when considering heat radiation exchanged between four nanoparticles in a C$_4$ symmetric configuration as in Fig.~\ref{Halleffectconf}, and applying a temperature difference $\Delta T = T_L - T_R$ between particle $L$ (left) and $R$ (right), then in the steady state of the system a temperature difference $T_B^{(st)} - T_T^{(st)}$ between particle $B$ (bottom) and $U$ (up) can build up when using non-reciprocal InSb nano-particles and applying a magnetic field perpendicular to the particle plane. Hence, one observes a Righi-Leduc or Hall effect for thermal radiation~\cite{PBA2016}. Again, the effect can be understood by the Lorentz force acting on the electrons in the nanoparticles. However, here the electrons do not serve as the heat carriers but introduce a circular heat flux leading to an asymmetric heat flow and finally to the Righi-Leduc effect. Its magnitude and directionality can be measured by the relative Hall temperature difference or Righi-Leduc-like coefficient
\begin{equation}
    R_T = \frac{T_B^{(st)} - T_T^{(st)}}{T_L - T_R},
    \label{Eq:RighiLeducCoeff}
\end{equation}
which is shown in Fig.~\ref{RighiLeducCoeff}. Written in terms of thermal conductances, this coefficient reads~\cite{PBA2016}
\begin{equation}
 R_T = \frac{G_{31}G_{42}-G_{41}G_{32}}{\underset{j\neq 3}{\sum}G_{3j}\underset{j\neq 4}{\sum}G_{4j}-G_{34}G_{43}}
	\label{Eq:relative2}.
\end{equation}
It can be seen that depending on the magnitude of the magnetic field the effect will change its directionality and there is a maximum for a magnetic field amplitude of about $0.5\,{\rm T}$ for the considered configuration.The effect is not very strong and high field amplitudes are needed to have a maximum effect. However it highly depends on the configuration and material parameters~\cite{AOetal2019} and therefore its magnitude can certainly be optimized by changing the spatial distribution of particles or their physical properties. 
To date an experimental proof of photon thermal Hall effect remains a challenging problem. However a direct measurement of the Hall temperature difference with measurements of electrical resistance variations with a very high accuracy~\cite{SGetal2014} in magneto-optical nanowires networks seems feasible.


Beside the ``normal'' thermal Hall effect, anomalous effects also called anomalous thermal Hall effects~\cite{Ferreirosetal2017,Huangetal2020} thermal analog of anomalous Hall effect~\cite{Karplus1954,Nagaosaetal2010} have also been described for the heat transport with electron or phonons in ferromagnetic materials and in semimetals. Very recently a similar effect in Weyl semi-metal nanoparticles networks for thermal photons has been predicted~\cite{AOetAl2020}. Since the Weyl semi-metals can exhibit a strong nonreciprocal response in the infrared, this effect allows for a directional control of heat flux by simply locally tuning the magnitude of temperature field without changing the direction of temperature gradient. 

\subsubsection{Heat flux rectification with non-reciprocal surface waves}
\label{rectification}

For most of the non-reciprocal effects discussed so far the environment does not play a decisive role. Now, instead of using only the intrinsical non-reciprocal properties of the nanoparticles to achieve a directional heat flux, also the non-reciprocity of the environment can be exploited as first shown in \cite{AOetal2019b}. As we have seen in Sec.~\ref{Sec:Guiding} before, the heat flux between two nanoparticles or more generally between two objects brought in close vicinity to an interface of a sample can be enhanced by transporting the heat via the surface modes of the interface~\cite{KSetal2014,KAetal2017,JDetal2018,RMetal2018,YZetal2019,HeEtAl2019}. If the material properties of the planar sample are non-reciprocal then the presence of a magnetic field will affect the surface modes~\cite{Chiu1972}. 

\begin{figure}
	\centering
	\includegraphics[width=0.45\textwidth]{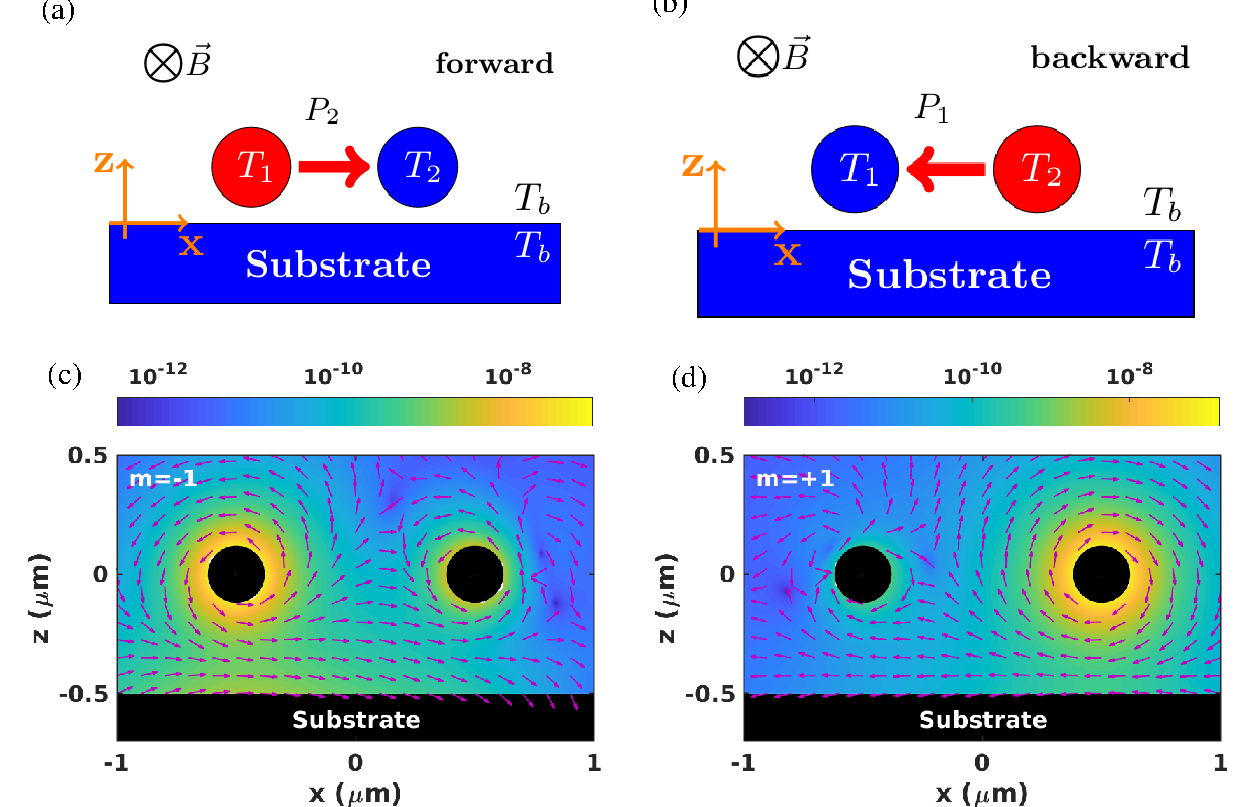}
	\caption{(a) Sketch of the diode in forward direction. Two InSb nanoparticles above an InSb substrate. The left particle is heated with respect to the other particle and the environment. (b) Sketch of the diode in backward direction. (c) Normalized mean spectral in-plane Poynting vector and its amplitude (${\rm J} {\rm m}^{-2}$, colorbar) for the $m = +1$ particle resonance for the diode in forward direction. (d) as in (c) but for the backward case and for the $m = -1$ particle resonance. See also~\cite{Ott2020}.}
	\label{DiodeNonRec}
\end{figure}

To be more specific, within the Voigt configuration as in Fig.~\ref{DiodeNonRec}(a) and (b) the dispersion relation for the surface modes at the interface of the substrate traveling to the right and left will be different~\cite{Chiu1972}. Similar to the localized mode inside an InSb nanoparticle the degeneracy of the surface modes for $k_x > 0$ and $k_x < 0$ is lifted and there is a splitting of the surface mode resonance frequency~\cite{Chiu1972}. Since the spin associated with the surface modes~\cite{Bliokh2012} shows a spin momentum locking~\cite{TvM2016}, meaning that the waves with $k_x > 0$ and $k_x < 0$ have a different spin direction, the splitting can again be understood as a Zeeman splitting~\cite{TvM2016,Khandekar2019}. 

\begin{figure}
	\centering
	\includegraphics[width=0.4\textwidth]{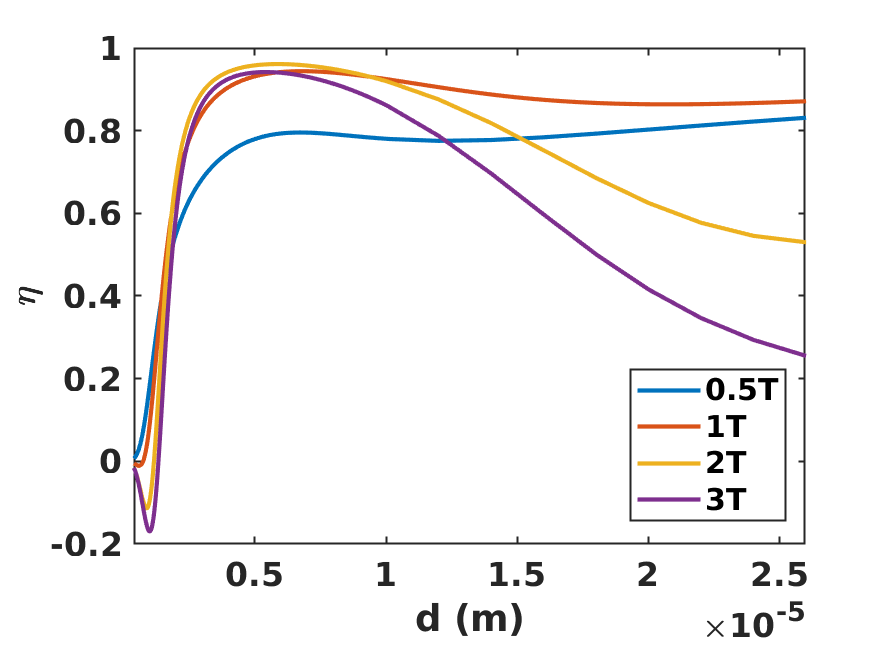}
	\caption{Rectification coefficient from Eq.~(\ref{Eq:RectCoeff}) for two InSb nanoparticles with 100\,nm radius in 500\,nm distance above an InSb substrate as sketched in Fig.~\ref{DiodeNonRec} as a function of the interparticle distance $d$ for different magnetic field amplitudes. See also~\cite{AOetal2019b,Ott2020}.}
	\label{RectCoeff}
\end{figure}

Now, considering the situation in Fig.~\ref{DiodeNonRec}(a) and (b) the thermally excited localized modes of the hot nanoparticle can directly couple to the localized modes of the cold nanoparticle leading to a direct heat transfer between the particles. The thermally-excited localized modes of the hot particle can couple to the surface modes of the substrate, travel along the interface of the substrate and then couple to the localized modes of the cold nanoparticle so that in this case the heat is transferred between the two nanoparticles via the surface modes. Due to the non-reciprocity of the substrate the heat flow $\mathcal{P}_2$ in the forward direction in Fig.~\ref{DiodeNonRec}(a) and the heat flow $\mathcal{P}_1$ in the backward direction in Fig.~\ref{DiodeNonRec}(b) will be different, leading to a rectification effect~\cite{AOetal2019b}. A detailed analysis shows~\cite{Ott2020} that there is a spin-selective coupling so that the localized modes couple preferably to the surface modes with the spin in the same direction. For example, the $m = -1$ ($m = +1$) resonance couples preferably to the surface modes with $k_x > 0$ ($k_x < 0$) providing the main heat flux channel in forward (backward) direction as shown in Fig.~\ref{DiodeNonRec}(c) (Fig.~\ref{DiodeNonRec}(d)). This can be also understood by a matching of the circularity of the particle resonances and the directionality of the interface resonances. The resulting rectification coefficient
\begin{equation}
    \eta = \frac{\mathcal{P}_1 - \mathcal{P}_2}{\mathcal{P}_1}
    \label{Eq:RectCoeff}
\end{equation}
shown in Fig.~\ref{RectCoeff} can be rather high even for relatively small magnetic fields. It should be kept in mind that when bringing the nanoparticles close to a substrate most of the heat will go to the substrate rather than to the other nanoparticle. Nonetheless, the rectification effect can result in a measurable heating of the cold nanoparticle~\cite{Ott2020}. 

\section{Outlook and open questions}

While the heat transport mechanisms mediated by thermal photons in 1D and 3D systems have been intensively studied during the last decade they remain today unknown in 2D systems. Can we observe a diverging radiative conductivity with respect to system size as has already been predicted for the phononic conductivity in 2D anharmonic lattices? To answer this question and also identify different heat transport regimes in these systems, the scaling laws of radiative thermal conductance must be analyzed. Another fundamental problem is the crossover from 1D to 2D and from 3D to 2D systems. The spatial confinement of evanescent photons in these systems should play a key role in those transitions. 

So far, dense many-body systems and effects like weak and strong
localization for thermal radiation remain largely unexplored. In these
strongly correlated systems, heat is typically carried through
multiple connected channels associated with different heat carriers
like electrons, phonons, and photons, which raises the question: under
which conditions can one or more of these heat carries dominate heat
transport? As hilighted in the introduction of this review, progress
in unifying various transport mechanisms is beginning to be made, yet
a complete theory capable of describing multichannel heat exchange in
large many-body systems remains a challenge for understanding possible
transport effects associated with coupling across such different
channels.

 As the number of bodies in interaction becomes large, the general formalism described in this review becomes numerically prohibitive. This is a serious issue to investigate heat transport in many-body systems in presence of long range interactions. A continuous description of heat transport in these systems could make the study of these systems feasible and it could in the same time be a powerful tool to study the NFRHT in mesoscopic physics or to make calculations of NFRHT between objects of arbitrary shape. Using the Chapman-Kolmogorov equation for the local temperature field, a Fokker-Planck equation can be derived and written in the hydrodynamic limit as an advection-diffusion equation which depends on directly measurable macroscopic quantities like the effective diffusion coefficient and which could be easily solved with standard numerical methods.
 
When it comes to recent exploration of the spin and angular momentum
of thermally fluctuating fields, nearly all investigations have
focused on single particles or semi-infinite materials. However, a
corresponding general N-body theory should be straight forwardly
derived using the general framework presented in this review. This
extension could pave the way to studies of thermal-field spin and
angular momentum transport in atomic and molecular systems. Since
magneto-optical effects based on the use of magneto-optical materials
or Weyl semi-metals reported thus far have been relatively small,
further studies aimed at enhancing these effects should be considered
in the future, for instance by exploiting ferromagnetic or more
strongly magnetic materials.

Non-Hermitian physics has attracted tremendeous interest during the last decade from a variety of fields in classical physics due to their mathematical equivalence with the Schrodinger equation, thus allowing one to mimic non-Hermitian wave physics with classical systems. Bipartite plasmonic and phonon-polaritonic many-body systems provide a natural platform to investigate such physics. Among their many peculiarities, one might point to the existence of original topological states that give rise to Berry-like phases and which may lead to the development of new materials such as topological insulators. These states and their consequences for the thermal management (active control of heat flux, heat pumping, heat flux focusing) remain largely unexplored in many-body systems.

Out-of equilibrium thermodynamics of many-body systems and its
connections with information theory is also a future field of
investigation. In systems with long-range interactions, the classical
thermodynamic theory fails to describe the evolution of state
variables since they cannot be sequenced in small independent
parts. Normally, to calculate thermodynamic properties it is necessary
to determine the microscopic states of a given system. However a
phenomenological approach analogous to Landau's transition theory may
be employed to study the thermodynamic behavior of these systems by
considering macroscopic quantities. Hence, mechanisms such as phase
transitions in magneto-optical systems could be investigated by
analyzing the dependence of quantities like the thermal conductance or
the entropy flux with order parameters such as the magnitude or
orientation of a magnetic field.

The peculiarities of heat transfer in many-body systems has given rise to numerous development in the emerging field of thermotronics to manipulate heat flux in analogy with electric currents in electric circuits. This radical change of paradigm opens the way to a new generation of devices for active thermal management, innovative wireless sensors using heat as their primary source of energy, and to “low-electricity” technologies capable of information processing. In these devices, infrared emission coming from various systems (people, machines, electric devices…) may for instance be captured by active thermal components to launch a sequence of logical operations in order to either control the heat propagation (modulate, amplify, split), trigger specific actions (opto-thermo-mechanical coupling with MEMS, thermal energy storage…) or even process information. Hence the development of thermal logical circuits such as neural networks could open the door to a low-power and even zero-power communication technology for the Internet of Things, allowing machine-to-machine communication with heat. The design of thermal metamaterials such as thermal insulators, topological insulators or superdiffusive solids is also a promising challenge.

Finally, building experimental platforms based on multi-tip SThM setups, suspended membranes or even networks of electromechanical systems interacting at the nanometre-scale is one of the most important challenges for the next few years to measure the NFRHT in many-body systems, prove all already predicted effects and develop operational devices. In order to be able to have an access to conductance variations of few nWK$^{-1}$, high-sensitive heat flux sensors must be developed. This will require fabrication of thermometers working at the nanoscale and able to measure temperatures with an accuracy $< 10 \,$mK.

\begin{acknowledgments}
We thank all our colleagues within the nanoscale heat transfer community that engaged us in many fruitful interactions and spirited discussions. S.-A.\ B. acknowledges support from Heisenberg Programme of the Deutsche Forschungsgemeinschaft (DFG, German Research Foundation) under the project No. 404073166. J.C.C. acknowledges funding from the Spanish Ministry of Economy and Competitiveness (MINECO) (Contract No. FIS2017-84057-P). P. B.-A. Acknowledges support from Natural Sciences and Engineering Research Council of Canada through the RGPIN-2017 program-No.05445 and from the Agence Nationale de la Recherche in France through the ComputHeat project ANR-19-MRS1-0009.
\end{acknowledgments}

\bibliography{NFMB}

\end{document}